\documentclass[11pt]{elsarticle}
\usepackage{lineno,hyperref}
\usepackage[T1]{fontenc}
\usepackage[shortlabels]{enumitem}
\modulolinenumbers[5]
\usepackage{float}
\usepackage{graphicx}
\usepackage{subfig}
\usepackage{float}
\usepackage{amsmath}
\usepackage{amsfonts}
\usepackage{xcolor}
\usepackage[subnum]{cases}
\journal{Transportation Research Part C: Emerging Technologies}
\usepackage[letterpaper, portrait, margin=1in]{geometry}

\usepackage{xcolor}

\usepackage{newtxtext,newtxmath}

\hypersetup{
    colorlinks,
    linkcolor={red!50!red},
    citecolor={blue!50!blue},
    urlcolor={blue!80!blue}
}
\usepackage{color}

\newcommand{\bc}[1]{{\color{red}}}

\makeatletter
\let\@afterindenttrue\@afterindentfalse
\makeatother

\makeatletter
\newcommand*{\centerfloat}{%
  \parindent \z@
  \leftskip \z@ \@plus 1fil \@minus \marginparwidth
  \rightskip \leftskip
  \parfillskip \z@skip}
\makeatother

\newcommand{\tabincell}[2]{\begin{tabular}{@{}#1@{}}#2\end{tabular}}  
\usepackage{setspace}
\usepackage{natbib}
\usepackage{algorithm}
\usepackage[noend]{algpseudocode}
\algdef{SE}[DOWHILE]{Do}{doWhile}{\algorithmicdo}[1]{\algorithmicwhile\ #1}

\usepackage{appendix}

\usepackage{array}
\usepackage{threeparttable}
\usepackage{multirow}
\usepackage{rotating}
\usepackage{makecell}

\biboptions{authoryear}

\begin{document}
\begin{frontmatter}

\title{Competition between Shared Autonomous Vehicles and Public Transit: A Case Study in Singapore}


\author[label1]{Baichuan Mo}
\author[label2]{Zhejing Cao}
\author[label4,smart]{Hongmou Zhang}
\author[label3]{Yu Shen}
\author[label4]{Jinhua Zhao\corref{mycorrespondingauthor}}
\address[label1]{Department of Civil and Environmental Engineering, Massachusetts Institute of Technology, Cambridge, MA 02139}
\address[label2]{School of Architecture, Tsinghua University, Beijing, China 100084}
\address[label4]{Department of Urban Studies and Planning, Massachusetts Institute of Technology, Cambridge, MA 02139}
\address[smart]{Future Urban Mobility IRG, Singapore--MIT Alliance for Research and Technology Centre, Singapore 138602}
\address[label3]{Key Laboratory of Road and Traffic Engineering of the Ministry of Education, Tongji University, Shanghai, China 201804} 
\cortext[mycorrespondingauthor]{Corresponding author}
\ead{jinhua@mit.edu}

\begin{abstract}
Emerging autonomous vehicles (AV) can either supplement the public transportation (PT) system or compete with it. This study examines the competitive perspective where both AV and PT operators are profit-oriented with dynamic adjustable supply strategies under five regulatory structures regarding whether the AV operator is allowed to change the fleet size and whether the PT operator is allowed to adjust headway. Four out of the five scenarios are constrained competition while the other one focuses on unconstrained competition to find the Nash Equilibrium. We evaluate the competition process as well as the system performance from the standpoints of four stakeholders---the AV operator, the PT operator, passengers, and the transport authority. We also examine the impact of PT subsidies on the competition results including both demand-based and supply-based subsidies. A heuristic algorithm is proposed to update supply strategies for AV and PT based on the operators' historical actions and profits. An agent-based simulation model is implemented in the first-mile scenario in Tampines, Singapore. We find that the competition can result in higher profits and higher system efficiency for both operators compared to the status quo. After the supply updates, the PT services are spatially concentrated to shorter routes feeding directly to the subway station and temporally concentrated to peak hours. On average, the competition reduces the travel time of passengers but increases their travel costs. Nonetheless, the generalized travel cost is reduced when incorporating the value of time. With respect to the system efficiency, the bus supply adjustment increases the average vehicle load and reduces the total vehicle kilometer traveled measured by the passenger car equivalent (PCE), while the AV supply adjustment does the opposite. The results suggest that PT should be allowed to optimize its supply strategies under specific operation goals and constraints, and AV operations should be regulated to reduce their system impacts, including potentially limiting the number of licenses, operation time, and service areas, which makes AV operate in a manner more complementary to the PT system. Providing subsidies to PT results in higher PT supply, profit, and market share, lower AV supply, profit, and market share, and increased passenger’s generalized cost and total system PCE.

\end{abstract}

\begin{keyword}
shared mobility-on-demand system; autonomous vehicles; public transportation; agent-based simulation; market competition; Nash Equilibrium.
\end{keyword}

\end{frontmatter}


\section{Introduction}\label{intro}
The emergence  of autonomous vehicles (AV) is expected to influence the future urban transportation systems in many different aspects such as traffic flow stability \citep{talebpour2016influence}, network congestion \citep{fagnant2015preparing}, land use patterns \citep{koh2012dawn}, and road safety \citep{zhang2016routing}. As a critical component of urban transportation, public transit (PT) systems cannot avoid the impact of AV. Due to the uncertainty of how the AV systems may evolve, many different scenarios of AV--PT interactions have been proposed \citep{lazarus2018shared}. Some studies argue that AVs will compete with the PT systems \citep{levin2015effects, chen2016management} or even replace them \citep{mendes2017comparison}, while others are optimistic about the AV--PT integration, stating that they could be complementary to each other \citep{lu2017data, wen2018transit}. 



AV--PT integration has been studied under various operation and regulation scenarios. \citet{salazar2018interaction} proposed a tolling scheme for an AV--PT intermodal system, which shows a significant reduction in travel time, cost, and emissions compared to a single-mode scenario. \citet{shen2018integrating} and \citet{wen2018transit} envisioned the transit-oriented autonomous mobility-on-demand (AMoD) systems in Singapore and London. 
However, the integrated AV--PT system requires significant regulatory interventions and operators’ willingness to collaborate. In many contexts, the AV and PT operators will likely behave as competitors by default. There are fewer studies on the competitive perspective of the relationship. Some prior research only evaluated the effect of AV on PT but did not specify the competition mechanisms between them \citep{levin2015effects, childress2015using}. Recently agent-based models (ABM) have been used to investigate the competition. Both \citet{liu2017tracking} and \citet{mendes2017comparison} argued that traditional PT services may not survive once the shared AV services become available; but they only considered the AV--PT competition to be a single-shot process---how PT systems are affected when AV is introduced---rather than a repeated process where both AV and PT can adjust their supplies dynamically during the competition. 


Our study examines the competitive perspective where both AV and PT operators are profit-oriented with dynamic adjustable supply strategies. In classic economics, market competition can ideally lead to optimal resource allocation. However, due to the common market failures in the transportation industry, regulations are often imposed, or even a certain level of cooperation is mandated. We examine five regulation scenarios (Table \ref{tab_scenarios}), which can be categorized by a two by two matrix regarding whether the AV operator is allowed to change the fleet size and whether the PT operator is allowed to adjust headway. The impact of different PT subsidies on the competition results of five scenarios is also evaluated. Other important regulation aspects, such as service fare, pricing, and bus routes, are assumed to be fixed across all five scenarios. 

\begin{table}[!htp]
\renewcommand{\arraystretch}{1.2}
\caption{Regulation scenarios examined in this paper}\label{tab_scenarios}
\footnotesize
\centering
\begin{tabular}{cc|c|c}
\hline
\multicolumn{2}{c|}{\multirow{2}{*}{\textbf{Regulation Scenarios}}} & \multicolumn{2}{c}{\textbf{PT supply}} \\ \cline{3-4} 
\multicolumn{2}{c|}{} & \multicolumn{1}{c|}{\textbf{Fixed}}   & \textbf{Adjustable}  \\ \hline
\multicolumn{1}{c|}{\multirow{2}{*}{\textbf{AV supply}}} & \textbf{Fixed}  & Both-fixed              & {PT-adjustable-only}                  \\ \cline{2-4} 
\multicolumn{1}{c|}{}                                      & \textbf{Adjustable} & {AV-adjustable-only}                 & Both-adjustable / NE                 \\ \hline
\end{tabular}
\end{table}

\begin{itemize}
    \item The ``\emph{Both-fixed}'' scenario corresponds to a fully-regulated transportation market without competition \citep{shen2018integrating}. The supplies, fare structure, route network, and service areas are all set by transport authorities. Neither AV nor PT is not allowed to change their supply.
    \item The ``\emph{AV-adjustable-only}'' scenario allows AV to adjust supplies but with all other operation parameters (e.g. price) fixed. The PT operators cannot adjust service headways. This is close to the situation where the ride-sourcing services, such as Uber and Lyft, are operated independently from transit agencies. 
    \item The ``\emph{PT-adjustable-only}'' scenario allows PT operators to adjust the service headways to maximize their profits. This gives the PT operators even more flexibility than the well-known ''Scandinavian'' or ''London'' organization models \citep{costa1996organisation,van1999organisational}, where the transport authority sets the service goals and then contracts out transit services to private operators.
    \item When both operators are allowed to adjust supplies, we have two different scenarios: ``\emph{Both-adjustable}'' and ``\emph{Nash Equilibrium}'' (\emph{NE}). These two scenarios have the weakest regulation compared to the other three and are close to the situation in the UK (except for Greater London) \citep{wilson1991organizational}, where all operators are private and compete against each other. The difference between these two scenarios can be looked at from two aspects: 1) Conceptually, the \emph{both-adjustable} scenario can be seen as a ``constrained'' competition while \emph{NE} ``non-constrained'' competition. \emph{Both-adjustable} scenario allows both PT and AV to adjust supplies according to a ``constrained'' profit-maximization strategy. Specifically, due to incomplete market information and the regulation from the government, the supply updating at each step may not be optimal (see Section \ref{sim_platform} for details). Therefore, the system may not converge to the NE. However, for the \emph{NE} scenario, we use the iterated best response (IBR) algorithm to find the NE. A NE is a profile of strategies such that each strategy is the best response to the rest of the profile. That is, each agent maximizes its own utility given the strategies played by the others. The supply updating at each iteration is the agent's best response to the environment and the competitor's strategy. 2) Algorithmically,  since AV and PT's actions at each step are not optimal in the \emph{both-adjustable} scenario, the \emph{both-adjustable} scenario can be seen as a premature convergence result of IBF due to the non-optimal responses of competitors. 
    
\end{itemize}

We selected the first-mile connection to the subway station \citep{lesh2013innovative} in Tampines, Singapore as the study area where there are multiple bus feeder services, and implemented an agent-based model to simulate the competition process between AV and PT including a heuristic algorithm to update the AV and PT supplies based on the operators' historical actions and profits. We evaluated the competition between AV and PT as well as the system performance from the standpoints of four stakeholders---the AV operator, the PT operator, passengers, and the transport authority. 

The remainder of this paper is organized as follows. Section \ref{lit} reviews the literature and identifies the research gaps. Section \ref{dgn} introduces system design, agent behavior, and performance metrics. Section \ref{res} presents the results, and Section \ref{con} concludes the paper.

\section{Literature Review}\label{lit}

AV is expected to reshape the future transportation system from different perspectives. Many advantages of AV, such as absolute compliance, expanded service hours, reduced labor forces, and human errors make it an efficient mode of urban transportation, which enables the reduction of operating costs and serves the travel demand for fewer fleet sizes \citep{de2008demand, alonso2017demand, spieser2014toward}.

Given the promising application of AV, recent studies have examined the impact of AV in different aspects, such as traffic flow \citep{talebpour2016influence}, road congestion \citep{fagnant2015operations}, land use \citep{koh2012dawn}, and road safety \citep{zhang2016routing}. \citet{talebpour2016influence} used a micro-simulation model to evaluate the impact of connected and autonomous vehicles on traffic flow and showed that AV can improve string stability and help prevent shockwave formation and propagation. \citet{azevedo2016microsimulation} applied an integrated agent-based micro-simulation model to design and evaluate the impact of AV on people's travel behavior. They found that the new AV technology can change people's travel patterns, in terms of mode shares, route choices, and activity destinations. \citet{fagnant2014travel} indicated that each shared AV can replace approximately 11 conventional vehicles, but it adds up to 10\% more travel distance than comparable non-shared AV trips. There are also empirical studies in Lisbon \citep{martinez2017assessing}, Toronto \citep{kloostra2019fully}, and New Jersey \citep{zhu2017interplay}, which reveal that the AV technology can reduce traveled vehicle kilometers and emissions \citep{martinez2017assessing}, average travel time \citep{kloostra2019fully}, and the number of vehicles on the road  \citep{martinez2017assessing}.

However, despite the large volume of recent research on the impact of AV, research focusing on the AV--PT interaction is limited. Two major relationships between AV and PT have been discussed in the literature.

First, AV and PT can be complementary and integrated. AV and PT can be cooperative in a way that AV is integrated as a part of the PT system for social welfare. In line with this assumption, \citet{ruch2018autonomous} used a simulation model to analyze whether AV can substitute the rural PT lines. They found such a service can operate at a lower cost and with a higher service level when the PT lines are short and are underutilized. Similarly, \citet{shen2018integrating} designed an AV--PT integration system with AMoD as an alternative to low-demand bus routes, which showed that the integrated system has the potential to enhance service quality, occupying fewer road resources and being financially sustainable. Another cooperative scenario is the multi-modality between AV and PT. For example, \citet{yap2016preferences} conducted a stated preference survey to analyze the use of automated vehicles as the egress mode of train trips. They found that AV is a good alternative for first-class train passengers. \citet{liang2016optimizing} used an integer programming model to study AV as a last-mile connection to train trips, which showed that using automated taxis can reduce the pick-up cost and improve profits. \citet{vakayil2017integrating} proposed a hybrid transit system where AMoD serves as the first- and last-mile feeder for the subway. The results show that an integrated system can provide up to 50\% reduction in total vehicle miles traveled. These studies provide sufficient analysis toward the AV--PT integration relationship. The co-operation strategies for AV and PT are also designed \citep{salazar2018interaction}.

Second, AV can be competitive with traditional PT when AV is operated privately and competes for market share and profit. The analysis of this scenario is very limited. Using a multiclass four-step model, \citet{levin2015effects} found that the transit ridership will decrease when AV is introduced. Similarly, \citet{childress2015using} used an activity-based model to evaluate the impact of AV, which revealed a reduction in PT's mode share. \citet{liu2017tracking} simulated the impact of shared AV based on a case study in Austin, Texas, and showed that conventional PT services may not survive once the shared AV services become available. \citet{mendes2017comparison} developed an event-based simulation model to compare the shared AV services with the proposed light rail services in New York City and found that AV is more cost-efficient in providing the same level of service. \citet{basu2018automated} applied an activity-driven agent-based simulation approach to test the impact of AV on mass transit and found that AMoD indirectly acts as a replacement to mass transit. 

Understanding how AV competes with PT is critical for managing future PT services and developing sound intervention on the private transportation market. However, existing studies addressing the AV--PT competition focused on the static interaction process, i.e., they only considered what will happen when AV is introduced as a one-shot game. However, as competitors, it is highly likely that AV and PT will dynamically adjust the operation strategies in the market as repeated games. These interactive dynamics have not been considered in the literature. Moreover, most existing studies only evaluate the AV's impact on the mode share of PT, without a comprehensive assessment of the cost and benefit of different stakeholders in the system, including operators, passengers, and transport authorities. 

In this study, we used an agent-based model to simulate the competition between AV and PT, with both parties trying to increase their profits. We also comprehensively evaluated the system from different stakeholders' perspectives, including the financial conditions, the level of service, vehicle kilometer traveled (VKT), and transport efficiency, aiming to fill the research gap in the literature. 

\section{System design} \label{dgn} 
We used the Tampines Planning Area in Singapore as the study area to apply our AV--PT competition model. We focused on the first-mile trips heading to the Tampines Mass Rapid Transit (MRT) station from surrounding residential blocks. Walking and buses are currently the dominant modes for first-mile trips \citep{mo2018impact}. Bus service in Singapore is highly regulated by the Land Transport Authority (LTA), which is responsible for its fare structure and route design. The LTA contracts out bus services to different bus operators on a five-year basis. Note that as walk and bus are currently dominant travel modes, AV may have a great potential to increase its ridership and compete with PT in this first-mile market.  

\subsection{Basic assumptions}
Based on the characteristics and the operational structure of the Singapore transport system, we make the following assumptions for the AV--PT competition model.
\begin{enumerate}[(a)]
\item System-level
\begin{itemize}
\item Travel mode: Before AV entering the market, walking, bus, and ride-hailing are the only travel modes available for the first-mile trips. After AV emerges, ride-hailing will be replaced by the AMoD. This assumption is based on the Singapore autonomous vehicle initiative policies \citep{LTA2017Selfdriving}. 

\item Traffic conditions: We perform a mesoscopic simulation; microscopic features such as signal system and driving behavior are not considered in this study. As AVs and buses in the first-mile market only account for a small proportion of traffic flows on roads, time-dependent exogenous congestion is considered. The speeds of AVs and buses on each road for different times of day are pre-obtained from the Google Map API.  

\item Demand and supply: The spatial and temporal distributions of travel demand are assumed to be fixed for all simulated days. The supply of AV and PT can be adjusted for competition. The fixed demand assumption holds when the residential population and infrastructures do not change substantially in an area (especially within a short-time period). It can reflect the early stage of the market when AV is introduced. 

\item Information: The AMoD and PT's supply information is complete and in real-time for passengers' mode choice decisions. 

\end{itemize}
\item Agent-level
\begin{itemize}
\item Ownership: PT and AMoD are both operated by private enterprises but under government regulation, i.e., they are in a \emph{constrained} competition  (but \emph{NE} scenario can be seen as an unconstrained competition).

\item Objective: PT and AMoD are able to adjust their supply to improve their profits.

\item Fare: PT and AMoD's fare structures are fixed by the government and cannot be changed.

\item Operating cost and subsidies: PT and AMoD's operating cost is proportional to their fleet size and driving distance. There is no operation subsidy for PT based on the current Singapore policy \citep{kuang2018lessons}. However, we test several hypothetical scenarios with different levels of subsidies to enrich the discussion. Details are shown in Section \ref{scenario_set} and \ref{subsidy}.

\item Constraints: PT has a lower bound and an upper bound for the bus headway. AMoD has an upper bound for the maximum fleet size. 

\item Dynamic intraday supply: AMoD and PT's supply can be different in different time periods in a day.

\item Supply updating frequency: AMoD can update the next day's supply strategy at the end of each day. PT can only update its supply strategy after a sufficiently long time (30 days in this study; more discussion in Section \ref{sim_platform}). For the \emph{NE} scenario, there is no definition of supply updating frequency (see Section \ref{sim_NE} for details). 

\end{itemize}
\end{enumerate}

To summarize, the AMoD updates the supply daily by adjusting the vehicle fleet size in each time interval of the next day and PT adjusts its supply by changing the headway for each bus route in different time intervals in a month. 
 
Price adjustment is also a common practice to increase profit in market competition. However, to isolate the impact of the supply change, we only focus on supply in this study, assuming that the prices are fixed for both PT and AMoD. Moreover, given the price regulation history of the LTA, it is very likely that the AMoD will be regulated on the fare structure to avoid price wars \citep{PTC2019bill}. Future research can explore a combined model incorporating supply change and pricing.

\subsection{Study area}
\begin{figure}[htb]
\centering
\subfloat[Study area]{\includegraphics[width=0.5\textwidth]{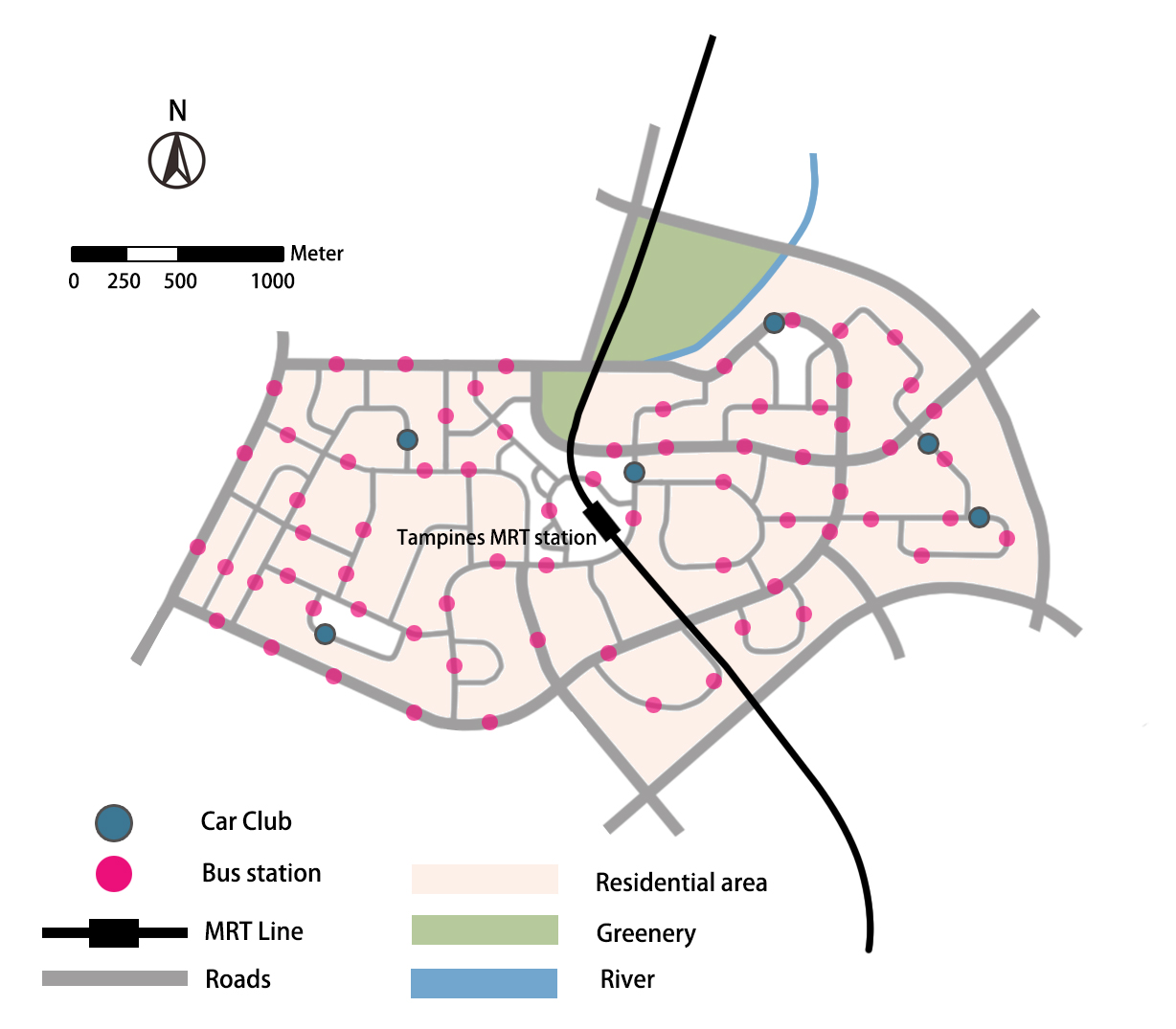}\label{fig_study_area}}
\hfil
\subfloat[Time-of-day distribution of first-mile demand to Tampines MRT station]{\includegraphics[width=0.8\textwidth]{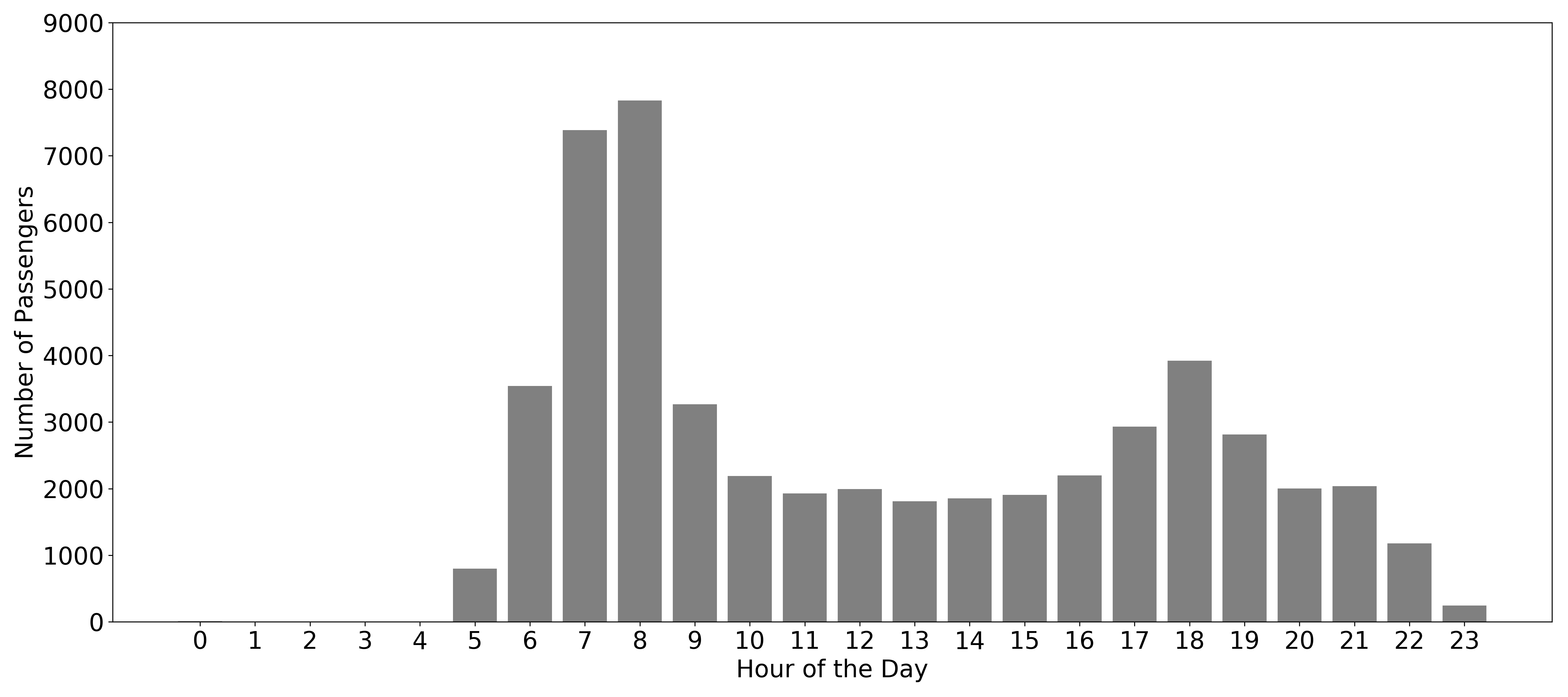}\label{fig_demand}}
\caption{Study area and demand}
\label{fig_PT_before_after_supply}
\end{figure}

Tampines is a 6.86-$\mbox{km}^2$ mixed residential and commercial area located in eastern Singapore (Figure \ref{fig_study_area}). It is centered around the Tampines MRT station, which is one of the busiest MRT stations in Singapore \citep{HDB}. Fifty-one bus stops serve Tampines, with 26 bus routes connecting to the MRT station. Three of them are dedicated feeder routes to the subway. The other 23 are passing-by routes. All 26 routes are included in the simulation model. 
We chose Tampines for the case study to illustrate the feasibility of our model because 1) it possesses a significant first-mile demand to the MRT station, and 2) it houses a large volume of bus supply, which provides a non-trivial testbed for competition analysis.

The first-mile travel demand was obtained from the transit smart card data. The dataset covers all PT trips in August 2014. A normal workday in August 2014 was selected as the study date. Since the smart card data in Singapore include both tap-ins and tap-outs, the accurate date and time of each entry and exit activity are recorded, as well as the boarding and alighting stops/stations, which allows us to extract complete first-mile demand by counting passengers who tap into the Tampines MRT station. The temporal distribution of first-mile travel demand on the study day is shown in Figure \ref{fig_demand}. A total of 51,850 passengers entered the Tampines MRT station on the study day.

\subsection{Stakeholders and system evaluation} \label{eval}
We evaluated the interests of multiple stakeholders in the AV--PT competition. Table \ref{tab_stakeholder} identifies four key stakeholders (passengers, AMoD operators, PT operators, and transport authority) and their evaluation indicators.

The main interests of passengers are the level-of-service and modal choice. The level-of-service indicators include travel cost, total travel time, waiting time, and generalized travel cost ---calculated as the sum of walking time, waiting time, and riding time multiplied by the corresponding value of time (VOT). It is a comprehensive value that incorporates both travel time and travel costs. The VOTs were derived from the estimated choice model in Table \ref{tab_mixed_logit}. In terms of modal choice, the number of travelers choosing walking, buses, and AV was recorded, respectively. 

The interests of AMoD and PT operators, based on our profit-orientation assumption, are financial viability and supply. The financial viability indicators include operating cost, revenue, profit, and market share, while the supply indicates the number of AV/bus supplied. 

For the transport authority, efficiency and vehicle kilometer traveled were considered. The average load per vehicle is one of the indicators of transport efficiency, which is calculated using the total passenger travel distance divided by the total vehicle travel distance. We considered the vehicle kilometers traveled (VKT) and total passenger car equivalent (PCE). The unit PCE for a bus was set as 3.5 in this study \citep{ahuja2007development}.

These indicators were recorded in the simulation process, which shows the entire changing profile during the competition between AV and PT. The time of the day distribution of some indicators was also recorded.

\begin{table}[htb]
\caption{Evaluated indicators of stakeholders}
\label{tab_stakeholder}
\centering
\begin{threeparttable}
\footnotesize
\begin{tabular}{c|c|c} \hline
\textbf{Stakeholder} & \textbf{Interests} & \textbf{Indicators} \\ \hline
\multirow{7}{*}{Passengers} & \multirow{4}{*}{Level of service} 
 & Travel cost \\ 
 & & Total travel time \\ 
 & & Waiting time \\
 & & Generalized travel cost \\\cline{2-3}
 & \multirow{3}{*}{Mode choice} & Walking demand \\
 & & Bus demand\\ 
 & & AV demand \\ \hline
\multirow{5}{*}{AMoD Operators} & \multirow{4}{*}{Financial viability} 
 & Operation cost \\ 
 & & Revenue \\ 
 & & Profit \\
 & & Market share \\\cline{2-3}
 & \multirow{1}{*}{Supply} & Average number of AV provided per hour\\\hline
 \multirow{5}{*}{PT Operators} & \multirow{4}{*}{Financial viability} 
 & Operation cost \\ 
 & & Revenue \\ 
 & & Profit \\
 & & Market share \\\cline{2-3}
 & \multirow{1}{*}{Supply} & Number of bus dispatched per day\\\hline
  \multirow{5}{*}{Transport Authority} & \multirow{2}{*}{Transport efficiency} 
 & AV average load per vehicle \\ 
 & & Bus average load per vehicle \\\cline{2-3}
 & \multirow{3}{*}{Vehicle kilometer traveled} & AV VKT\\
 & & Bus VKT \\ 
 & & Total PCE (AV and PT) \\
 \hline
\end{tabular}
\end{threeparttable}
\end{table}

\subsection{Agent behavior}
The overall simulation system is composed of three main types of agents: buses, AMoD vehicles, and passengers. The PT system is derived from \citet{shen2018integrating}'s study, which has been calibrated and validated with real-world data. Here we give an overview of the behavior of the agents, with detailed formulation given in \ref{agent_behavior}.

\begin{itemize}
    \item Passengers' mode choices are based on mixed logit models using trip-specific variables (waiting time, in-vehicle time, and monetary cost) and individual-specific variables (household income); if a passenger chooses the bus or AMoD, a maximum waiting time is considered to avoid a shortage of services. The passenger changes to other modes if the maximum waiting time is reached
    \item AMoD dispatching is based on the first-come-first-serve principle upon passengers' requests. The vehicles follow the shortest paths from pick-up locations to the MRT station. When there is no empty vehicle available, the dispatching center will then look for vehicles with passengers who agree to share their trips. Every day, the AMoD operator updates the fleet size of each time period of the day to increase its profit.
    \item Buses follow the existing routes given by the LTA. Each month, the operator \emph{for each bus route} independently optimizes its profit by changing the headway of each time period of the day.
\end{itemize}

\subsection{Competition process}\label{sim_platform}
An agent-based simulation model is used to perform the AV PT competition process. There are four sets of input variables/parameters: simulation setting parameters $\boldsymbol{\theta}$, initial supply strategies of the bus $\boldsymbol{S}_B^{(0)}$, initial supply strategies of AMoD $\boldsymbol{S}_A^{(0)}$, and the first-mile demand $\boldsymbol{D}$. The detailed notations and values used are given in Table \ref{tab_notation}. The values of these constants were chosen based on the Singapore context (details in \ref{agent_behavior}).

The supply, profit, and supply changing units are all time-specific for AV, which means that for different days and different time periods they may have different values. For buses, these variables are time- and route-specific. The pseudo-code for the competition process (\emph{both-adjustable} scenario) is shown in Algorithm \ref{alg_simulation}.

\begin{table}[p]
\caption{Notations and values of system parameters}\label{tab_notation}
\centering
\centerfloat
\begin{threeparttable}
\begin{tabular}{c| c| c} \hline
\textbf{Categories} & \textbf{Parameters/Variables} & \textbf{Value} \\ \hline
\multirow{20}{*}{\tabincell{l}{Simulation setting \\parameters $(\boldsymbol{\theta})$}} 
 & Duration of simulation  ($T$) & 365 days \\
 & Supply changing unit reduced factor  ($\gamma$) & 0.75 \\
 &  Passenger choice lag factor ($\alpha$) & 0.5 \\
 & Passenger mode choice parameters ($\boldsymbol{\beta}$) & See Table \ref{tab_mixed_logit} \\
 & Passenger maximum AV waiting time & 10 min \\
 & Passenger maximum bus waiting time & 30 min \\
 & Passenger ride-sharing-agreement rate & 50\% \\
 & AV supply time interval ($i_A$) & 1 hour \\
 & AV supply updating frequency ($N_A$) & 1 day \\
 & AV fixed operation cost & 4 SGD/hour$\cdot$veh \\
 & AV variable operation cost & 0.12 SGD/km\\
 & AV base distance of fare ($d_b$) & 1 km \\
 & AV base fare ($f_b$) & 3.4 SGD \\
 & AV distance-based fare ($f$) & 0.22 SGD/400 m \\
 & AV detour discount rate of fare ($\lambda$) & 2 \\
  & AV supply lower bound ($S_A^L$) & 0\\
    & AV supply upper bound ($S_A^U$) & 150\\
  & AV initial supply changing unit (${C}_A^{\text{ini}}$)& 10 vehicles \\
 & Bus supply time interval ($i_B$) & 2 hour \\
 & Bus supply updating frequency ($N_B$)& 30 days  \\
 & Bus operation cost & 2.71 SGD/km \\ 
 & Bus dwell time & 30 sec \\
 & Bus fare & 0.77 SGD/trip\\
 & Bus headway upper bound ($S_B^U$)& 2100 sec\\
 & Bus headway lower bound ($S_B^L$) & 210 sec\\
 & Bus initial headway changing unit (${C}_B^{\text{ini}}$)& 3 min \\
 \hline

  \multirow{3}{*}{Bus supply $\boldsymbol{S}_B$} 
   & Bus headway of route $l$ in time interval $t$ on day $d$ ($S_B^{l,t,d}$) & Intermediate\\
 & Bus headway arrangement on day $d$ ($\boldsymbol{S}_{B}^{(d)} = [S_B^{l,t,d}]_{l,t}$)
 & Intermediate  \\
 & Bus initial headway arrangement ($\boldsymbol{S}_B^{(0)}$)& See \ref{pt_behavior} \\
  \hline
 
 \multirow{3}{*}{AV supply $\boldsymbol{S}_A$} 
  & Number of AV supplied in time interval $t$ on day $d$ ($S_A^{t,d}$) & Intermediate\\
 & AV supply strategies on day $d$ ($\boldsymbol{S}_{A}^{(d)}=[S_A^{t,d}]_t$)
 & Intermediate  \\
 & AV initial supply strategies ($\boldsymbol{S}_A^{(0)}$)& See \ref{av_behavior} \\
 \hline
  \multirow{1}{*}{Demand $\boldsymbol{D}$} 
 & Demand of first-mile trips $\boldsymbol{D}$
 & See Figure \ref{fig_demand}  \\
  \hline
 
 \multirow{8}{*}{\tabincell{l}{Supporting \\variables}} 
 &AMoD profit in time interval $t$ on day $d$ ($p_A^{t,d}$) & Intermediate\\
 &AMoD profit vector on day $d$ ($\boldsymbol{p}_A^{(d)}=[p_A^{t,d}]_t$) & Intermediate\\
  & AV supply changing unit in time interval $t$ for day $d$ (${C_A^{t,d}}$) & Intermediate \\
 & AV supply changing unit vector on day $d$ ($\boldsymbol{C}_A^{(d)}=[C_A^{t,d}]_t$) & Intermediate \\

 & Bus profit of route $l$ in time interval $t$ on day $d$ ($p_B^{l,t,d}$)& Intermediate\\
  & Bus profit vector day $d$ ($\boldsymbol{p}_B^{(d)} = [p_B^{l,t,d}]_{l,t}$) & Intermediate\\

  & Bus headway changing unit of route $l$ in time interval $t$ for day $d$ (${C_B^{l,t,d}}$) & Intermediate \\
 & Bus headway changing unit vector on day $d$ ($\boldsymbol{C_B^{(d)}} = [C_B^{l,t,d}]_{l,t}$) & Intermediate \\
 \hline
 
\end{tabular}
\begin{tablenotes}\footnotesize
\item[*] Values of ``intermediate'' means the intermediate variables in the model.
\end{tablenotes}
\end{threeparttable}
\end{table}

\begin{algorithm}[htb]
\caption{AV-PT competition (\emph{both-adjustable} scenario)} \label{alg_simulation}
\begin{algorithmic}[1]
\Procedure{AV-PT competition}{$\boldsymbol{\theta}$, $\boldsymbol{S}_B^{(0)}$, $\boldsymbol{S}_A^{(0)}$, $\boldsymbol{D}$}
\State Initialize $\boldsymbol{p}_A^{(0)}=0$, $\boldsymbol{p}_B^{(0)}=0$, $\boldsymbol{S}_A^{(1)}=\boldsymbol{S}_A^{(0)}$, $\boldsymbol{S}_B^{(1)}=\boldsymbol{S}_B^{(0)}$
\State Initialize ${\boldsymbol{C}_A^{(1)}}={C}_A^{\text{ini}}$, ${\boldsymbol{C}_B^{(1)}}={C}_B^{\text{ini}}$
\State Let day counter $d = 0$
\While{$d < T$}
\State $d = d+1$
\State $\boldsymbol{p}_A^{(d)}, \boldsymbol{p}_B^{(d)}=
\textsc{One-Day-Sim}(\boldsymbol{S}_B^{(d)}, \boldsymbol{S}_A^{(d)}, \boldsymbol{D} \mid \boldsymbol{\theta})$
\State $\boldsymbol{S}_A^{(d+1)}, \boldsymbol{C}_A^{(d+1)}=\textsc{SupplyUpdate}(\boldsymbol{p}_A^{(d)},  \boldsymbol{p}_A^{(d-1)}, \boldsymbol{S}_A^{(d)}, \boldsymbol{C}_A^{(d)})$
\If {$d \mod N_B==0$}
\State $\boldsymbol{S}_B^{(d+1)}, \boldsymbol{C}_B^{(d+1)}=\textsc{SupplyUpdate}(\boldsymbol{p}_B^{(d)},  \boldsymbol{p}_B^{(d-N_B)}, \boldsymbol{S}_B^{(d)}, \boldsymbol{C}_B^{(d)})$
\State Initialize $\boldsymbol{C}_A^{(d)}={C}_A^{\text{ini}}$
\Else
\State $\boldsymbol{S}_B^{(d+1)} = \boldsymbol{S}_B^{(d)}$
\State $\boldsymbol{C}_B^{(d+1)} = \boldsymbol{C}_B^{(d)}$
\EndIf
\EndWhile
\State \Return the system evaluation indicators (Table \ref{tab_stakeholder})
\EndProcedure
\end{algorithmic}
\end{algorithm} 

As shown in Algorithm \ref{alg_simulation}, the overall framework aims to simulate the interaction between AV and PT over time period $T$. $\textsc{One-Day-Sim}$ is a pseudo function of running the simulation for a single day given the supply, demand, and $\boldsymbol{\theta}$, which is considered as the engine of the simulation model. In this function, each agent will follow the behaviors mentioned in \ref{agent_behavior}. 

The profits of AMoD and PT can be obtained after running this function. The profit, supply, and supply change unit are then used as the input of function $\textsc{SupplyUpdate}$ (shown in Algorithm \ref{alg_supplyupdate}), deriving the updated supply strategies and new supply change units for AV and PT. Since we assume that the AV supply updating frequency is daily, this function is executed for AV every day, while it is executed for PT every $N_B$ days, considering the limited flexibility of PT planning. 

In practice, the PT schedules do not adjust frequently. A reasonable period should be more than three to six months, which allows the transit operators to notify the passengers and gives people enough time to adapt to the new schedule. Based on our numerical test, however, given a specific bus schedule, the system will stabilize in one or two weeks when only the AMoD updates its supply. Therefore, any values of $N_B$ that are greater than two weeks are equivalent because the system state will not change after two weeks. In this study, $N_B = 30$ days was used to simulate the long-period updating frequency of the bus. The system evaluation indicators (Table \ref{tab_stakeholder}) were recorded during the simulation and returned at the end of the model.

It is worth noting that Algorithm \ref{alg_simulation} is designed for the \emph{both-adjustable} scenario. When AV or PT are not allowed to update their supplies, Algorithm \ref{alg_simulation} is revised with the corresponding supply updating steps removed. The IBR algorithm for estimating NE (i.e. the \emph{NE} scenario) is different from Algorithm \ref{alg_simulation} and thus elaborated in Section \ref{sim_NE}.

In terms of supply updates, we proposed a heuristic algorithm, which applied to both AMoD and PT. Specifically, for AMoD, at the beginning of a day, the algorithm determines the AV fleet size for each hour in the day. For PT, at the beginning of a month, the algorithm determines the PT headway for each route and each hour for all days in the month. 

In this study, we assume the profits across different time periods and routes are independent, which leads to a simple univariate optimization problem for each route and time period combination. For example, $p_B^{l,t,d}$, the profit of bus route $l$ at time interval $t$ on day $d$ only depends on the headway of route $l$ at time interval $t$ on day $d$. Then, we can adjust the supply of $l$ in the same time interval $S_B^{l,t,d}$ to improve the corresponding profit as a single-variable optimization problem. Similarly for AMoD, if the profit $p_A^{t,d}$ is greater than $p_A^{t,d-1}$, the last change in supply $C_A^{t,d-1}$ will lead to the profit increase. Thereupon, we can continue increasing the fleet size of the same time period on the next day. 

If the change in profits between two supply strategies becomes sufficiently small (the threshold was set to be 5\%), we reduce the size of the changing unit by $\gamma$ (i.e., $C^{t,d+1} = \gamma \cdot C^{t,d+1}$). This setting ensures the convergence of the results. 

We acknowledge that the heuristic algorithm is a simplification for the profit maximization problem because the profits of different time periods and routes are likely to be dependent on each other, especially when the routes have overlapping stations. Nonetheless, we apply this simplified algorithm for the following reasons: 1) The concept of this algorithm is in line with the reality, where information is incomplete and every adjustment is a trial of a new supply strategy based on previous experience; 2) This heuristic algorithm indeed yields an improved profit, which satisfies the initial purpose of adjusting the supply by the operators; and 3) Capturing the dependency will require a much more complicated optimization problem, which we suggest as a future research topic. 

We implemented the model on AnyLogic 8.1. The simulation is executed for $T=365$ days to ensure that the competition process converges. 

\begin{algorithm}[htb]
\caption{Supply updating} \label{alg_supplyupdate}
\begin{algorithmic}[1]
\Procedure{SupplyUpdate}{$\boldsymbol{p}^{(d)}$, $\boldsymbol{p}^{(d-N)}$, $\boldsymbol{S}^{(d)}$, $\boldsymbol{C}^{(d)}$}
\State Let the index of elements in $\boldsymbol{p}^{(d)}$, $\boldsymbol{p}^{(d-N)}$, $\boldsymbol{S}^{(d)}$ and $\boldsymbol{C}^{(d)}$ correspond with each other. (i.e., the $k$-th element of $\boldsymbol{p}^{(d)}$, $\boldsymbol{p}^{(d-N)}$, $\boldsymbol{S}^{(d)}$ and  $\boldsymbol{C}^{(d)}$ represents the information of same time interval and same route.)
\State Initialize $k=0$
\While{$k < (\text{length of $\boldsymbol{p}^{(d)}$})$}
\State $k=k+1$
\State Let ${p^{k,d}}$, ${p^{k,d-N}}$, ${S^{k,d}}$  ${C^{k,d}}$ be the $k$-th element in $\boldsymbol{p}^{(d)}$, $\boldsymbol{p}^{(d-N)}$, $\boldsymbol{S}^{(d)}$ and $\boldsymbol{C}^{(d)}$, respectively.
\If{$p^{k,d}> p^{k,d-N}$} 
\State $C^{k,d+1} = C^{k,d}$
\Else
\State $C^{k,d+1} = -C^{k,d}$
\EndIf
\State $S^{k,d+1} = S^{k,d}+C^{k,d+1}$
\If{$S^{k,d+1}$ violates the upper or lower bound constraints}
\State Set $S^{k,d+1}$ equal to the upper or lower bound
\State $C^{k,d+1} = \gamma \cdot C^{k,d+1}$
\EndIf 
\If {the difference between $p^{k,d}$ and $p^{k,d-N}$ is small enough}:
\State $C^{k,d+1} = \gamma \cdot C^{k,d+1}$
\EndIf
\EndWhile
\State Let $\boldsymbol{S}^{(d+1)} = [S^{(k,d+1)}]_{k}$ and $\boldsymbol{C}^{(d+1)} = [C^{(k,d+1)}]_{k}$
\State \Return $\boldsymbol{S}^{(d+1)}$, $\boldsymbol{C}^{(d+1)}$
\EndProcedure
\end{algorithmic}
\end{algorithm}

\subsection{Iterated best response for solving Nash Equilibrium}\label{sim_NE}
For a competition game, one of the most important tasks is to find the NE if it exists. In this study, as the environment (e.g., daily demand patterns and infrastructure layouts) does not evolve over time, the competition between AV and PT with adjustable supplies can be seen as a static game\footnote{Note that though this is a static \emph{game}, the competition can still be seen as a dynamic \emph{process}, where AV and PT react to each other's strategies as opposed to previous studies without supply updating.}. One way to solve the pure NE in a static game is the IBR algorithm. In the IBR, each player in the game chooses strategies iteratively, and every action they have selected is the best response (or one of the best responses) to the other players' strategies \citep{nash1950equilibrium}. The success of IBR depends on the context of a game. It has been shown that IBR can always converge to a pure NE for any finite potential game, i.e., games in which the payoffs of all players to change their strategies can be expressed using a single global function called the potential function \citep{monderer1996potential}. There are also some counterexamples of coordination games where IBR does not converge. However, in this study, it is hard to provide a theoretical analysis of the pure NE existence and the performance of IBR because the payoff (i.e. profit) of each player is calculated from an agent-based simulation without analytical forms. Therefore, we first implement the IBR and discuss its convergence with numerical examples in \ref{convergence}. 

The details of the IBR algorithm is shown in Algorithm \ref{alg_IBR} and \ref{alg_BR}, where $I$ is the maximum number of iterations and $\Bar{\epsilon}$ is a predetermined threshold for the termination of the best response algorithm. The best response at each iteration is solved by using the \textsc{SupplyUpdate} algorithm for a specific player many times (while fixing the supply for another player) until the profit increment is small enough. Compared to the \emph{both-adjustable} scenario where \textsc{SupplyUpdate} is only implemented once per step, the supply strategies at each iteration in the \emph{NE} scenario is closer to the optimal. Therefore, \emph{both-adjustable} scenario can be seen as a premature convergence result of IBF due to non-optimal responses of PT. However, we have to admit that Algorithm \ref{alg_BR} is still a heuristic way to find the best response, in which the results may still not be optimal. Algorithm \ref{alg_BR} can be seen as uni-variate adaptive step size random searching, which is capable of finding an optimal solution for convex function but may converge to a local optimum for a non-convex function \citep{schumer1968adaptive}. As the profit functions of AV and PT have no analytical form, we cannot provide more theoretical analysis here. In \ref{convergence}, we test the convergence results of the proposed IBR with respect to different initial AV and PT supplies numerically. And it is found that the system can always converge to a similar supply level (which can be seen as an approximated pure NE). This validates the effectiveness of the IBR algorithm proposed in this study.

\begin{algorithm}[H]
\caption{Iterated best response for solving the Nash Equilibrium} \label{alg_IBR}
\begin{algorithmic}[1]
\Procedure{IBR}{$\boldsymbol{\theta}$, $\boldsymbol{S}_B^{(0)}$, $\boldsymbol{S}_A^{(0)}$, $\boldsymbol{D}$}
\State Initialize $\boldsymbol{p}_A^{(1)}=0$, $\boldsymbol{p}_B^{(1)}=0$, $\boldsymbol{S}_A^{(1)}=\boldsymbol{S}_A^{(0)}$, $\boldsymbol{S}_B^{(1)}=\boldsymbol{S}_B^{(0)}$
\State Initialize ${\boldsymbol{C}_A^{(1)}}={C}_A^{\text{ini}}$, ${\boldsymbol{C}_B^{(1)}}={C}_B^{\text{ini}}$
\State Let iteration ID $i = 0$
\While{$i < I$}
\State $i = i+1$
\State $\boldsymbol{S}_A^{(i+1)}$, $\boldsymbol{S}_B^{(i+1)}$ =\textsc{BestResponse}($\boldsymbol{S}_A^{(i)}$, $\boldsymbol{S}_B^{(i)}$, $\boldsymbol{C}_A^{(i)}$, $\boldsymbol{C}_B^{(i)}$, $\boldsymbol{\theta}$, $\boldsymbol{D}$)
\State $\boldsymbol{p}_A^{(i+1)}, \boldsymbol{p}_B^{(i+1)}=
\textsc{One-Day-Sim}(\boldsymbol{S}_B^{(i+1)}, \boldsymbol{S}_A^{(i+1)}, \boldsymbol{D} \mid \boldsymbol{\theta})$
\For{$k$= 1:length of $\boldsymbol{p}_A^{(i+1)}$}
\If {the difference between ${p}_A^{(k,i+1)}$ and ${p}_A^{(k,i)}$ is small enough}:
\State $C_A^{k,i+1} = \gamma \cdot C_A^{k,i}$
\EndIf
\EndFor

\For{$k$= 1:length of $\boldsymbol{p}_B^{(i+1)}$}
\If {the difference between ${p}_B^{(k,i+1)}$ and ${p}_B^{(k,i)}$ is small enough}:
\State $C_B^{k,i+1} = \gamma \cdot C_B^{k,i}$
\EndIf
\EndFor

\EndWhile
\State \Return the system evaluation indicators (Table \ref{tab_stakeholder})
\EndProcedure
\end{algorithmic}
\end{algorithm} 

\begin{algorithm}[htb]
\caption{Best response} \label{alg_BR}
\begin{algorithmic}[1]
\Procedure{BestResponse}{$\boldsymbol{S}_A$, $\boldsymbol{S}_B$, $\boldsymbol{C}_A$, $\boldsymbol{C}_B$, $\boldsymbol{\theta}$, $\boldsymbol{D}$}
\State Initialize $\boldsymbol{p}_A^{(0)}=0$, $\boldsymbol{S}_A^{(1)}=\boldsymbol{S}_A$, ${\boldsymbol{C}_A^{(1)}}=\boldsymbol{C}_A$
\State Let day counter $d = 0$
\While{$\epsilon > \Bar{\epsilon}$} \Comment{Find the best response for AV}
\State $d = d+1$
\State $\boldsymbol{p}_A^{(d)}, \boldsymbol{p}_B^{(d)}=
\textsc{One-Day-Sim}(\boldsymbol{S}_B, \boldsymbol{S}_A^{(d)}, \boldsymbol{D} \mid \boldsymbol{\theta})$
\State $\boldsymbol{S}_A^{(d+1)}, \boldsymbol{C}_A^{(d+1)}=\textsc{SupplyUpdate}(\boldsymbol{p}_A^{(d)},  \boldsymbol{p}_A^{(d-1)}, \boldsymbol{S}_A^{(d)}, \boldsymbol{C}_A^{(d)})$
\State $\epsilon = |(\sum_k {p}_A^{(k,d)} - \sum_k {p}_A^{(k,d-1)})/\sum_k {p}_A^{(k,d-1)}|$
\EndWhile
\State $\boldsymbol{S}_A^{\text{BR}} = \boldsymbol{S}_A^{(d+1)}$

\State Initialize $\boldsymbol{p}_B^{(0)}=0$, $\boldsymbol{S}_B^{(1)}=\boldsymbol{S}_B$, ${\boldsymbol{C}_B^{(1)}}=\boldsymbol{C}_B$
\State Let day counter $d' = 0$
\While{$\epsilon > \Bar{\epsilon}$} \Comment{Find the best response for PT}
\State $d' = d'+1$
\State $\boldsymbol{p}_A^{(d')}, \boldsymbol{p}_B^{(d')}=
\textsc{One-Day-Sim}(\boldsymbol{S}_B^{(d')}, \boldsymbol{S}_A^{\text{BR}}, \boldsymbol{D} \mid \boldsymbol{\theta})$
\State $\boldsymbol{S}_B^{(d'+1)}, \boldsymbol{C}_B^{(d'+1)}=\textsc{SupplyUpdate}(\boldsymbol{p}_B^{(d')},  \boldsymbol{p}_B^{(d'-1)}, \boldsymbol{S}_B^{(d')}, \boldsymbol{C}_B^{(d')})$
\State $\epsilon = |(\sum_k {p}_B^{(k,d')} - \sum_k {p}_B^{(k,d'-1)})/\sum_k {p}_B^{(k,d'-1)}|$
\EndWhile
$\boldsymbol{S}_B^{\text{BR}} = \boldsymbol{S}_B^{(d'+1)}$

\State \Return $\boldsymbol{S}_A^{\text{BR}} $, $\boldsymbol{S}_B^{\text{BR}}$
\EndProcedure
\end{algorithmic}
\end{algorithm}

\subsection{Scenarios settings}\label{scenario_set}
There is a rich space for all the value combinations of $\boldsymbol{\theta}$, as shown in Table \ref{tab_notation}. Each value combination represents a distinct assumption of competition markets---some are realistic and some are not. It is not the focus of this study to systematically explore all possible scenarios. From the market organization perspective, we only focus on the five regulation scenarios introduced in Section \ref{intro} (Table \ref{tab_scenarios}). 

In addition, we conduct another set of experiments with respect to different subsidy levels provided to the PT. Specifically, in the case study (Section \ref{subsidy}), we test two different subsidy schemes: demand-based and supply-based. The demand-based subsidy is provided to PT operators on a per passenger served basis, while supply-based subsidy on a  bus kilometer operated basis. The sensitivity analysis for subsidy can be seen from two aspects. First, it reflects different schemes of revenue and cost calculation. Demand-based subsidies can be seen as the sensitivity analysis for revenue calculation while supply-based for cost calculation. Second, subsidies can be seen as positive externalities of passengers using PT. From this angle, PT's objective is not only profit but also social welfare. Providing subsidies to PT can be seen as a way for the government to internalize the positive externalities. Note that we approximate the change of social welfare using the increase or reduction of government subsidy in this article, while acknowledging that social welfare involves a broader category of considerations, including equity, caring for disadvantaged social groups, fostering social interaction, etc. We leave the broader discussion of the impact of public transportation on social welfare for future research.

\section{Results and discussion} \label{res}


\subsection{PT perspective} 
\begin{figure}[H]
\centering
\subfloat[PT revenue]{\includegraphics[width=0.33\textwidth]{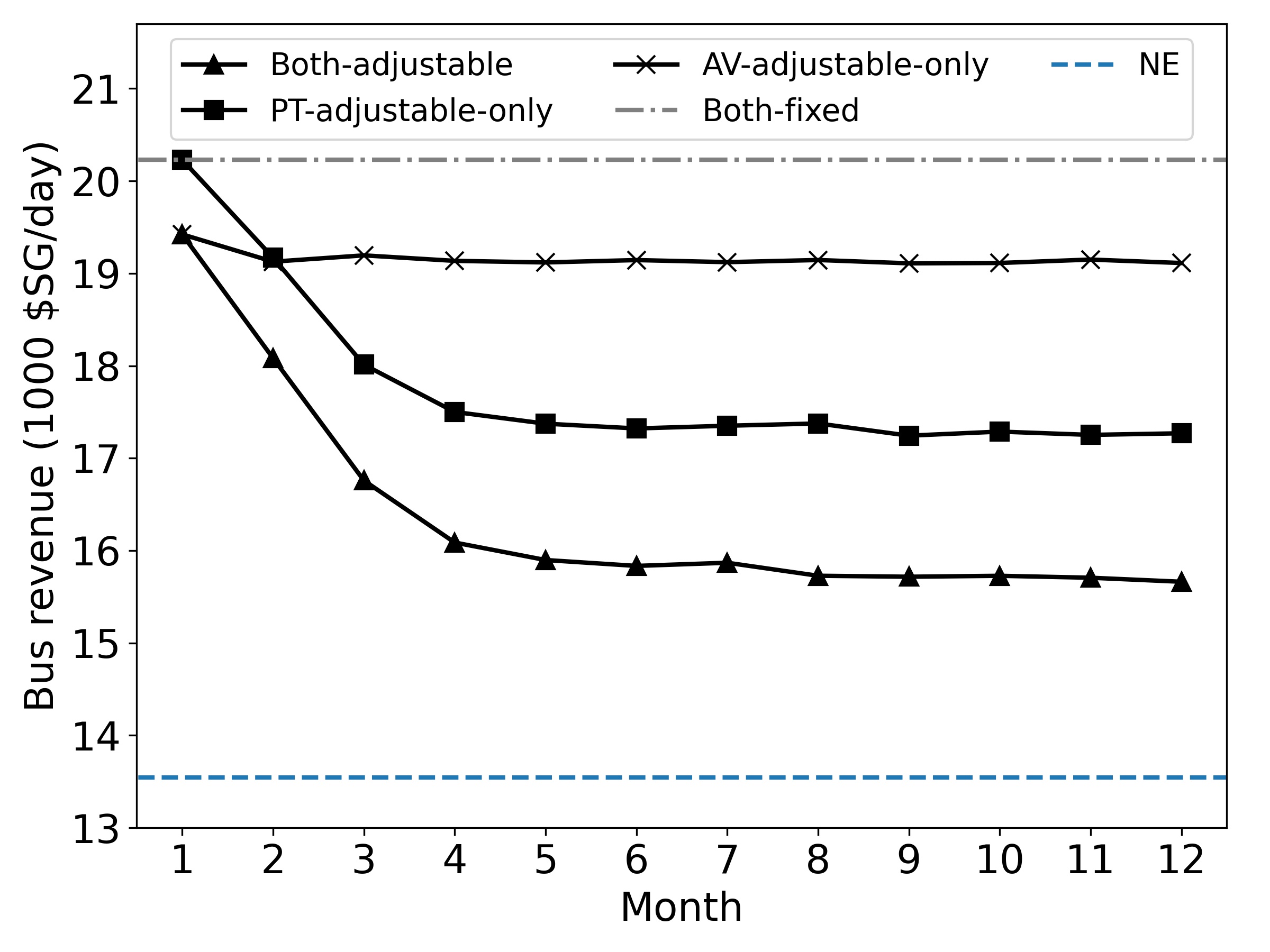}}
\hfil
\subfloat[PT operation cost]{\includegraphics[width=0.33\textwidth]{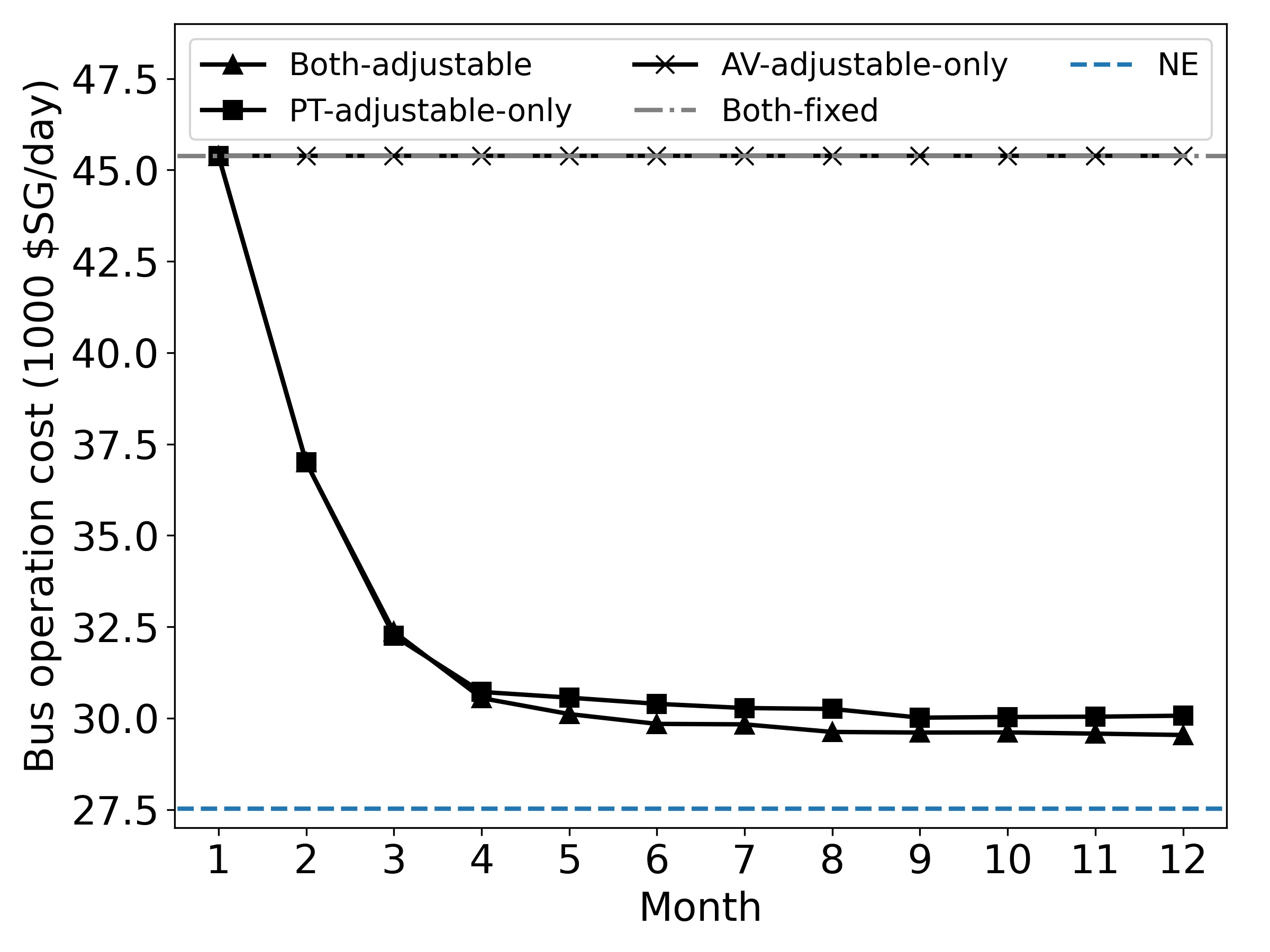}\label{fig_PT_operation_cost}}
\hfil
\subfloat[PT profit]{\includegraphics[width=0.33\textwidth]{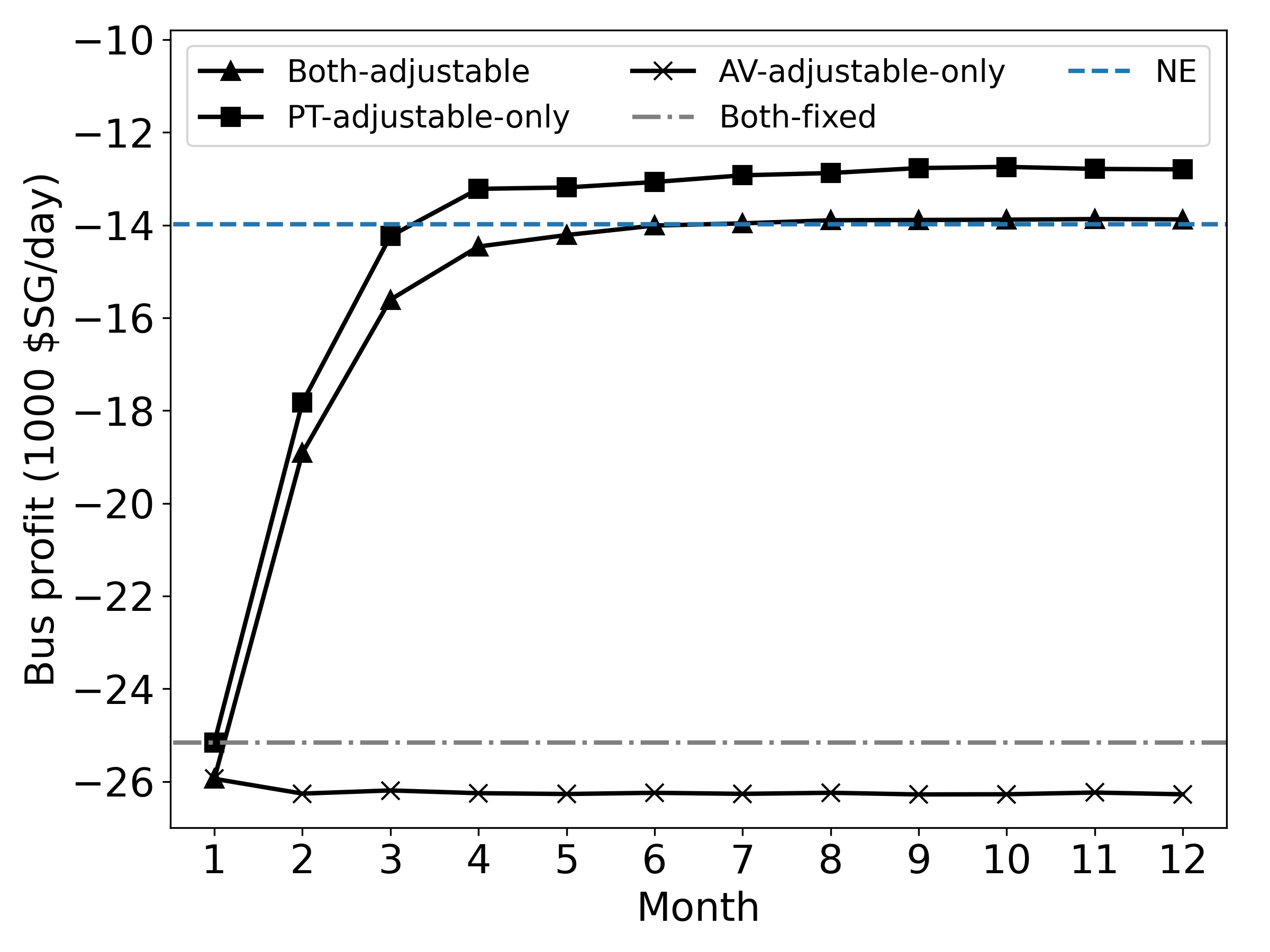}}
\hfil
\subfloat[PT supply]{\includegraphics[width=0.33\textwidth]{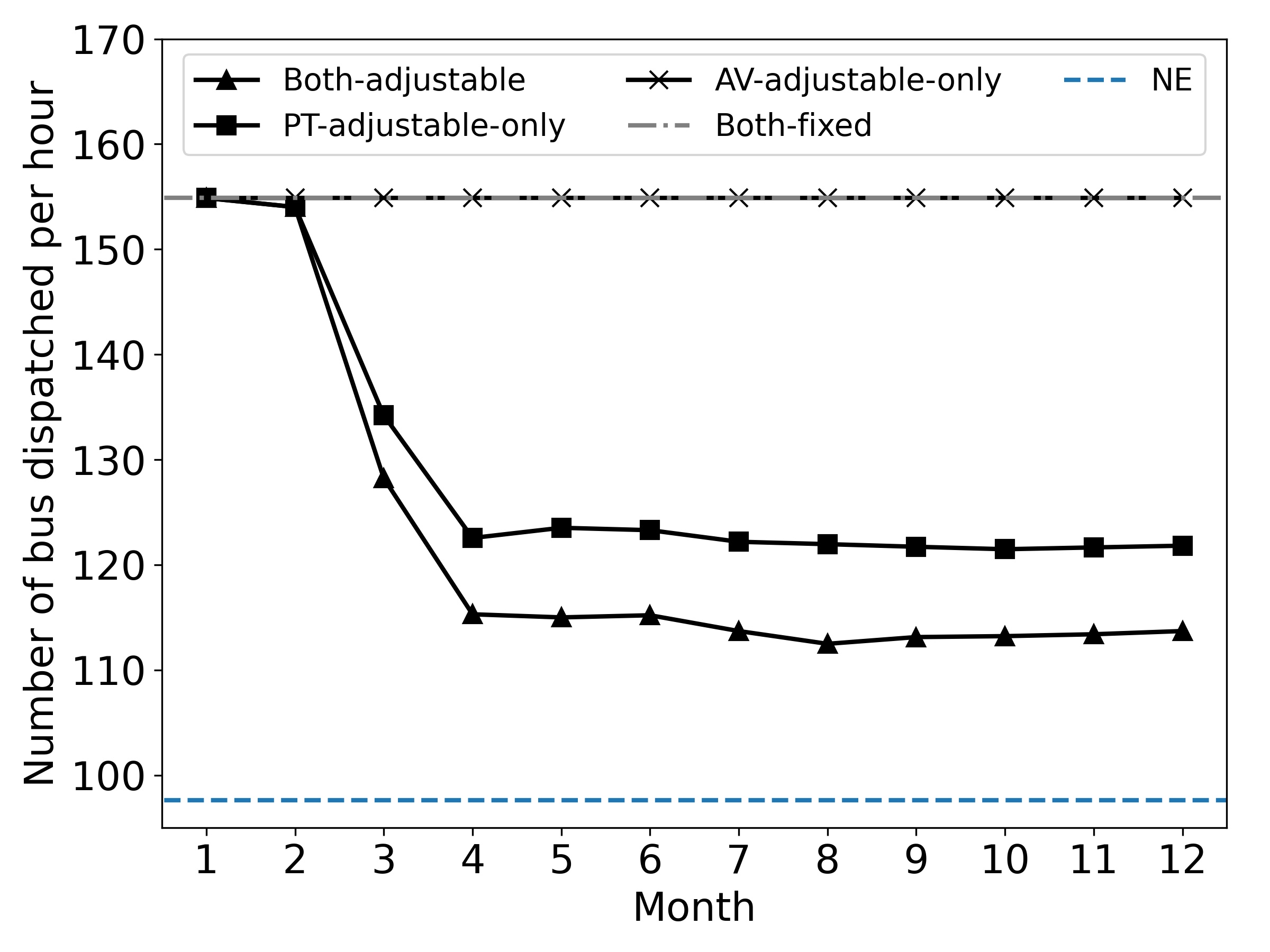}\label{fig_PT_supply}}
\hfil
\subfloat[PT market share]{\includegraphics[width=0.33\textwidth]{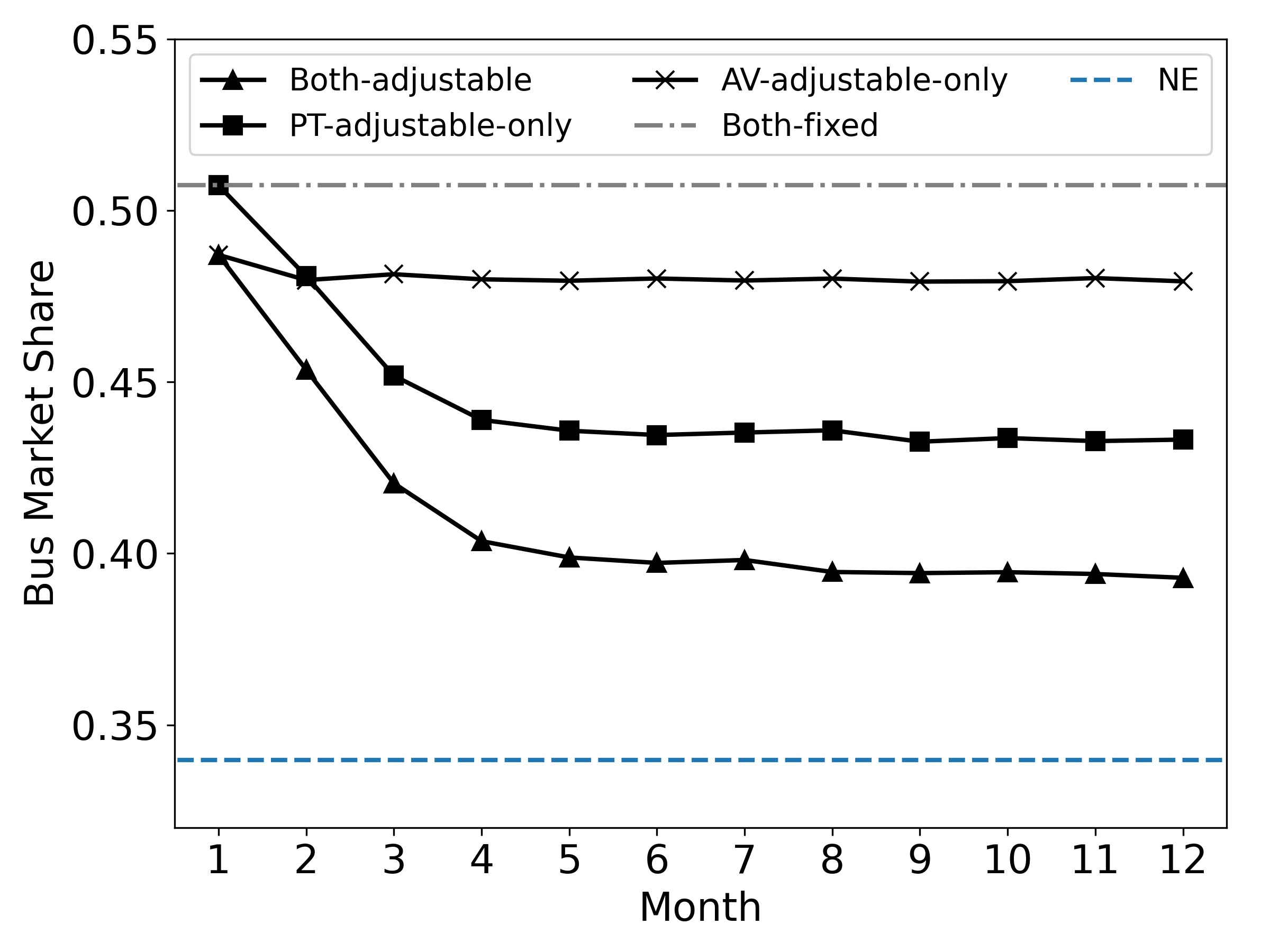}\label{fig_PT_MS}}
\caption{Indicators of PT's interest over the simulation process}
\label{fig_PT_interets}
\end{figure}

Figure \ref{fig_PT_interets} presents the indicators of interest for PT operators, including revenues, operating costs, profits, supplies, and market shares. Each point in the graph represents the average value of the corresponding month (same for all the following graphs).

The final PT profit of the \emph{PT-adjustable-only} scenario is higher than those in all other scenarios because there is no AV competition in this scenario, and is the lowest in the \emph{AV-adjustable-only} scenario, reflecting the competition of AV reducing the profit of PT. Both the revenue and operating cost of PT decrease over the simulation period, but the operating cost shows a sharper reduction. Both the revenue and the cost of the \emph{NE} scenario are the lowest among all scenarios, indicating that the market share of buses is reduced in this scenario. However, in terms of PT profit, the \emph{NE} scenario is roughly at the same level as the \emph{both-adjustable} scenario, which means that the bus supply is similarly optimized towards profit in the two scenarios.

The number of buses dispatched per day and PT's market share (Figure \ref{fig_PT_supply} and \ref{fig_PT_MS}) have a similar declining trend. Since buses do not change supply in \emph{AV-adjustable-only} and \emph{both-fixed} scenarios, the supply curves of these two scenarios are flat. The bus supply and market share decrease with AV competition and reach the lowest levels in the \emph{NE} scenario.

One important observation is the additive effect of bus and AV supply adjustments. Take the bus market share as an example. The curve of \emph{PT-adjustable-only} scenario roughly represents how the PT operator gradually "gives up" the unprofitable market share. The \emph{AV-adjustable-only} curve shows how much market share is grabbed by AV (i.e. the impact from AV supply adjustment). The \emph{both-adjustable} scenario shows the additive effect of two, which is nearly the sum of the two reductions. Moreover, the \emph{NE} scenario has the lowest PT market share among all, indicating a more "aggressive" bus supply optimization strategy. Later results suggest that although the bus supply is significantly reduced in the \emph{NE} scenario, the generalized travel cost is reduced on average.

\begin{figure}[H]
\centering
\subfloat[PT supply temporal distribution]{\includegraphics[width=0.42\textwidth]{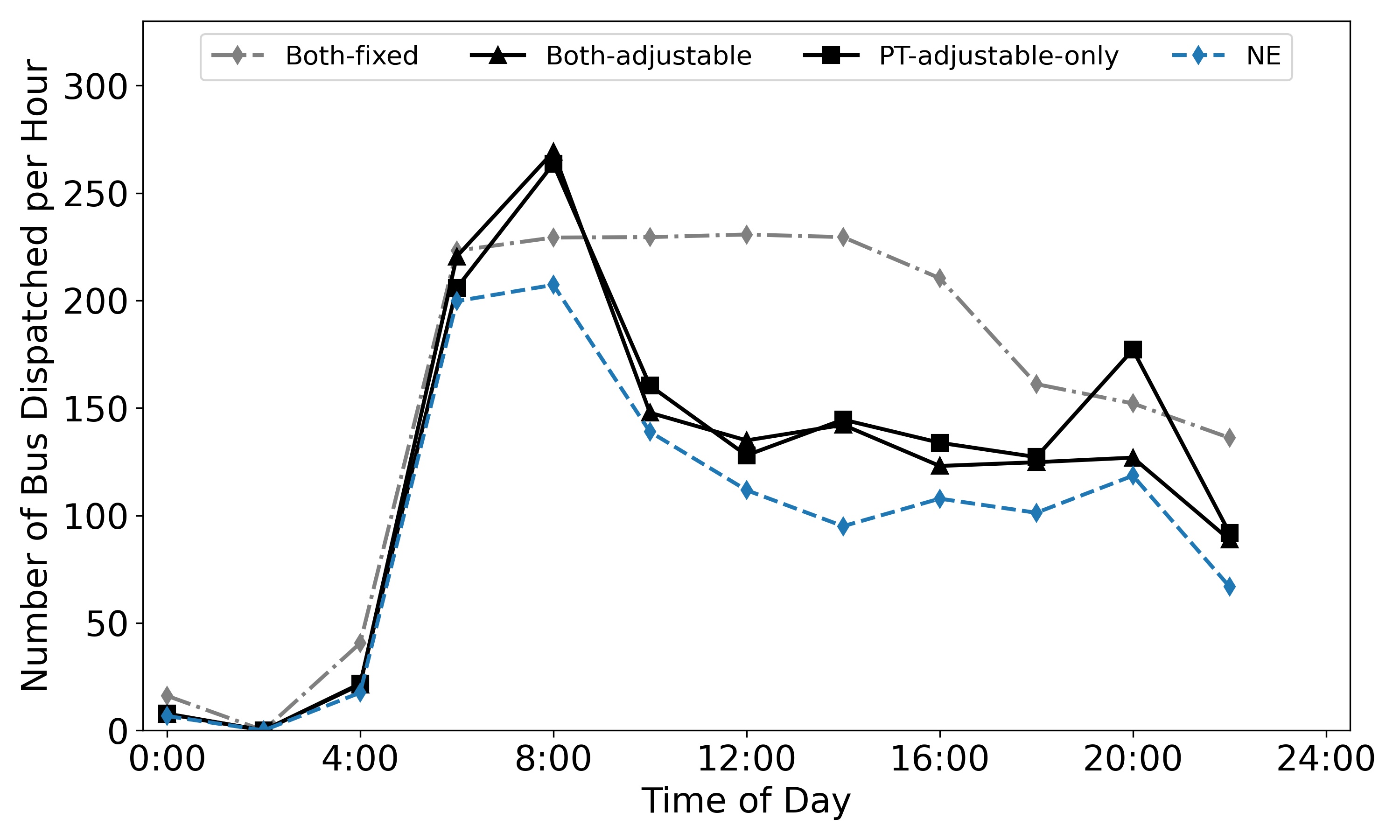}\label{fig_PT_before_after_supply_temporal}}
\hfil
\subfloat[PT supply spatial distribution]{\includegraphics[width=0.63\textwidth]{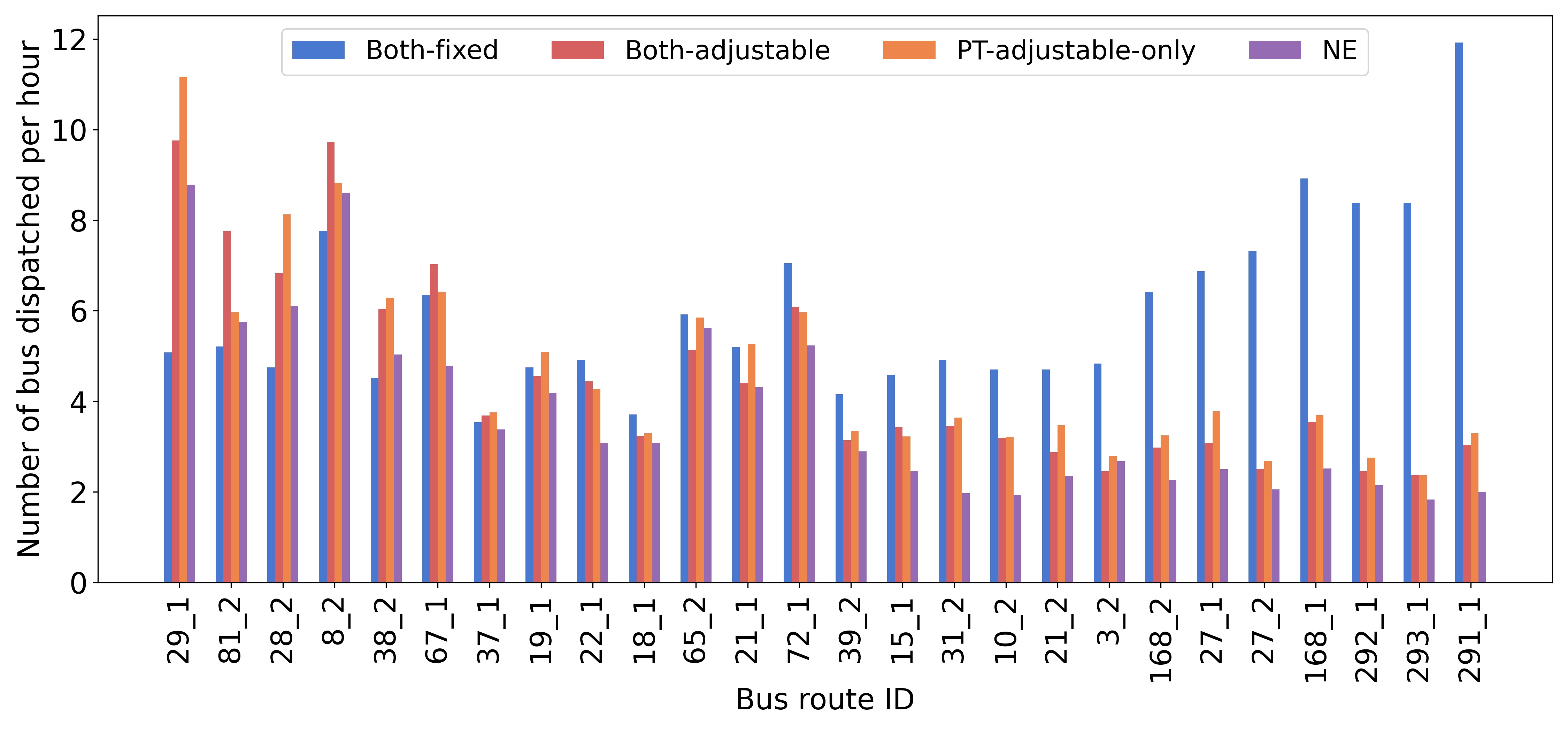}\label{fig_PT_before_after_supply_spatial}}
\hfil
\subfloat[Routes with increased supply (\emph{NE})]{\includegraphics[width=0.4\textwidth]{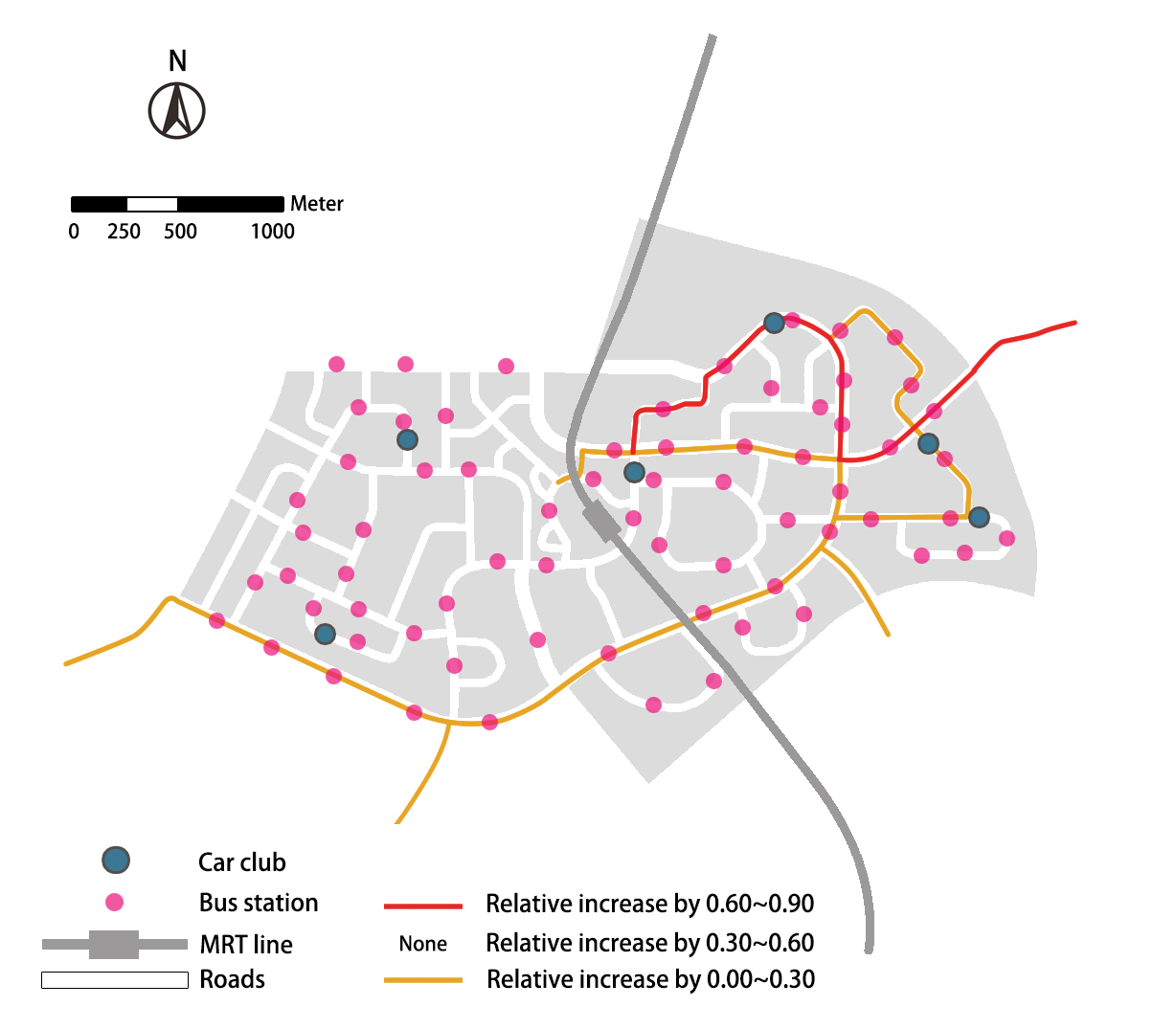}\label{fig_route_increase}}
\hfil
\subfloat[Routes with reduced supply (\emph{NE})]{\includegraphics[width=0.4\textwidth]{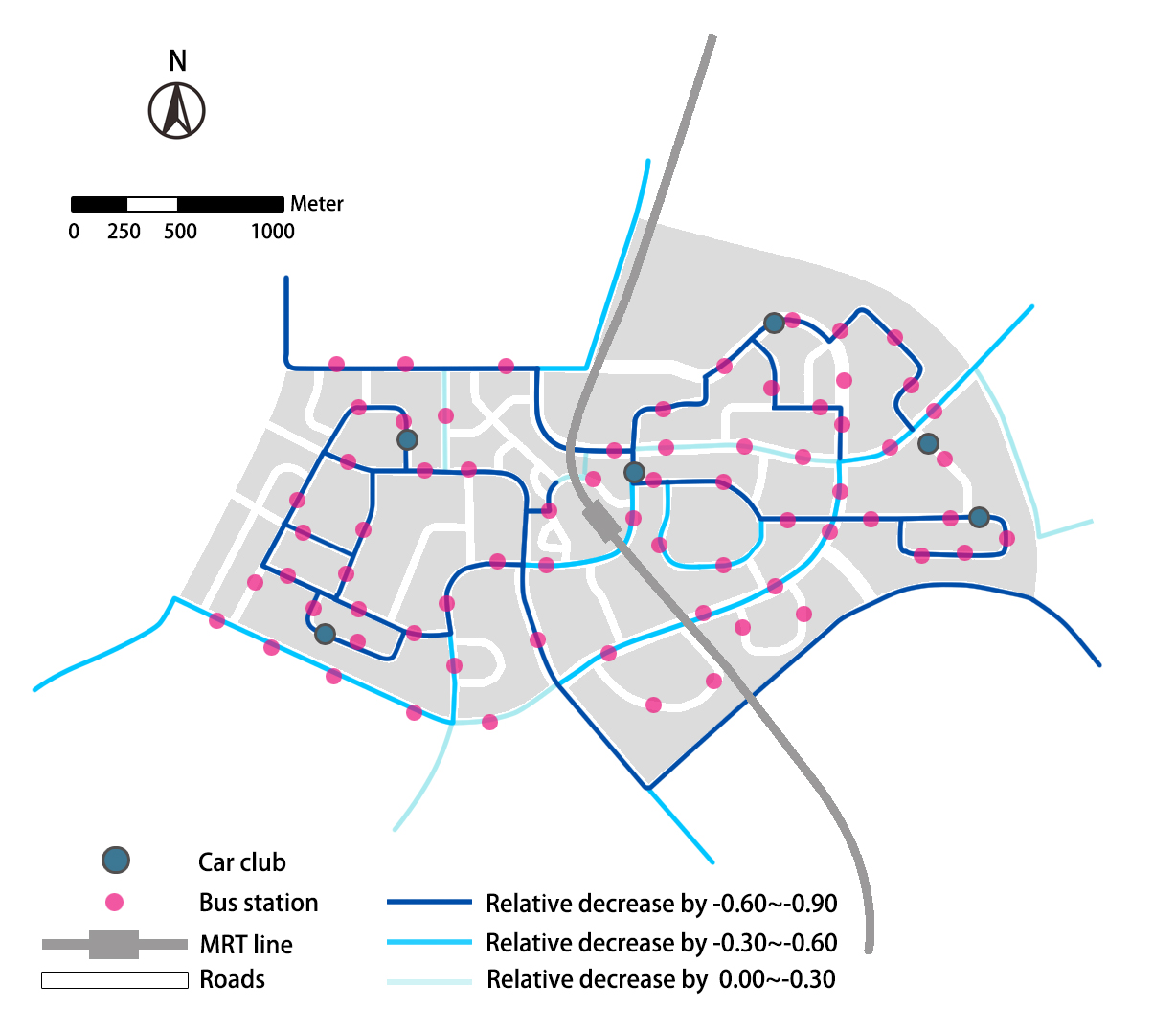}\label{fig_route_decrease}}
\caption{Spatial and temporal distribution changes in PT supply. Note that \emph{both-fixed} scenario represents the initial distribution before the supply adjustment.}
\label{fig_PT_supply_spatial_temporal}
\end{figure}

In addition to the total supply, we consider the change in temporal and spatial distributions of the supply before and after adjustment (Figure \ref{fig_PT_supply_spatial_temporal}). In the temporal dimension, the numbers of dispatched buses for most time periods except for the morning peak (6:00--10:00) are reduced. This implies that the PT operator \emph{concentrates} the supplies to the morning and evening peak periods, which have higher demands and are more profitable. In the spatial dimension, some routes (e.g., 29$\_$1, 28$\_$2) are allocated with higher frequencies, while some routes (e.g., 291$\_$2, 292$\_$1) are adjusted to a lower service rate. The routes with increased and reduced supplies are shown in Figure \ref{fig_route_increase} and \ref{fig_route_decrease}, respectively. Overall, the increased-supply routes are short and cost-efficient and they cross the residential areas and are directly connected to the MRT station. The routes with decreased supply are long and sinuous and greater reductions in the operating cost are achieved by decreasing the supply of these routes. Therefore, the change in the spatial distribution is related to an \emph{implicit} \emph{coordination} within the bus routes (even if our algorithm does not consider the inter-route coordination). To summarize, the PT operator reduces the supply for higher-cost routes and transfer the service to the lower-cost routes, resulting in a more profitable operation scheme. On both the spatial and temporal distributions, the \emph{NE} scenario resembles the results of \emph{both adjustable}, but with each route and time period slightly lower.

\subsection{AMoD perspective}
\begin{figure}[htb]
\centering
\subfloat[AMoD revenue]{\includegraphics[width=0.33\textwidth]{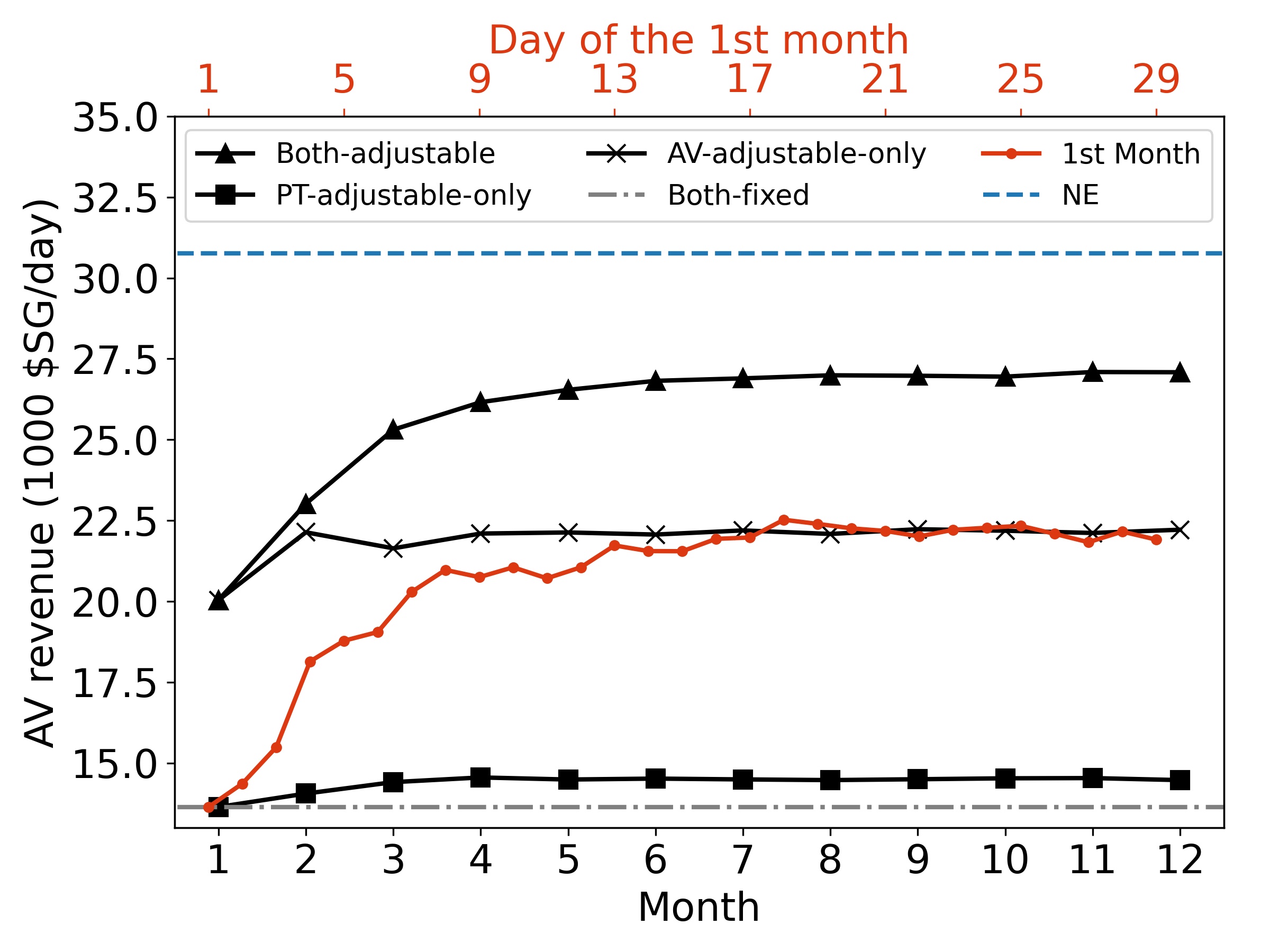}}
\hfil
\subfloat[AMoD operation cost]{\includegraphics[width=0.33\textwidth]{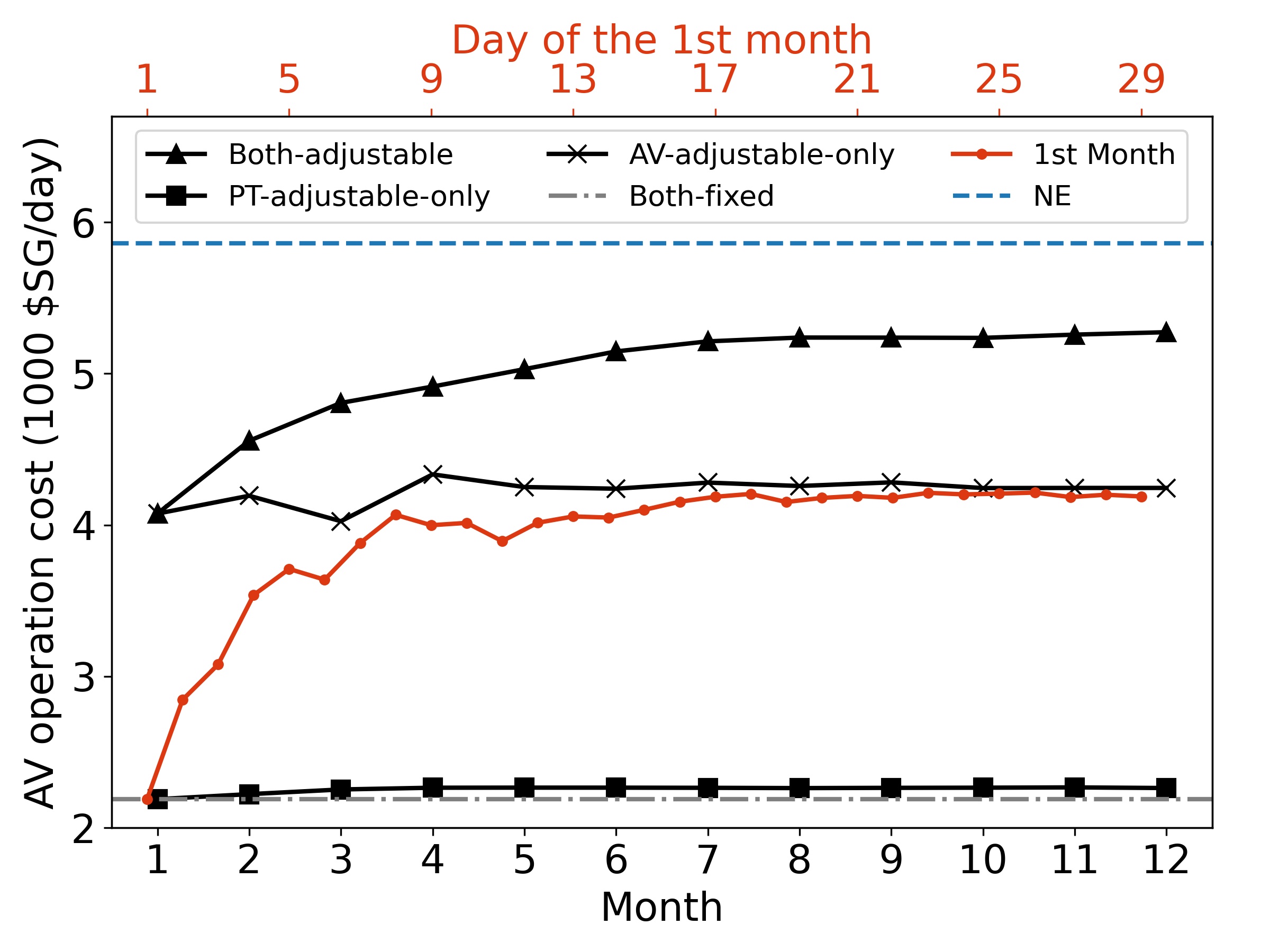}\label{fig_AMoD_cost}}
\hfil
\subfloat[AMoD profit]{\includegraphics[width=0.33\textwidth]{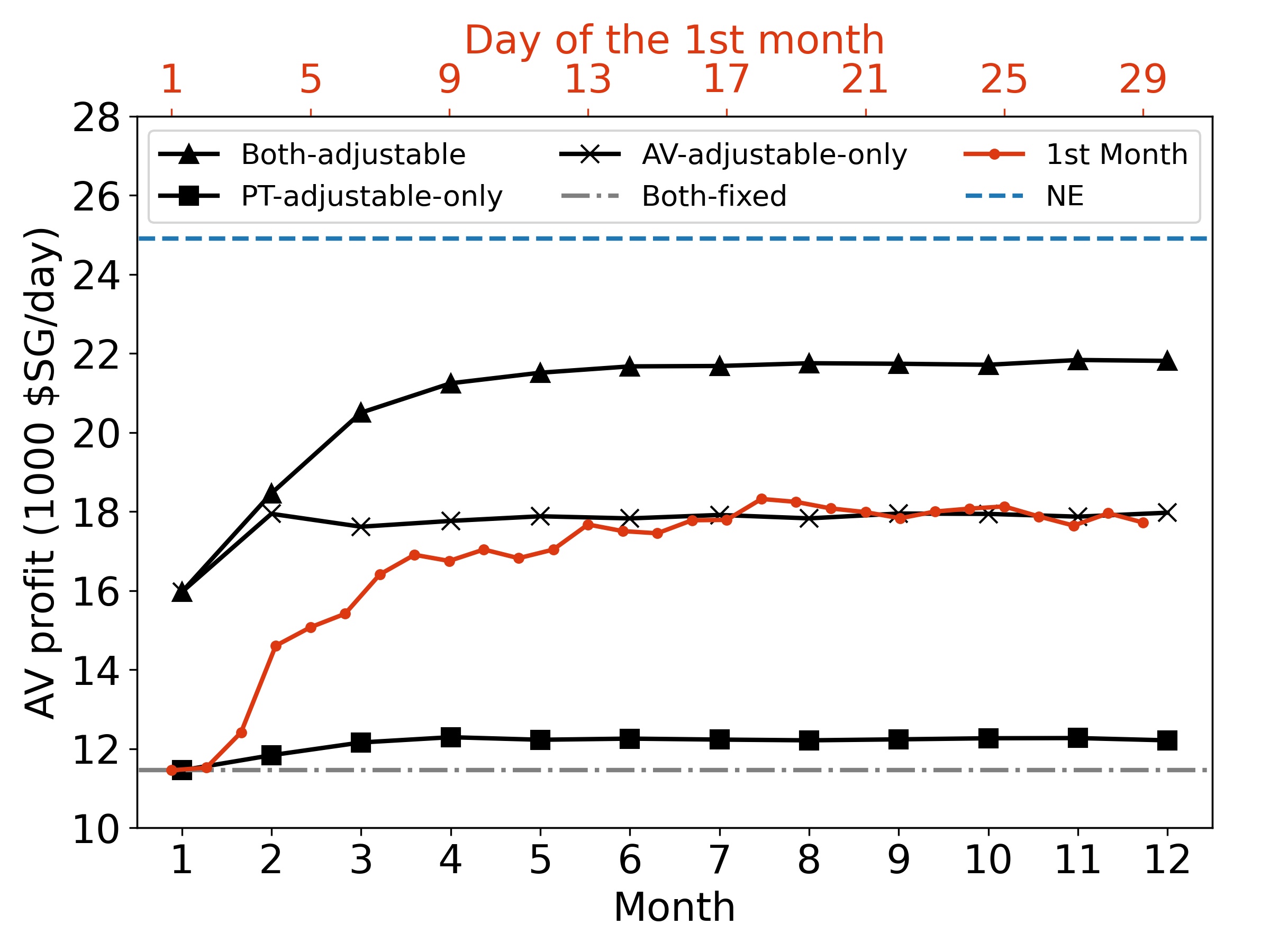}}
\hfil
\subfloat[AMoD supply]{\includegraphics[width=0.33\textwidth]{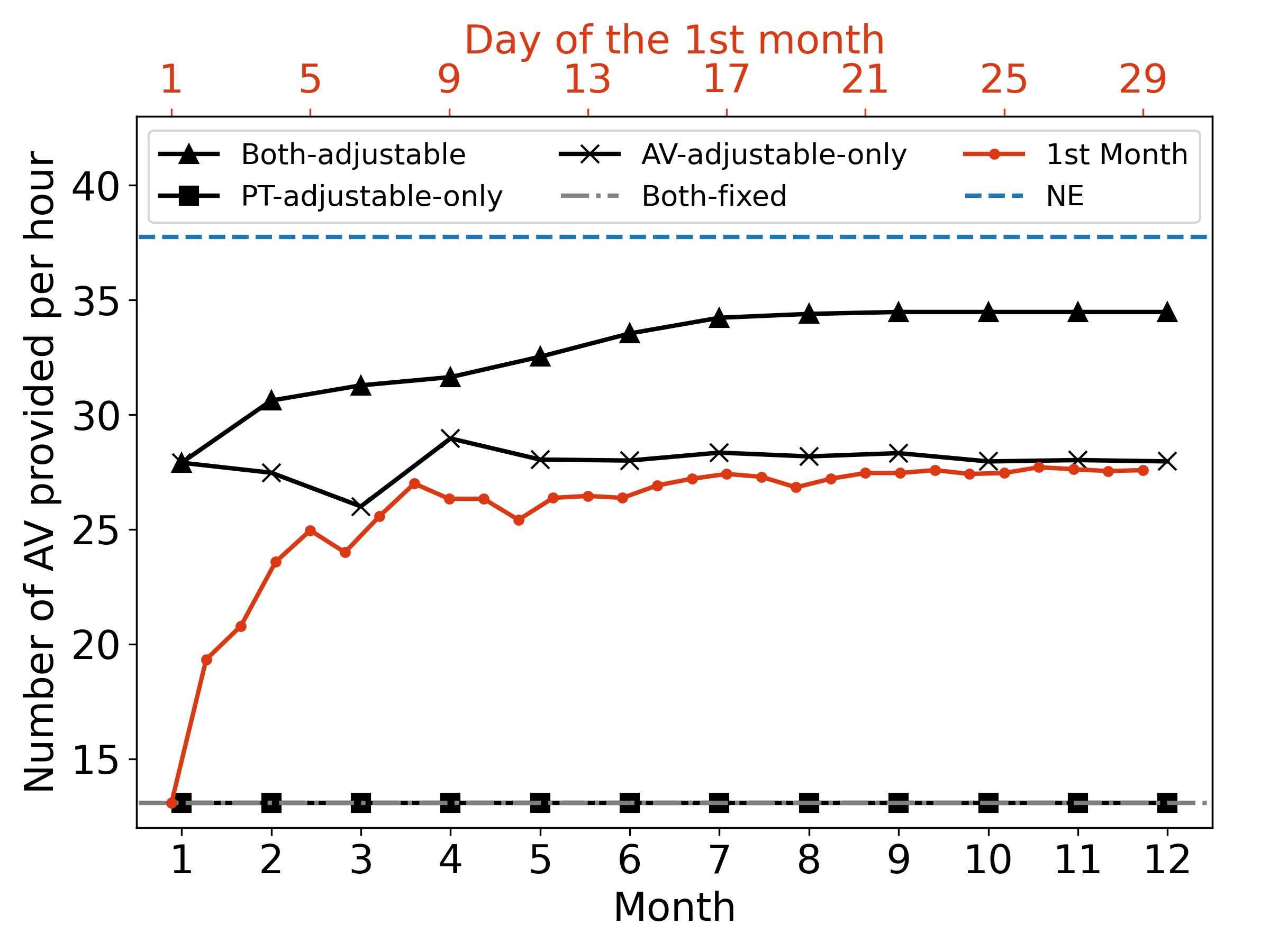}\label{fig_AMoD_supply}}
\hfil
\subfloat[AMoD share]{\includegraphics[width=0.33\textwidth]{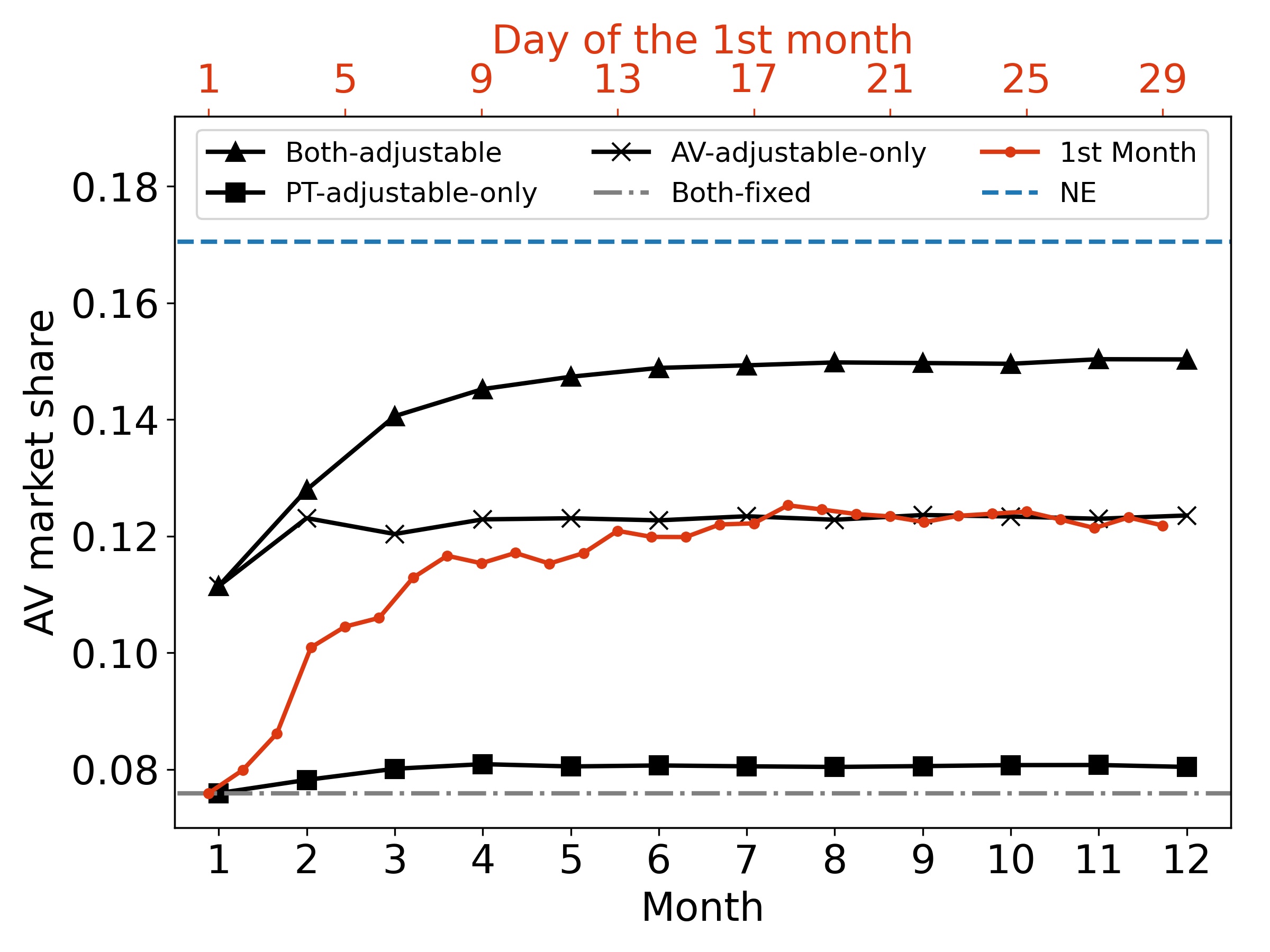}\label{fig_AMoD_MS}}
\caption{Indicators of AMoD's interest over the simulation process. The first-month curve (red, with upper x-axis) for the \emph{AV-adjustable-only} and \emph{both-adjustable} scenarios are the same because the bus supplies stay the same, and AV does not change in the other two scenarios. Therefore, only one curve is shown for the first month.}
\label{fig_AMoD_interest}
\end{figure}

As Figure \ref{fig_AMoD_interest} shows, AMoD’s revenue, operating cost, and profit increase rapidly at the beginning and then stabilize. The AMoD operator provides more services to improve its profit in the competition. As shown in Figure \ref{fig_AMoD_supply} and \ref{fig_AMoD_MS}, the AMoD's supply and market share keep increasing during the first few months and then become stable. The major change in the AMoD's supply happens in the first month. When both AV and PT can adjust---including the \emph{both-adjustable} and \emph{NE} scenarios---the profit of AV is higher. The \emph{NE} scenario can achieve the highest AV profit and the largest market share, and the \emph{both-adjustable} scenario rank the second. The bus supply adjustment not only improves the profit of the bus but also benefits that of AV. This is because the initial bus service is over-supplied for the first-mile market. When the bus operator is allowed to adjust its supply, some of the yielded unprofitable travel demand is served by AV. This observation is in line with the previous research on the AV--PT integration system \citep{shen2018integrating}. Therefore, when PT is oversupplied, although we assume that AV and PT compete with each other, they still implicitly show some extent of cooperative behavior, resulting in higher profits for both. However, this cooperative feature is only manifested when the bus changes its supply. The AV supply adjustment only unilaterally benefits itself. 

For the AMoD's operating cost (Figure \ref{fig_AMoD_cost}), the value in the \emph{NE} scenario is the highest, followed by the \emph{both-adjustable} and \emph{AV-adjustable-only} scenarios. This corresponds to the supply levels shown in Figure \ref{fig_AMoD_supply}. We also observe the additive effect of supply adjustment on the AV profit and market share. Different from the effect on bus supply, the two changes both benefit AV. Thus, the \emph{both-adjustable} scenario has higher profit and market share than the \emph{AV-adjustable-only} and the \emph{PT-adjustable-only} scenarios. The \emph{NE} scenario has the highest AV revenue, operation cost, and also the highest profit and market share among all scenarios. 

\begin{figure}[htb]
\centering
\includegraphics[width=0.5\textwidth]{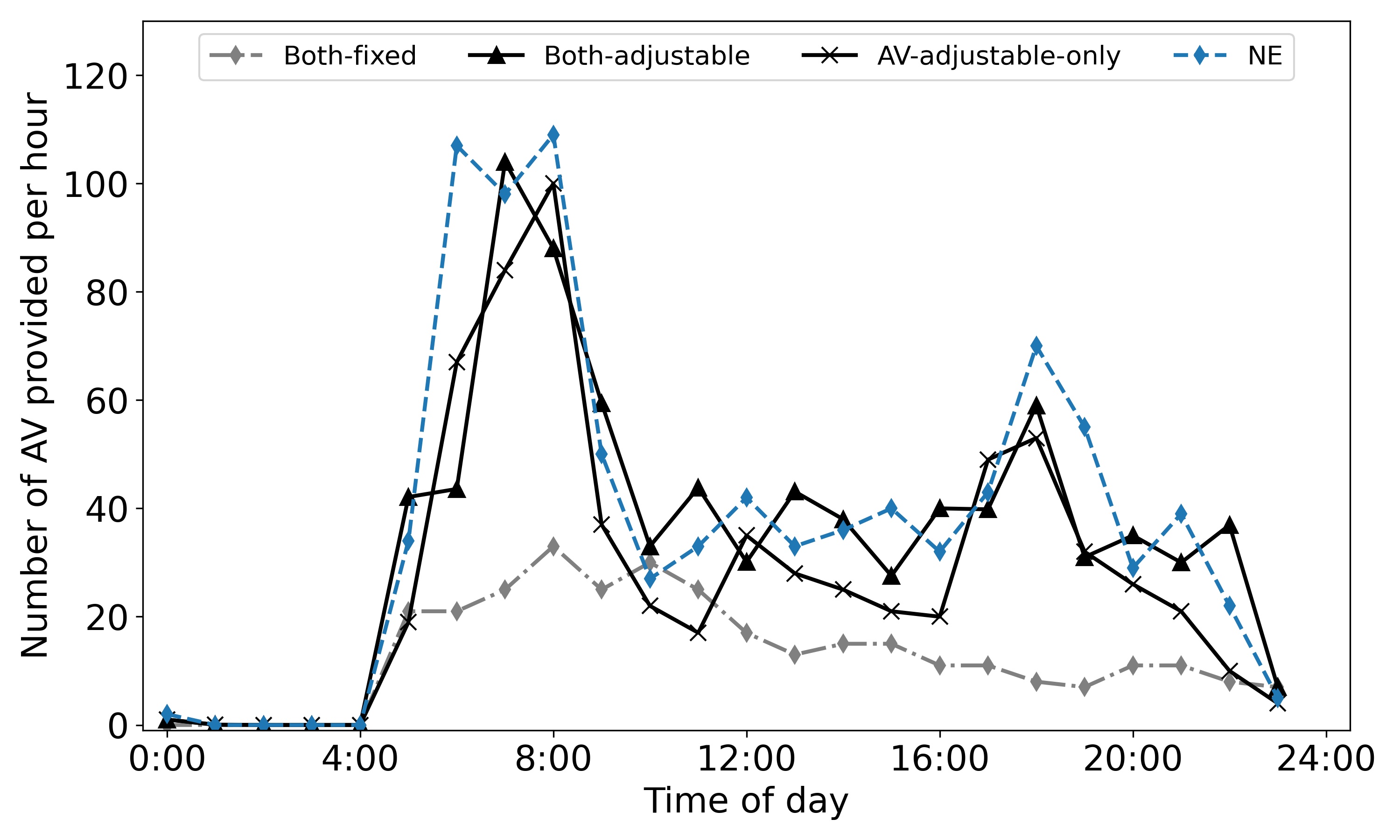}
\caption{Time of day distribution of 
AV supply}
\label{fig_time_day_AV_supply}
\end{figure}

Similar to our analysis for PT, we also examine the time-of-day distribution of the AMoD supply before and after the adjustment. As shown in Figure \ref{fig_time_day_AV_supply}, the final supply patterns of \emph{NE}, \emph{both-adjustable} and \emph{AV-adjustable-only} scenarios are similar. After 12 months of adjustment, the supply was re-distributed across time. More supply was provided in the morning and evening peak hours, which is considered more profitable. The supply in off-peak hours (e.g., 11:00--13:00) stayed the same or became even lower. These temporal changes correspond to the demand distribution in Figure \ref{fig_demand}.

\subsection{Passenger perspective} 
\begin{figure}[H]
\centering
\subfloat[Travel cost]{\includegraphics[width=0.4\textwidth]{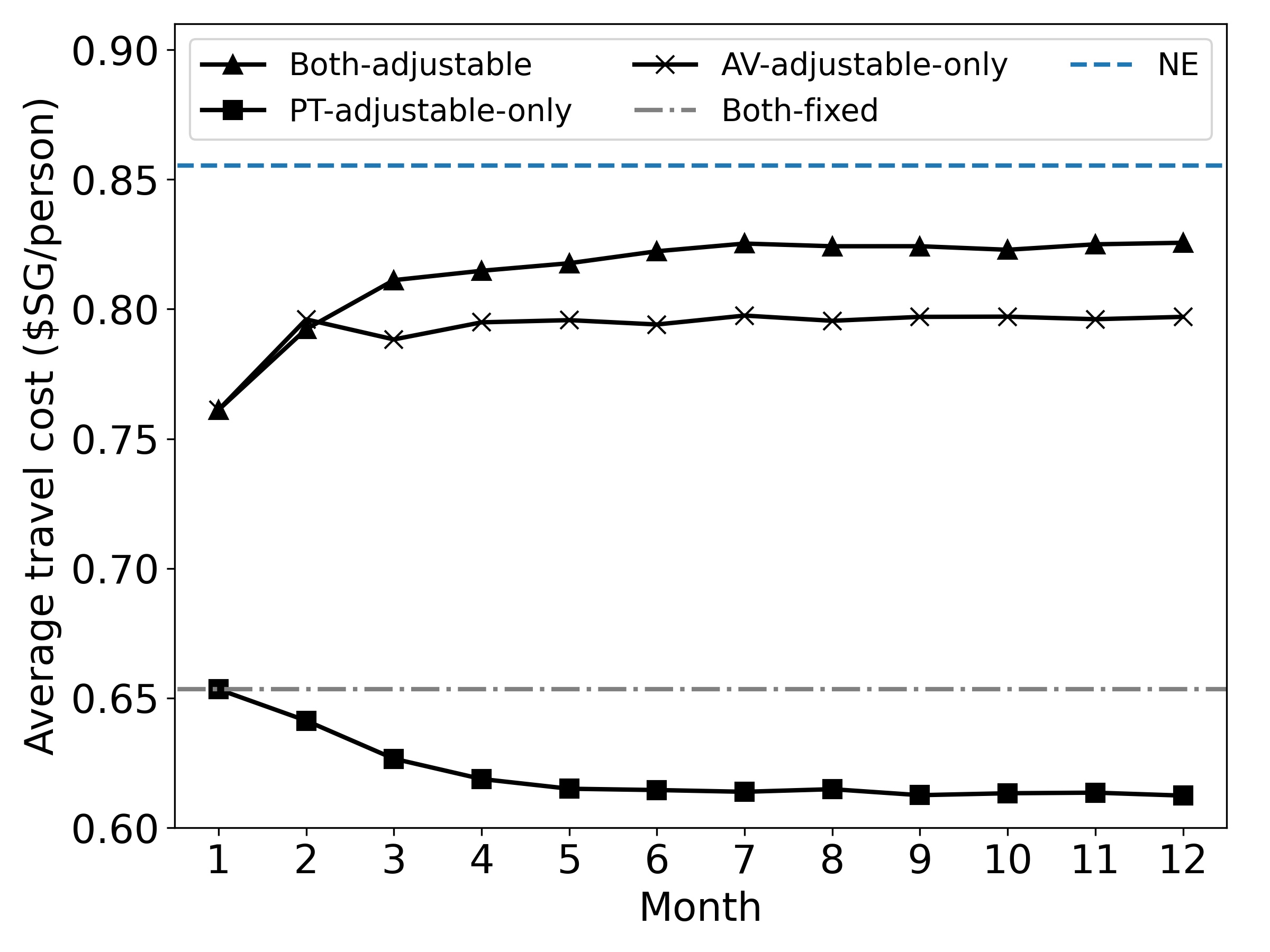}}
\hfil
\subfloat[Total travel time]{\includegraphics[width=0.4\textwidth]{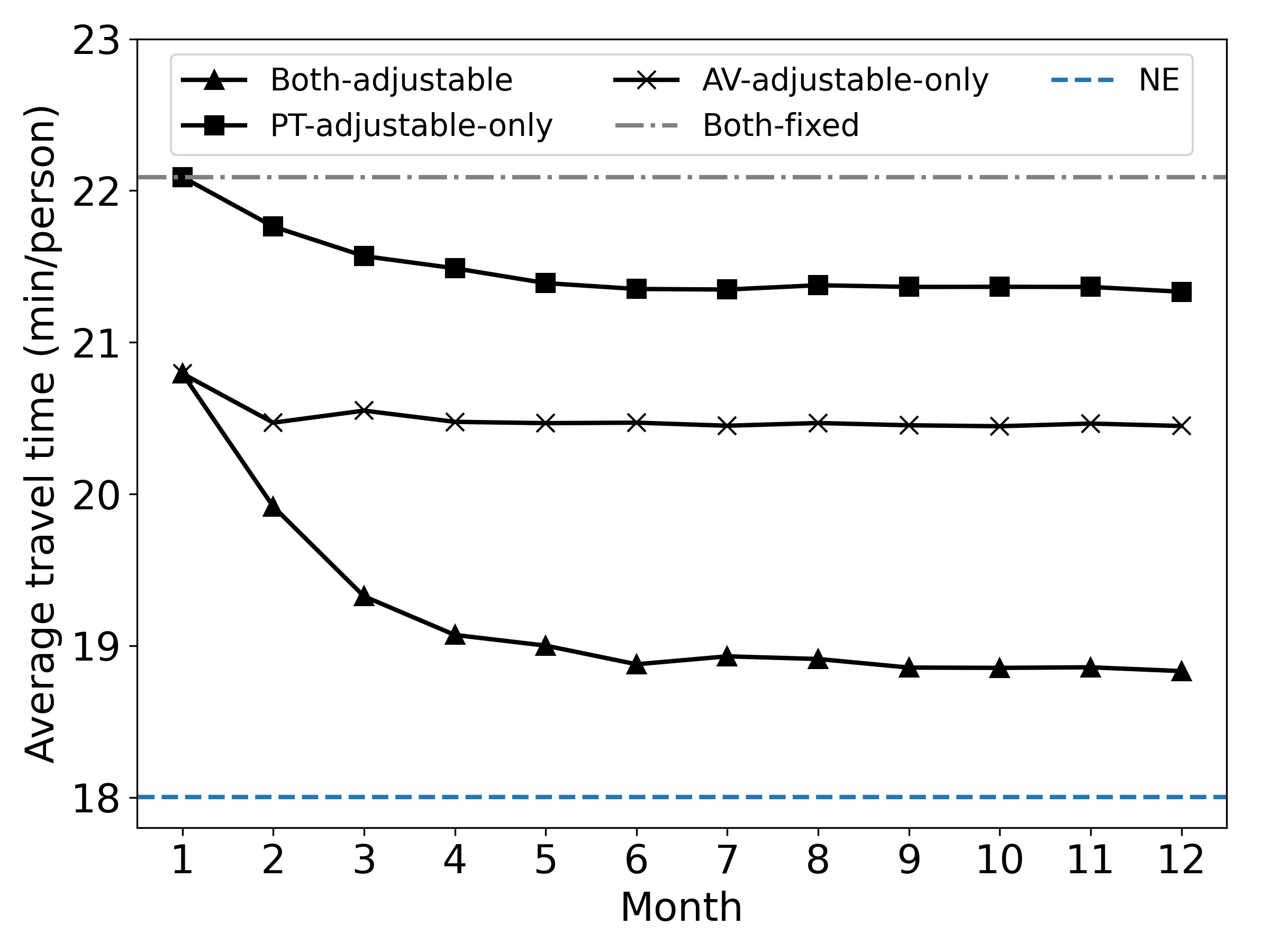}}
\hfil
\subfloat[Waiting time]{\includegraphics[width=0.4\textwidth]{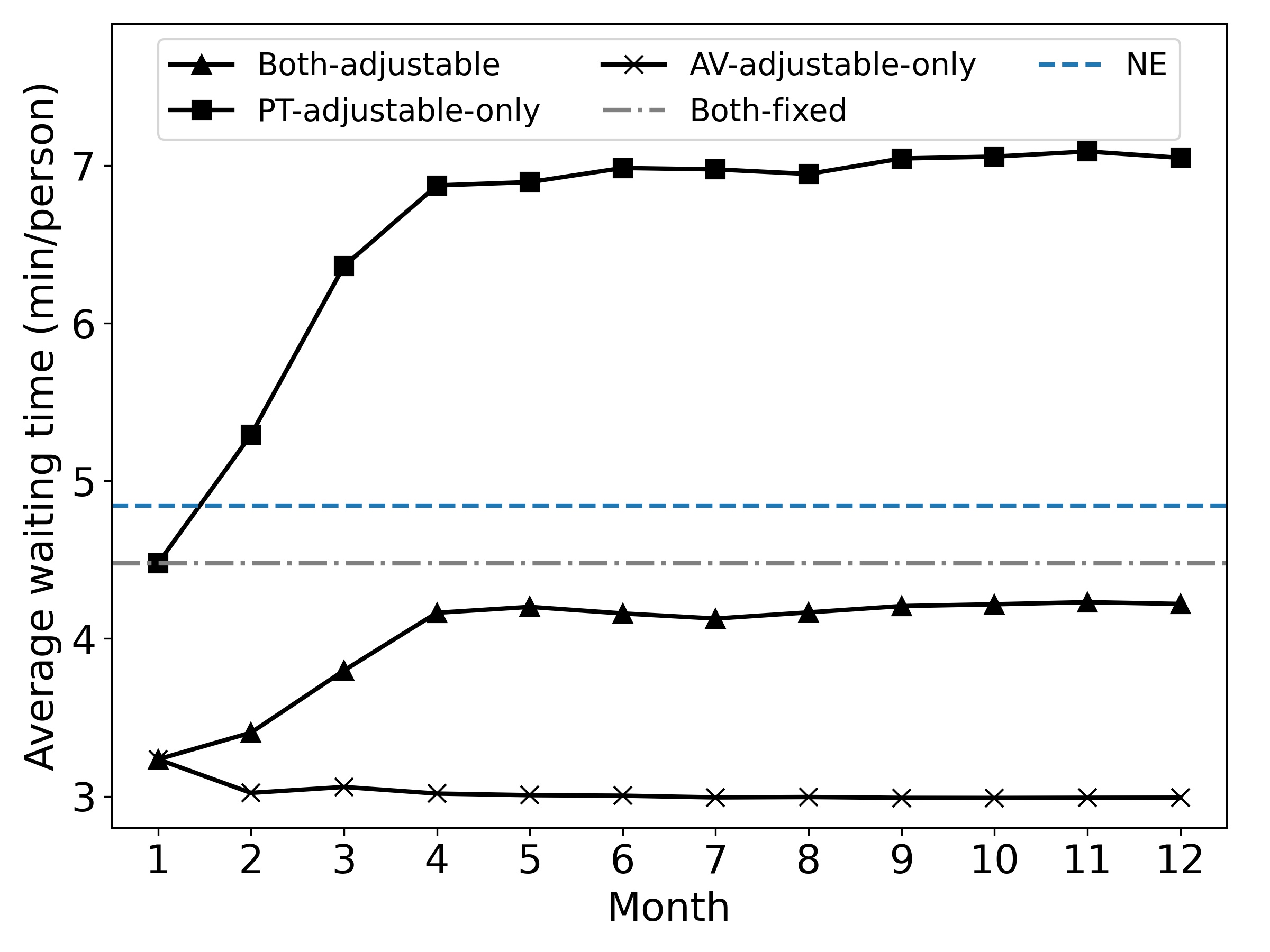}\label{fig_waiting_time}}
\hfil
\subfloat[Generalized travel cost]{\includegraphics[width=0.4\textwidth]{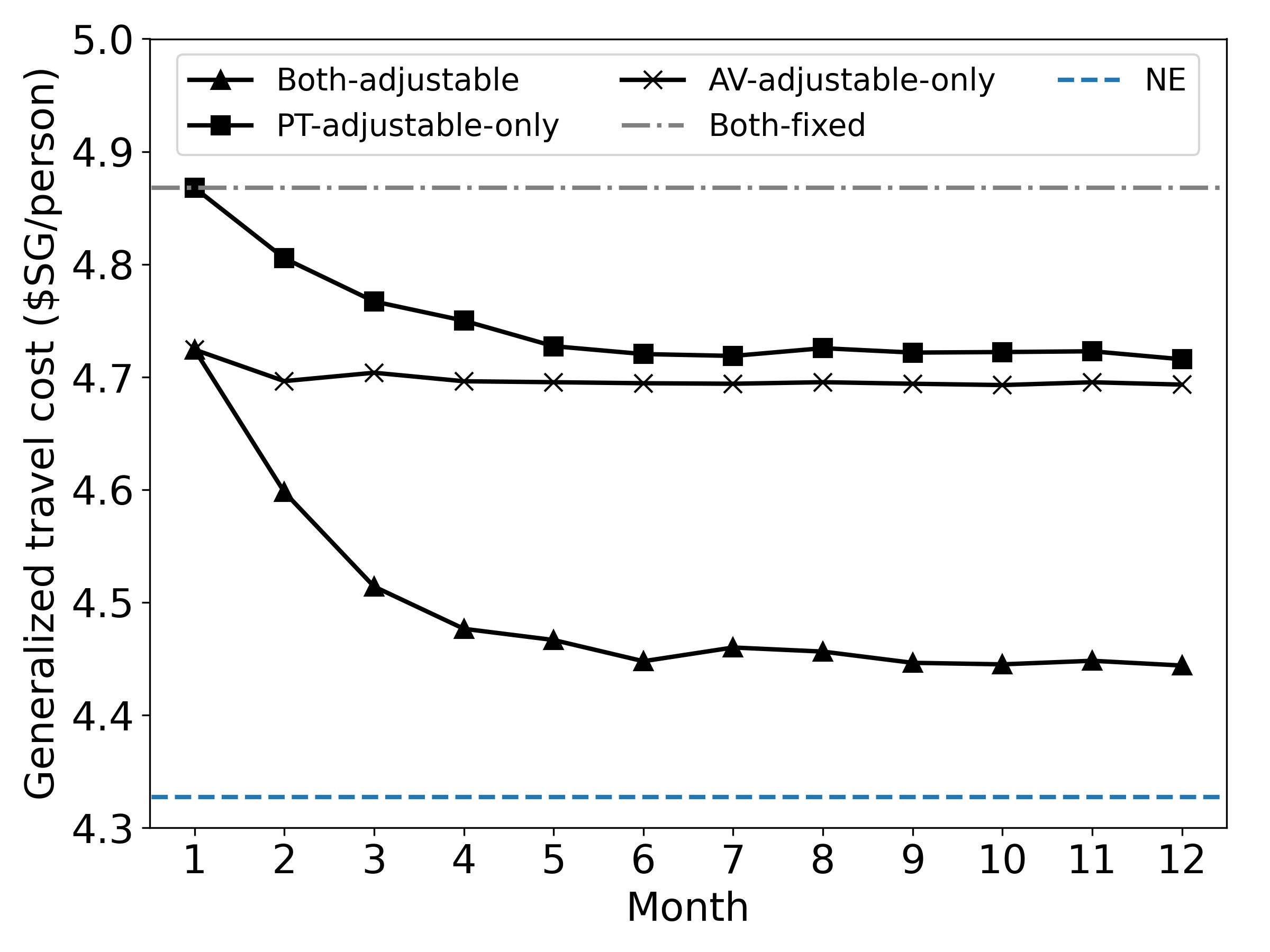}\label{fig_generalzied_cost}}
\caption{Level of service change}
\label{fig_level_service}
\end{figure}

\begin{figure}[htb]
\centering
\subfloat[AV demand]{\includegraphics[width=0.33\textwidth]{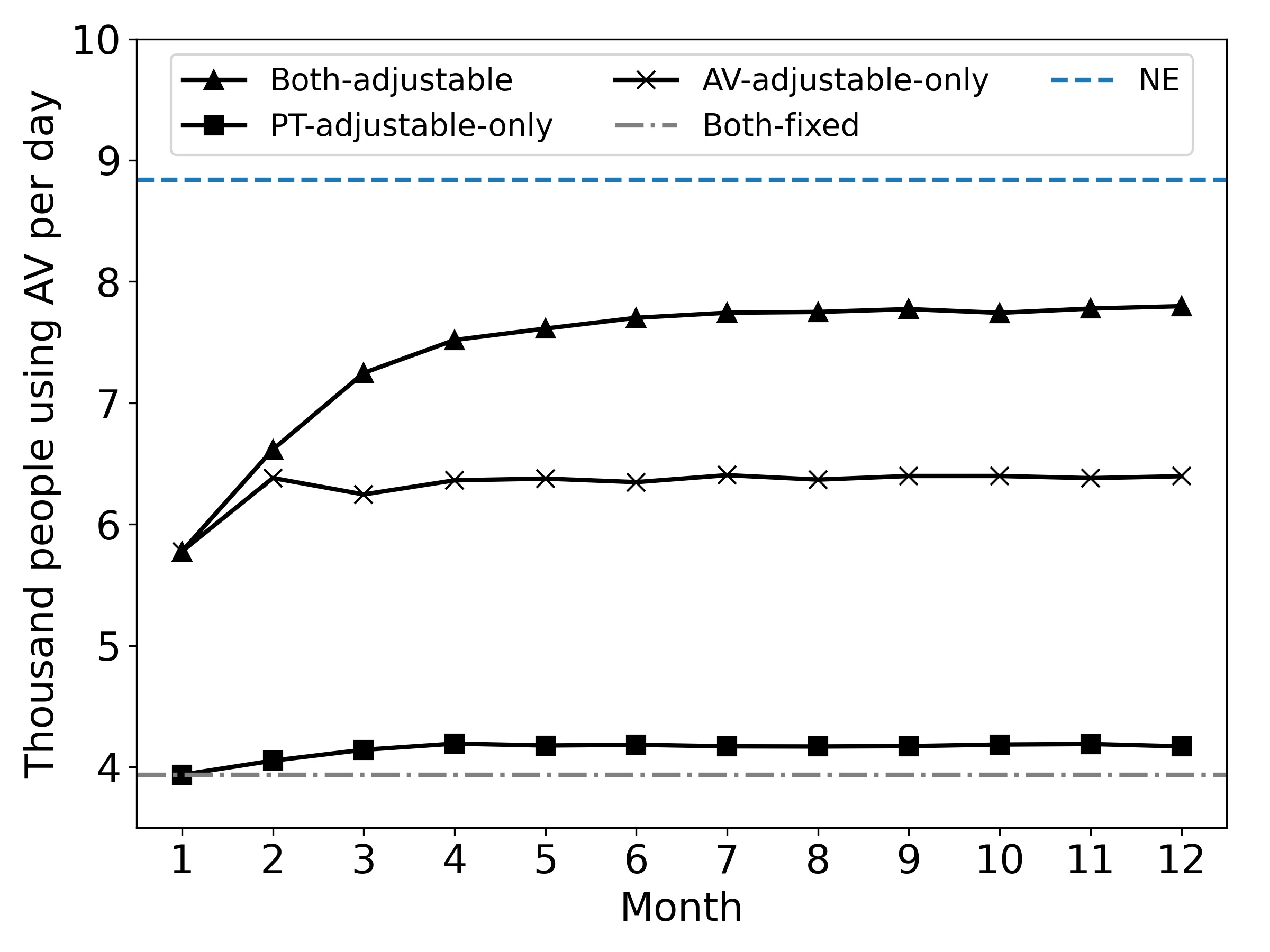}}
\hfil
\subfloat[Bus demand]{\includegraphics[width=0.33\textwidth]{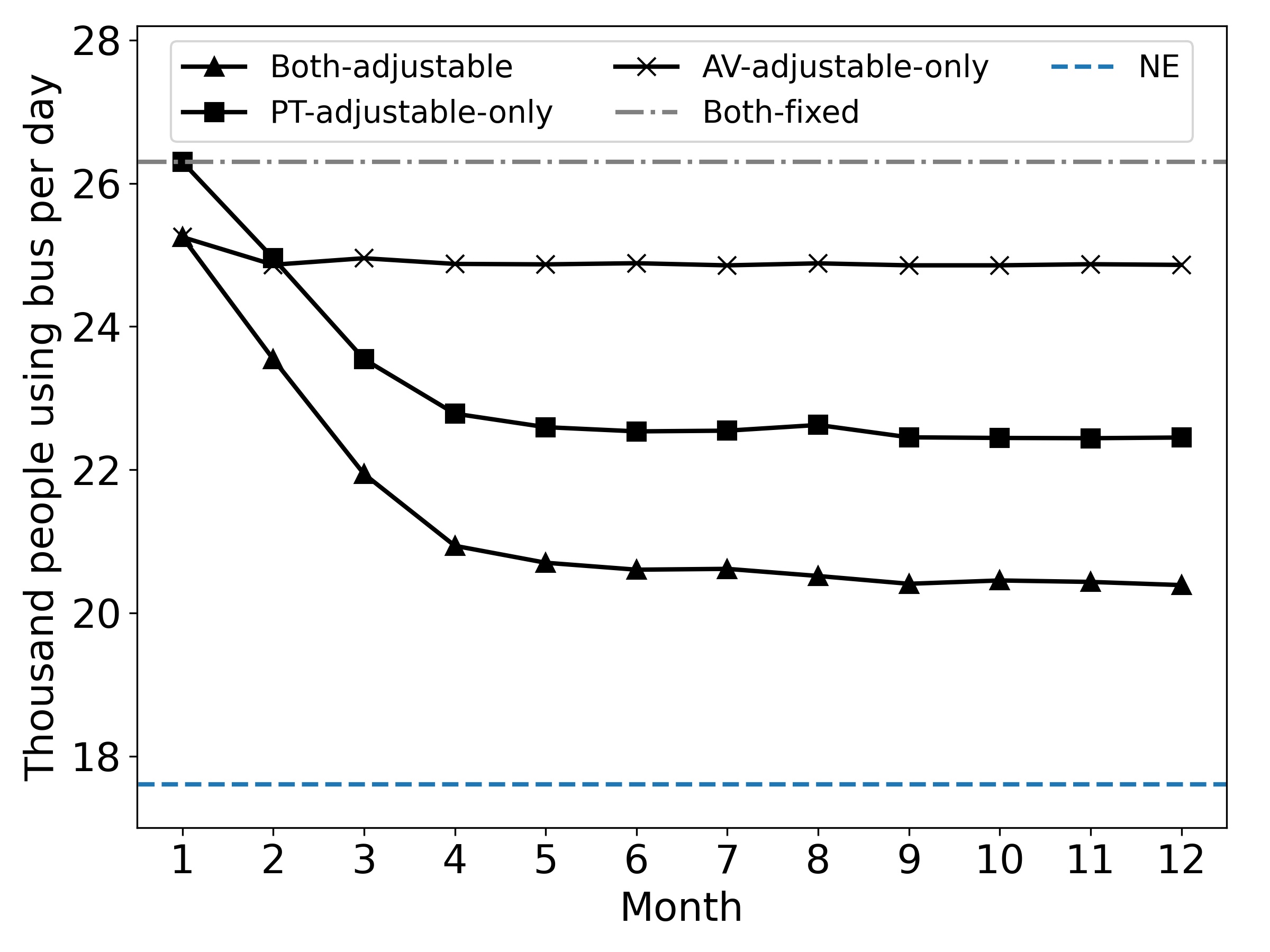}}
\hfil
\subfloat[Walking demand]{\includegraphics[width=0.33\textwidth]{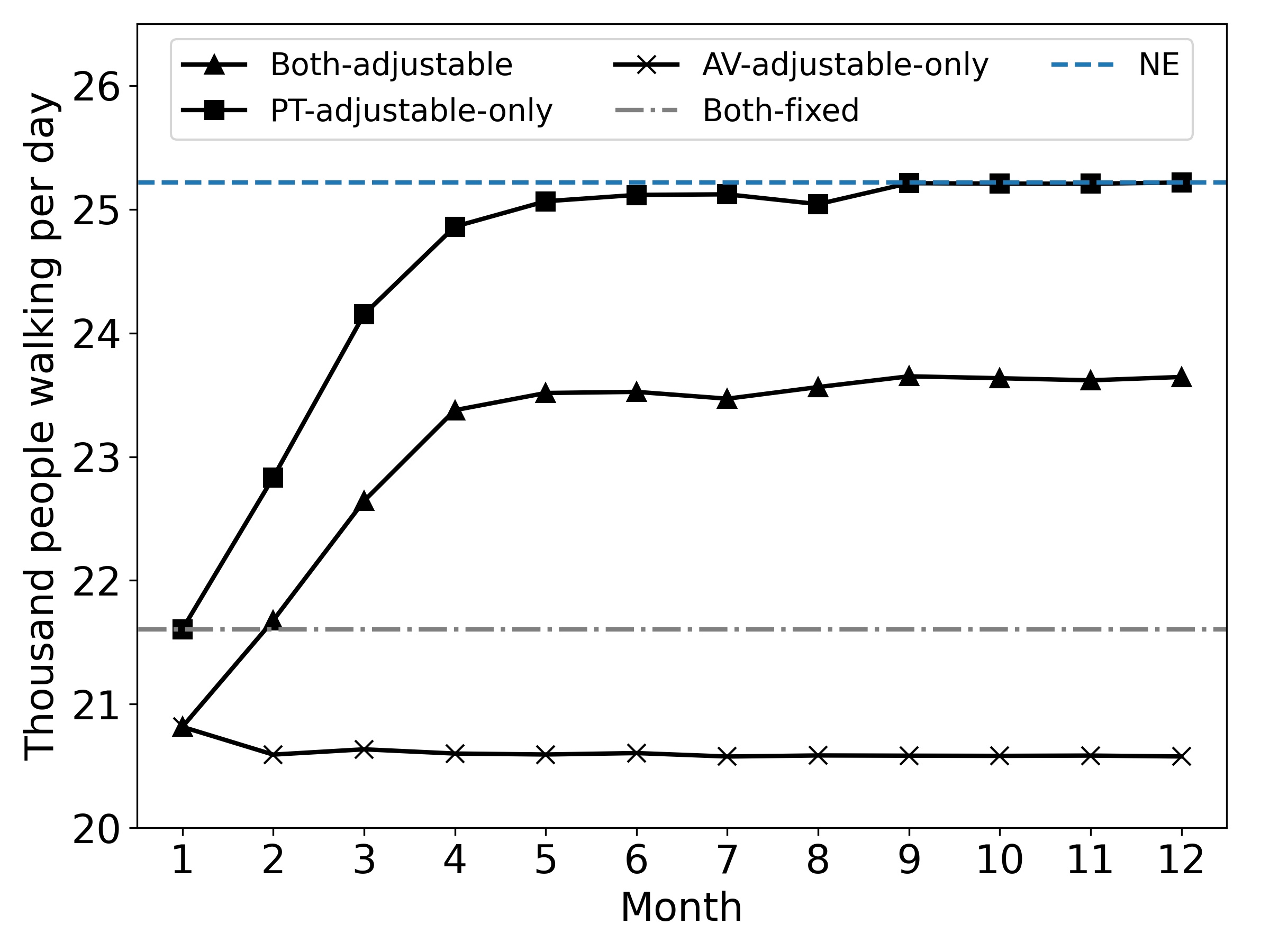}}
\caption{Change of passenger mode choices}
\label{fig_Passenger_Mode_choice}
\end{figure}

Figure \ref{fig_level_service} plots the changes in the level-of-service indicators: travel cost, total travel time, waiting time, and generalized cost. Passengers' travel cost shows different patterns: it decreases in the \emph{PT-adjustable-only} scenario and increases in all other scenarios, although the magnitude of the changes is small. Because AV is more expensive than PT, when AV starts to compete and serve more demand, the average travel cost will increase. When there are only buses adjusting the supply, a greater number of people will switch to walking and only a small proportion of people will convert to AV (Figure \ref{fig_Passenger_Mode_choice})---the average travel cost will decrease. The \emph{NE} scenario observes the highest average travel cost due to the highest market share of AV.

All scenarios show decreasing trends for the total travel time. In the \emph{PT-adjustable-only} scenario, passengers' travel time and travel cost both decrease after the adjustment of the bus supply, which implies absolute benefits to passengers. Conversely, in the \emph{both-adjustable} and \emph{AV-adjustable-only} scenarios, the directions of the changes in the travel time and cost differ. To capture the combined effect of travel time and travel cost, we calculated the generalized travel cost based on the method described in Section \ref{eval}. 

As shown in Figure \ref{fig_generalzied_cost}, passengers' generalized travel cost decreases in all scenarios, among which the \emph{NE} scenario shows the largest decline and \emph{both-adjustable} scenario ranks the second. The supply adjustment of buses and AV not only benefits the operators but also the passengers. The major contributing factor to the changes in generalized travel cost is the decrease in travel time. 

Another important indicator for passengers is waiting time. In the \emph{PT-adjustable-only} scenario, passengers' waiting time increases as a direct consequence of the reduction of the bus supply (Figure \ref{fig_waiting_time}). In the \emph{AV-adjustable-only} scenario, passengers' waiting time decreases because of the increased AV supply. In the \emph{both-adjustable} scenario, we observe a combined effect of the two. The gap between the first month and \emph{both-fixed} scenario is caused by the significant increase in the AV supply in the first month. Starting from the first month, passengers' waiting time gradually increases because the effect of bus supply reduction starts to manifest. Nonetheless, due to the opposite effect from the AV side, the increase in waiting time is not as pronounced as in the \emph{PT-adjustable-only} scenario. The converged waiting time is still shorter than the \emph{both-fixed} scenario. The \emph{NE} scenario has a slightly higher waiting time than the \emph{both-fixed} scenario.

Figure \ref{fig_Passenger_Mode_choice} shows the passenger mode choice: the demand for AV increases, the demand for buses decreases and the demand for walking varies across scenarios. In the \emph{both-adjustable} and \emph{PT-adjustable-only} scenarios, more people turn to walk due to the reduced bus supply, whereas in the \emph{AV-adjustable-only} scenario, fewer people walk because of the increased supply of AV. 

In addition to the aggregate demand changes, we explore the attributes of the passengers who change their mode choices, i.e. from bus to AV or walking. For illustration purposes, we only show the analysis for the \emph{both-adjustable} scenario. We examine two passenger attributes, the household income and the distance from home to the MRT station. People who originally chose buses were set as the control group. People who changed their travel modes from bus to AV or walking were set as the experimental group. Our purpose is to test whether there is a statistically significant difference between the control groups and the experimental groups in their household income and distance to the MRT station. Table \ref{tab_KS_test} summarizes the results of the two-sample Kolmogorov-Smirnov (KS) test. For household income, the attributes of the people who changed their mode choice from bus to AV are significantly higher than those of the control group. Higher-income people tend to change from bus to AV after a decline in the bus supply. However, for people who chose walking, household income does not show a significant difference. In terms of the distance to the MRT station, the results for both experimental groups are significantly different from those for the control group. People who live near MRT stations tend to walk more, while people who live far from MRT stations tend to convert to AV. This implies that the interaction of AV and PT may sharpen people's travel mode choices and increase the dependence on cars for people living far from subway stations. 

\begin{table}[htb]
\caption{KS Test Results for the AV--PT Scenario}
\label{tab_KS_test}
\centering
\begin{threeparttable}
\small
\begin{tabular}{c|c|c|c|l} \hline
\textbf{Attributes} & \textbf{Groups} & \textbf{Mean (Std.)} & \textbf{KS Statistics} & \textbf{$p$-values}\\ \hline
\multirow{3}{*}{Household Income (SGD)} & Baseline & 4760.3 (3689.5) & N.A. & N.A.  \\ 
 & Experiment (Bus to AV) & 5134.9 (3697.0) & 0.061 & 0.000* \\ 
 & Experiment (Bus to Walk) & 4731.3 (3668.9) & 0.008 & 0.795   \\
 \hline
 \multirow{3}{*}{Distance to MRT Station (m)} & Baseline & 1065.0 (333.1) & N.A. & N.A. \\ 
 & Experiment (Bus to AV) & 1112.7 (320.4) & 0.064 & 0.000* \\ 
 & Experiment (Bus to Walk) & 933.1 (313.5) & 0.166 & 0.000* \\ 
 \hline
\end{tabular}
\begin{tablenotes}\footnotesize
\item[-] Control groups are set as reference, so they do not have KS statistics and p-value.
\item[-] The larger the KS Statistics, the greater the difference between the two groups.
\item[-] *: Significant at 99\% confidence level. 
\end{tablenotes}
\end{threeparttable}
\end{table}

\subsection{Transport authority perspective}\label{Transport_per}
\begin{figure}[htb]
\centering
\subfloat[AV average load]{\includegraphics[width=0.34\textwidth]{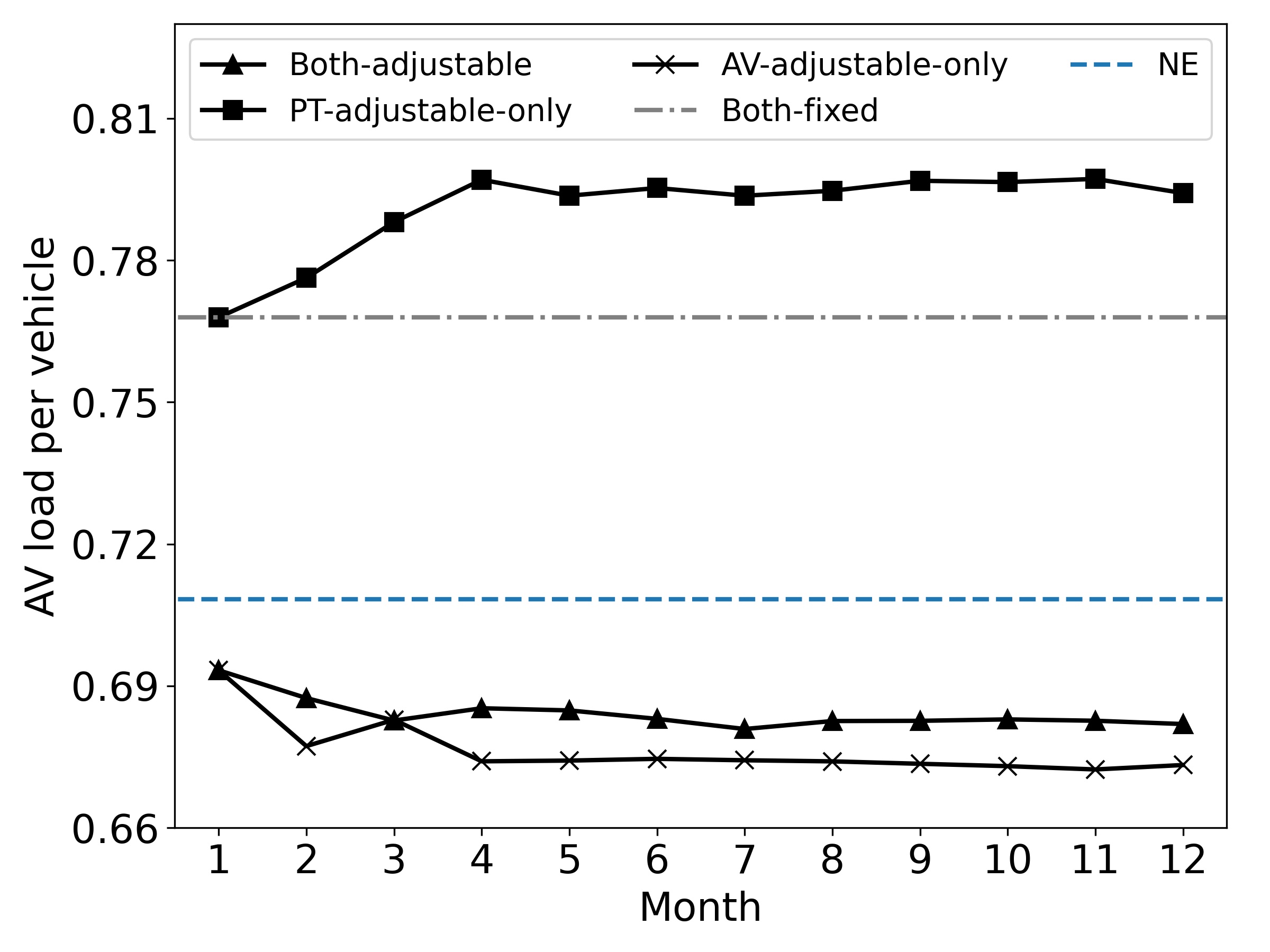}\label{fig_Transport_Eff_av}}
\hfil
\subfloat[Bus average load]{\includegraphics[width=0.34\textwidth]{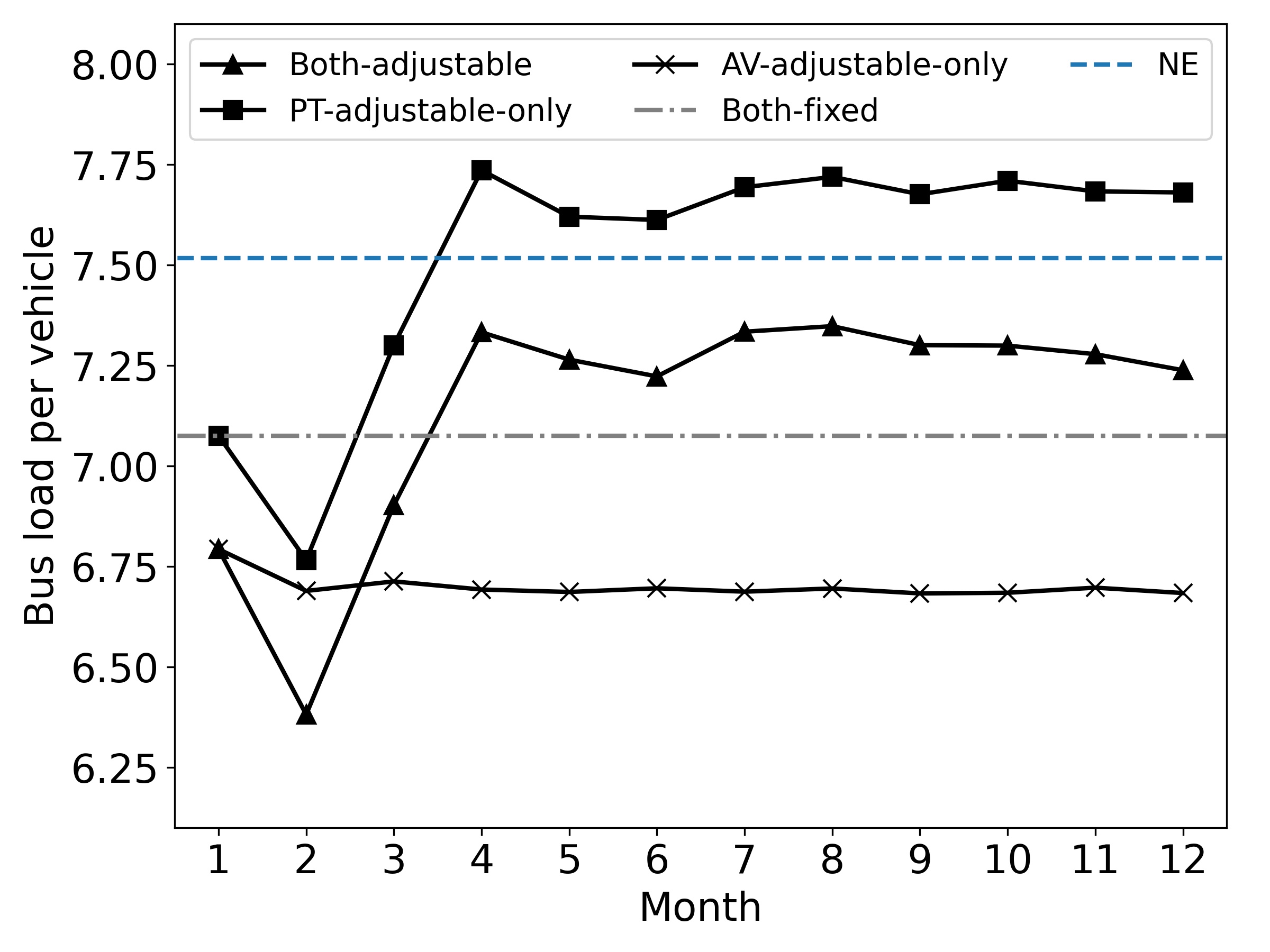}\label{fig_Transport_Eff_bus}}
\hfil
\subfloat[AV VKT]{\includegraphics[width=0.33\textwidth]{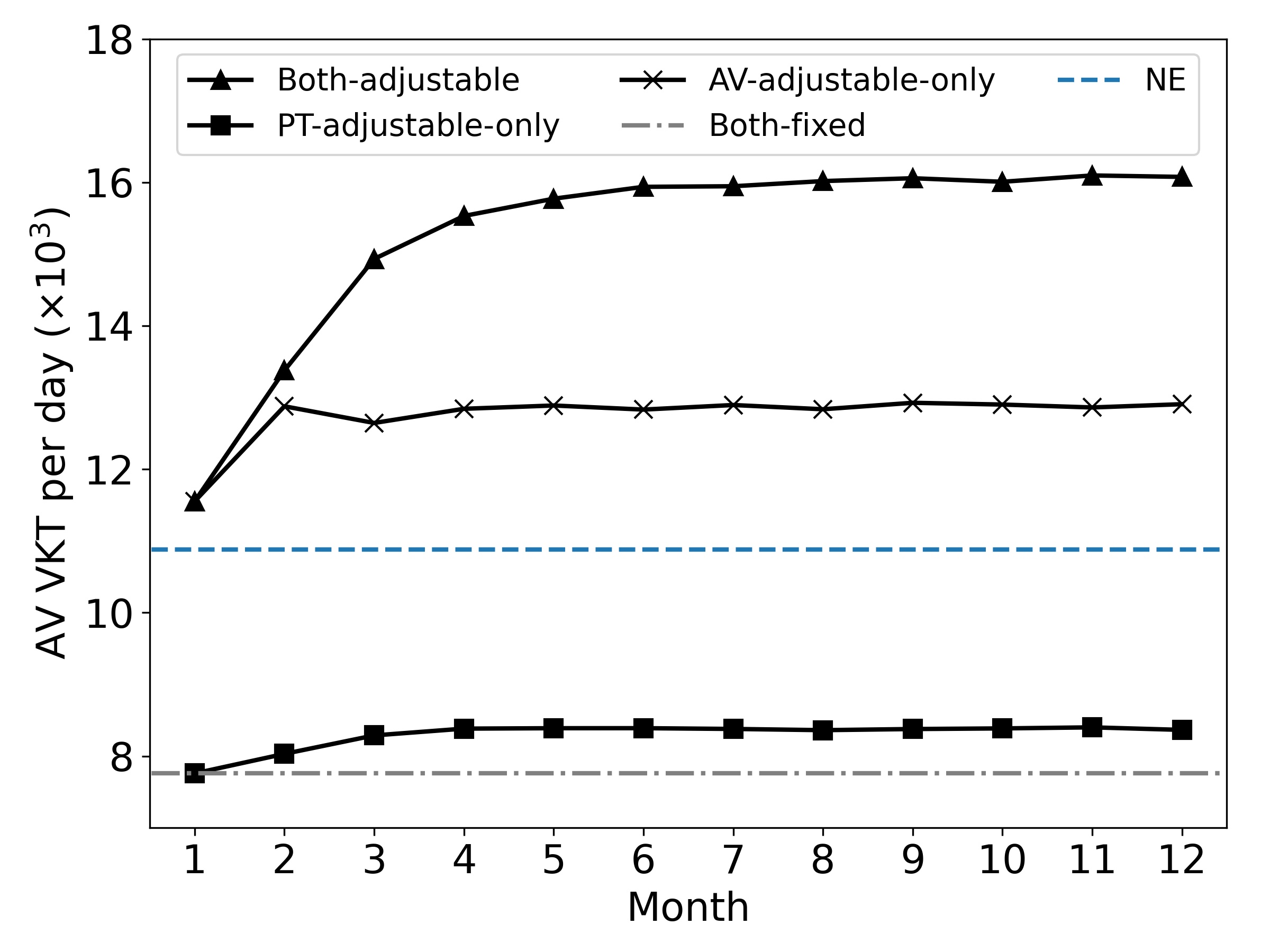}\label{fig_sustain_AV}}
\hfil
\subfloat[Bus VKT]{\includegraphics[width=0.33\textwidth]{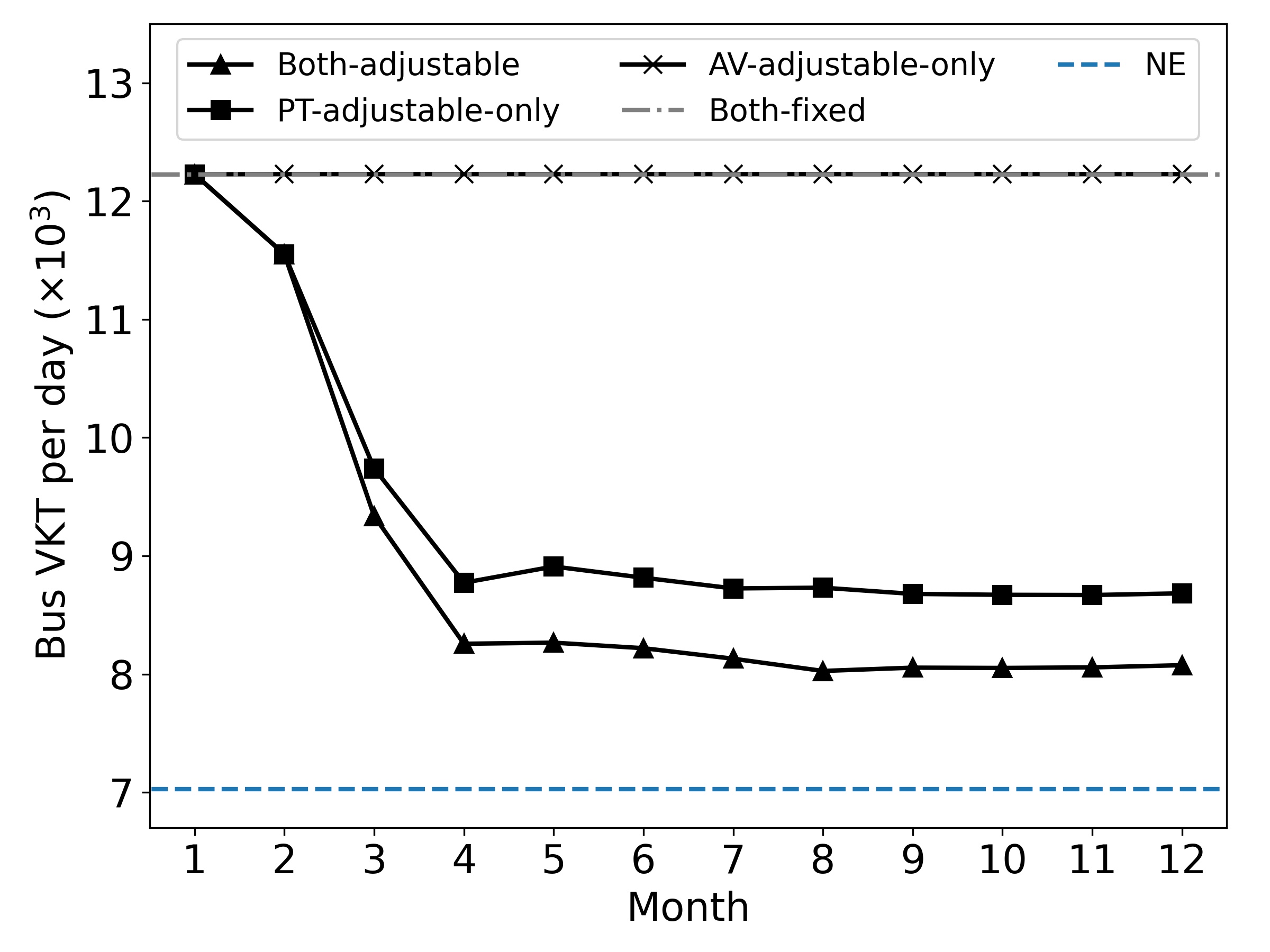}\label{fig_sustain_bus}}
\hfil
\subfloat[Total PCE (bus + AV)]{\includegraphics[width=0.33\textwidth]{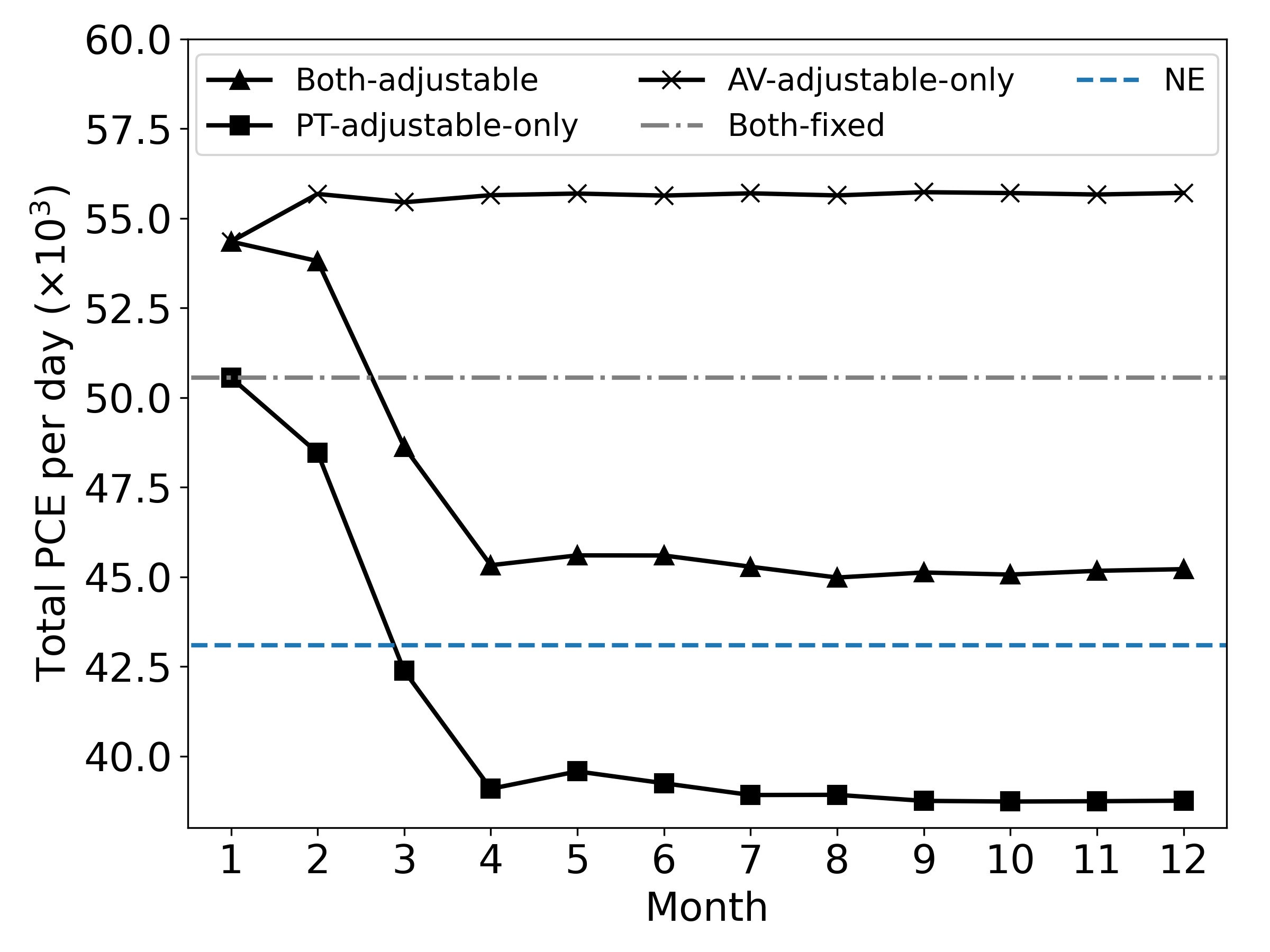}\label{fig_sustain_all}}
\caption{Change of transport efficiency indicators}
\label{fig_Transport_Eff}
\end{figure}

We consider efficiency and vehicle kilometers traveled from the transport authority perspective as shown. In Figure \ref{fig_Transport_Eff_av} and \ref{fig_Transport_Eff_bus}, the average load of AV slightly increases in the \emph{PT-adjustable-only} scenario. However, in the \emph{both-adjustable} and \emph{AV-adjustable-only} scenarios, the AV load decreases considerably in the first month (the gap between the \emph{both-fixed} scenario and the first month) and then stabilizes. This suggests that to maximize profit, ride-sharing behavior is inhibited in the model. Since travel distance is usually short in the first-mile scenarios, ride-sharing is less likely to happen or generate higher profits. Considering the proposed price structure of AMoD, serving two passengers with two vehicles separately may yield more profit. Therefore, although the AMoD operator earns more profit, the AV's operating efficiency is lower. 

In contrast to AV, the average bus load increases for the \emph{both-adjustable} and \emph{PT-adjustable-only} scenarios, which means that after the competition, the buses are operated more efficiently, with not only higher profit, but also higher average load.

The \emph{NE} scenario observes relatively low AV average load, and relatively high bus average load, but stays in the middle positions among all scenarios.

Overall, the VKT of AV increases, while the VKT of buses decreases, which is consistent with the change in supply (Figure \ref{fig_sustain_AV}-\ref{fig_sustain_all}). Since the unit PCE for the bus is large, the shape of the total PCE is similar to that of the bus VKT. The system total PCE decreases in the \emph{both-adjustable} and \emph{PT-adjustable-only} scenarios, which means that the deregulation of the bus operator has the potential to reduce the environmental impacts. When the bus system is not allowed to change its supply, the total PCE increases owing to the increase in the VKT of AV. The \emph{NE} scenario has the lowest bus VKT, relatively low VKT of AVs, and overall relatively low total PCE.

\subsection{Sensitivity analysis on PT subsidies}\label{subsidy}
In this section, we compare five scenarios of PT subsidies including demand-based subsidies (0, 1, and 2 SGD per passenger trip) and supply-based subsidies (0, 0.5, and 1 SGD per km operated). It turns out the two subsidy regimes show largely similar results and therefore we focus on the demand-based subsidies in this section and leave the discussion on the supply-based subsidies in \ref{sec_supply_subsidy}. 

\subsubsection{PT perspective}
\begin{figure}[H]
\centering
\subfloat[PT profit]{\includegraphics[width=0.33\textwidth]{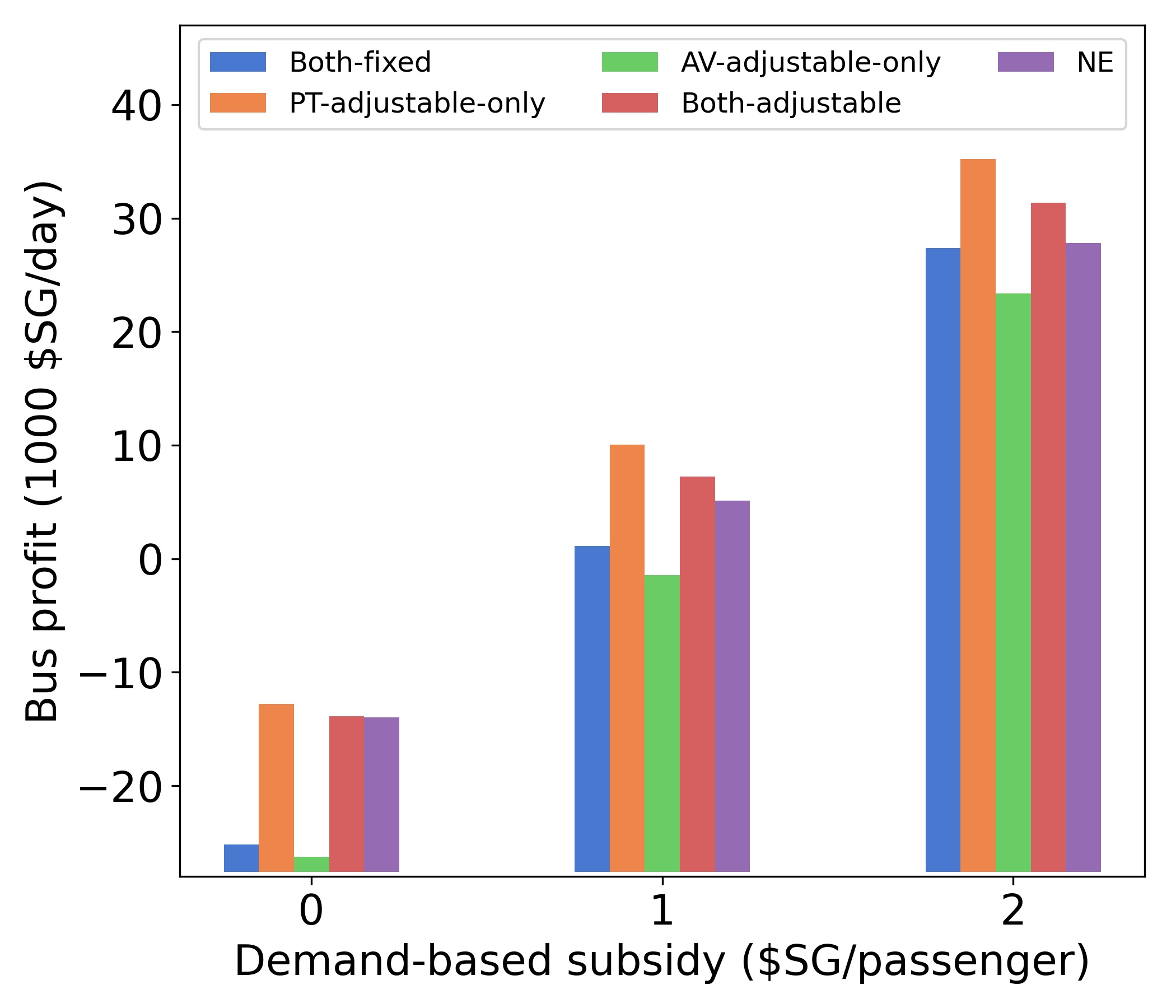}}
\hfil
\subfloat[PT supply]{\includegraphics[width=0.33\textwidth]{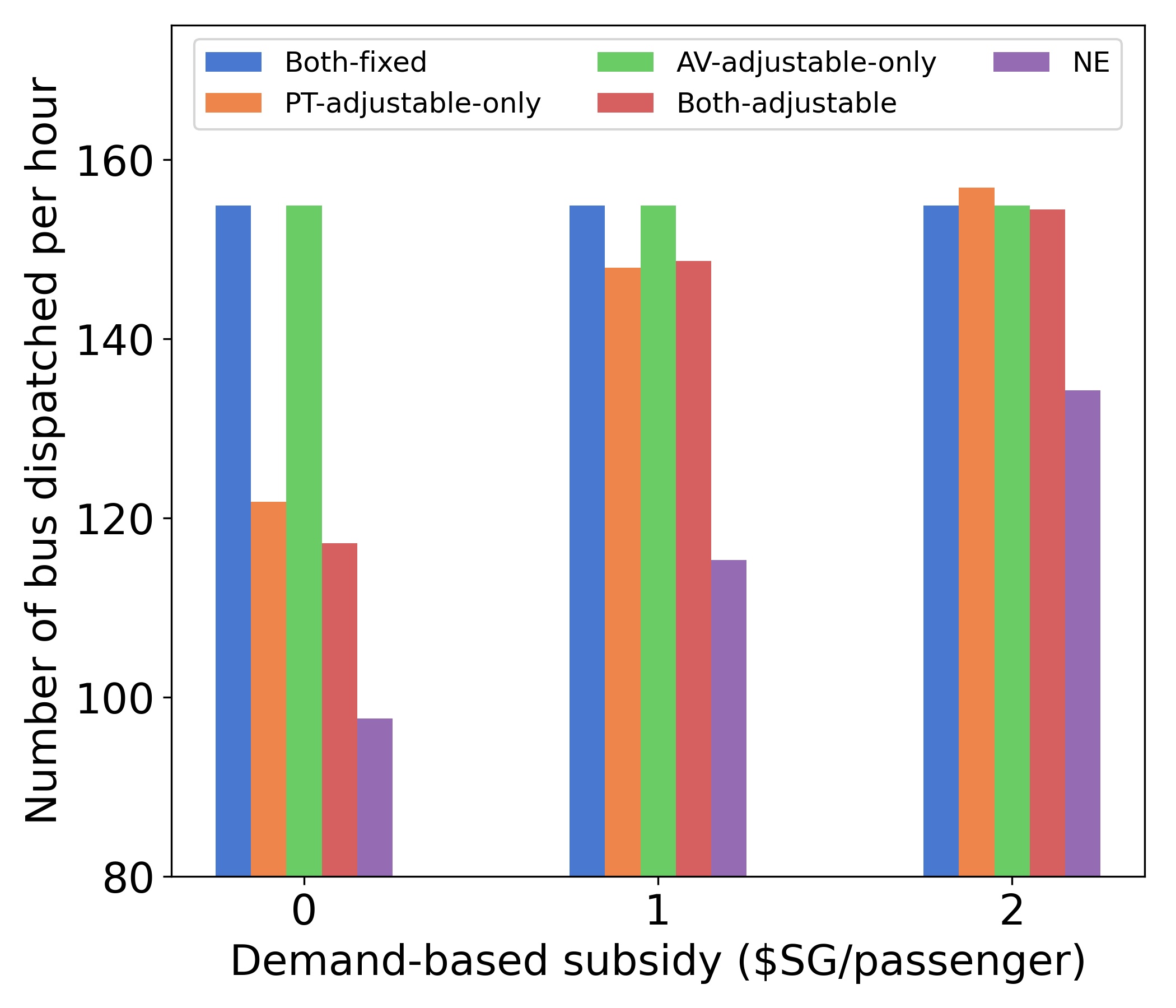}}
\hfil
\subfloat[PT market share]{\includegraphics[width=0.33\textwidth]{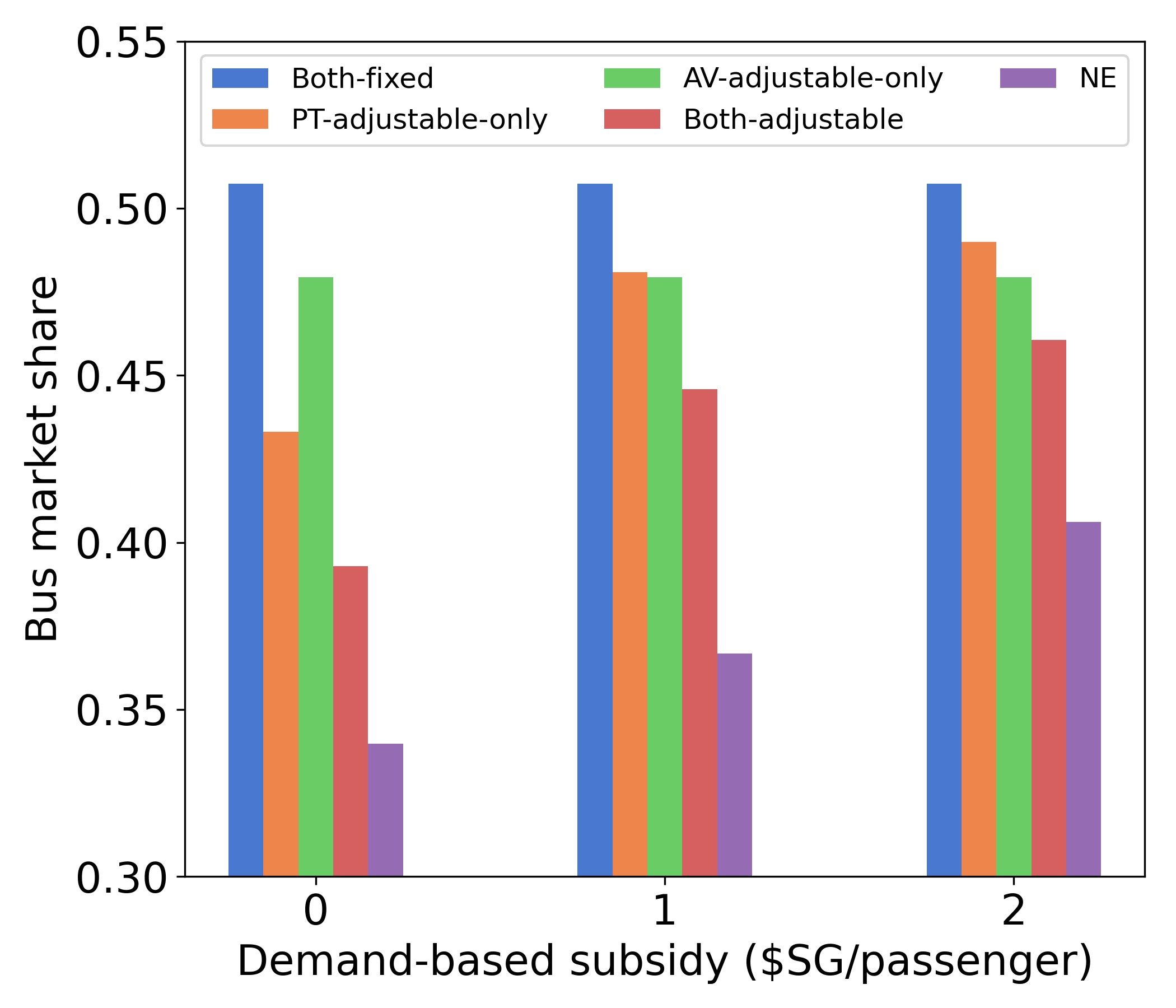}}
\caption{Impact of demand-based subsidies on PT operators. All bars represent the converged values of the corresponding scenarios.}
\label{fig_PT_demand_based_subsidy}
\end{figure}

Figure \ref{fig_PT_demand_based_subsidy} shows the results of PT's profit, supply, and market share under three different demand-based subsidies. As expected higher demand-based subsidies result in an increase in PT profit, supply and market share because the PT operator earns more money per passenger and thus maintains higher supply and market share. Critically the PT profit turns positive when the subsidy level is greater or equal to 1 SGD/passenger. The PT supply and market share of \emph{AV-adjustable-only} scenario are not affected by the PT subsidy. The change patterns of \emph{PT-adjustable-only}, \emph{both-adjustable}, and \emph{NE} scenarios are consistent and similar.

\subsubsection{AMoD perspective}
\begin{figure}[H]
\centering
\subfloat[AMoD profit]{\includegraphics[width=0.33\textwidth]{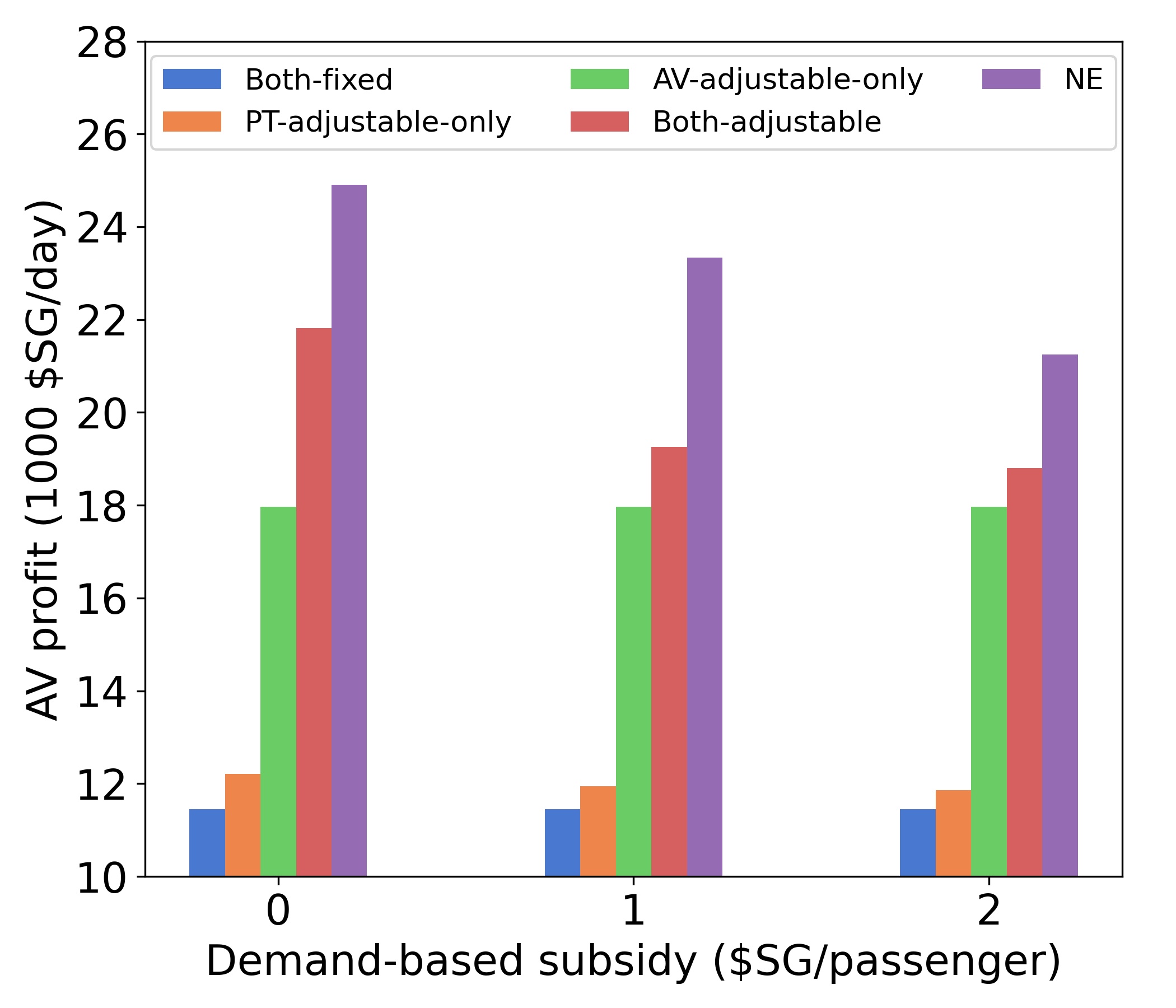}}
\hfil
\subfloat[AMoD supply]{\includegraphics[width=0.33\textwidth]{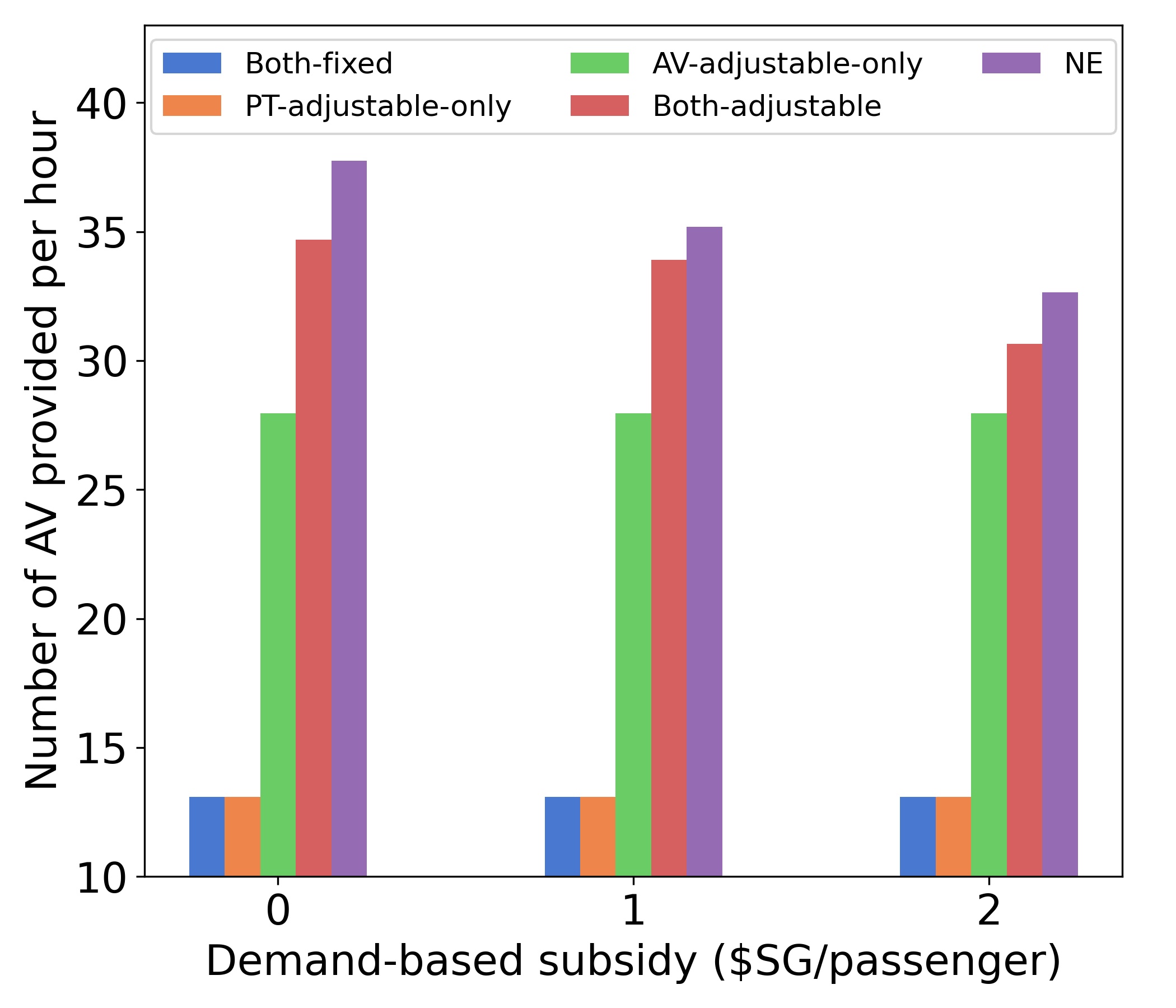}}
\hfil
\subfloat[AMoD market share]{\includegraphics[width=0.33\textwidth]{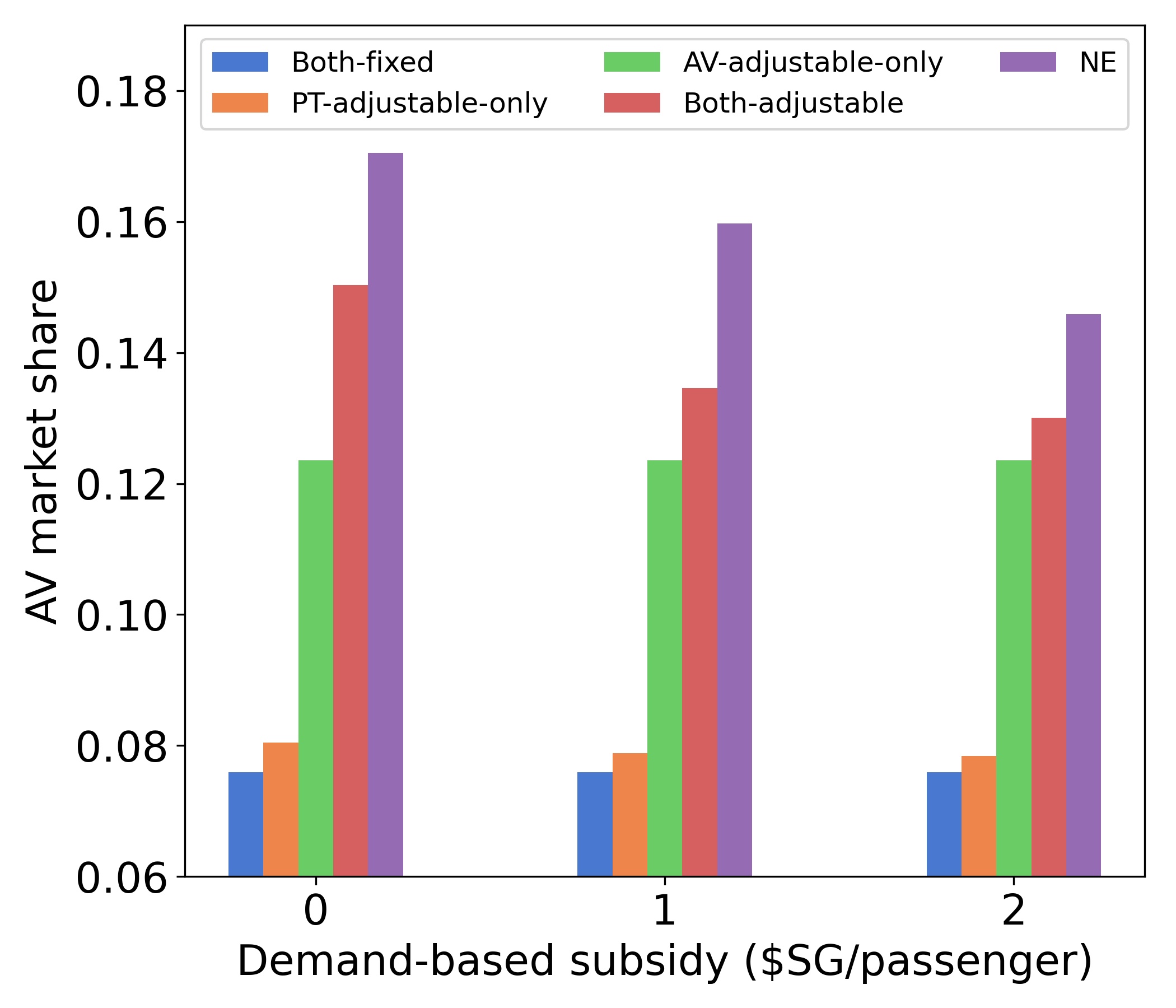}}
\caption{Impact of demand-based subsidies on AMoD operators}
\label{fig_AV_demand_based_subsidy}
\end{figure}

As shown in Figure \ref{fig_AV_demand_based_subsidy}, AV's profit, supply, and market share all decrease with the increased PT's subsidy. Those changes are more prominent in \emph{both-adjustable} and \emph{NE} scenarios. \emph{AV-adjustable-only} scenario is not affected because the updating of AV only depends on the environment and PT's supply, which is fixed by definition in this scenario. The impact on \emph{PT-adjustable-only} scenario is negligible. 

\subsubsection{Passenger perspective}
\begin{figure}[H]
\centering
\subfloat[Travel cost]{\includegraphics[width=0.4\textwidth]{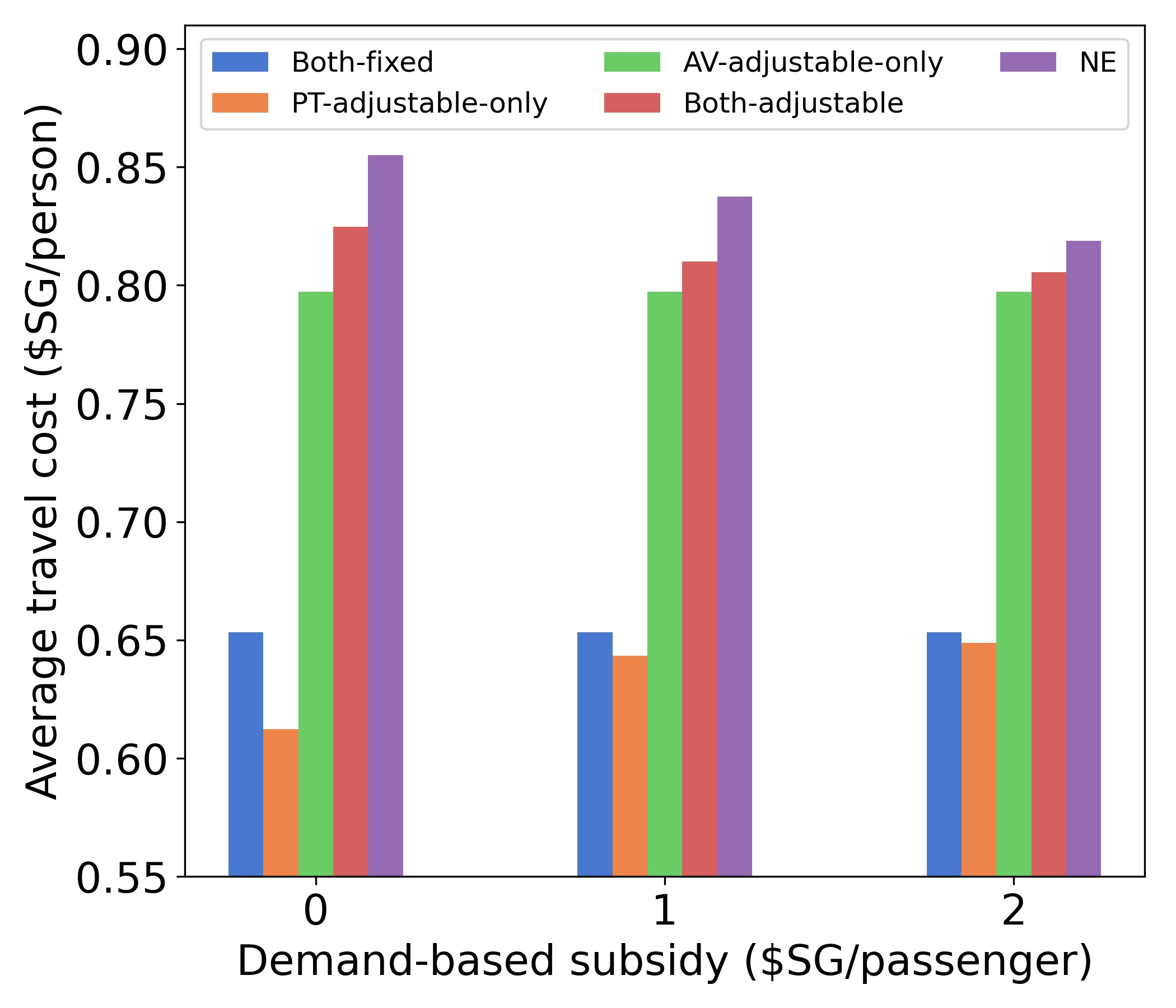}\label{fig_pax_demand_travel_cost}}
\hfil
\subfloat[Total travel time]{\includegraphics[width=0.4\textwidth]{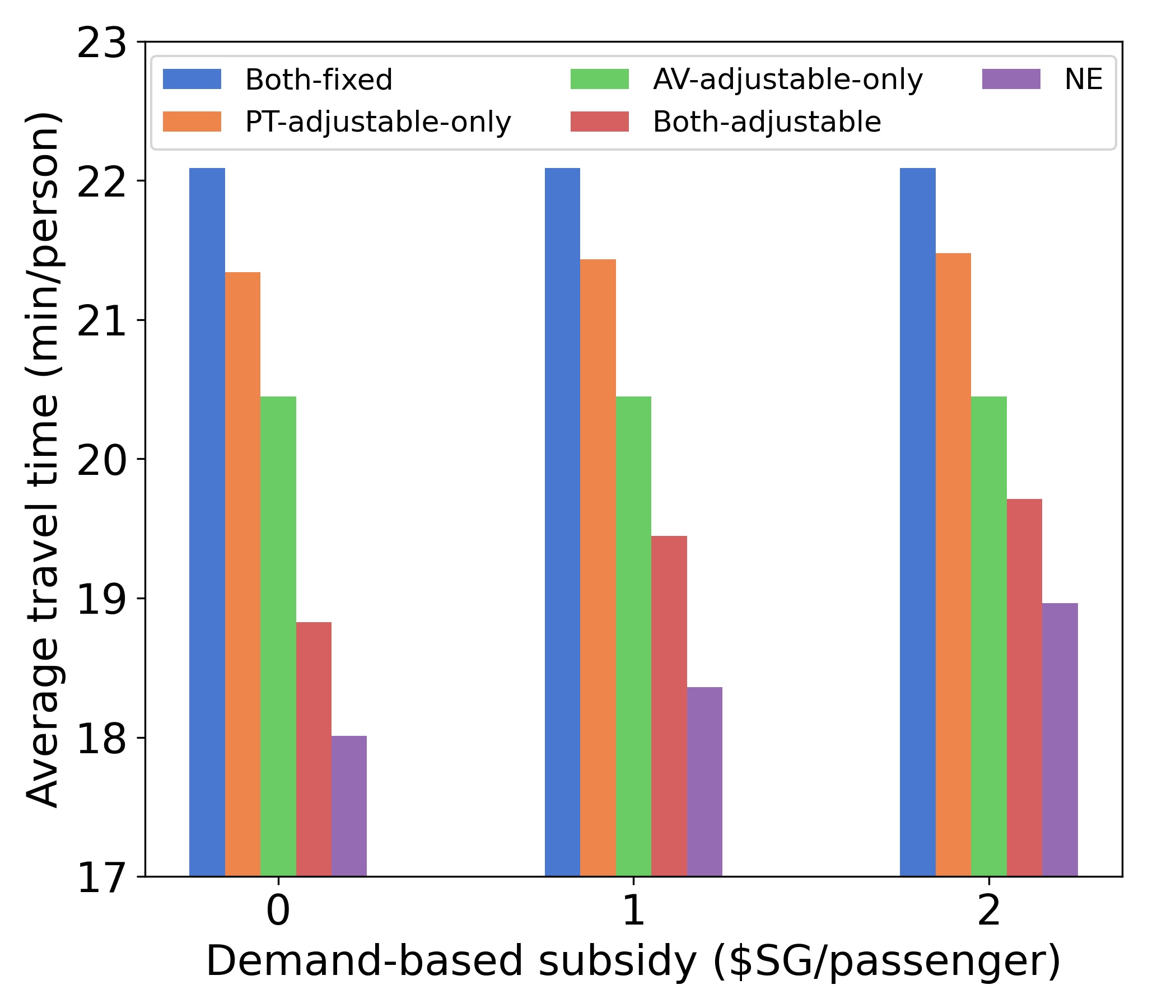}\label{fig_pax_demand_travel_time}}
\hfil
\subfloat[Waiting time]{\includegraphics[width=0.4\textwidth]{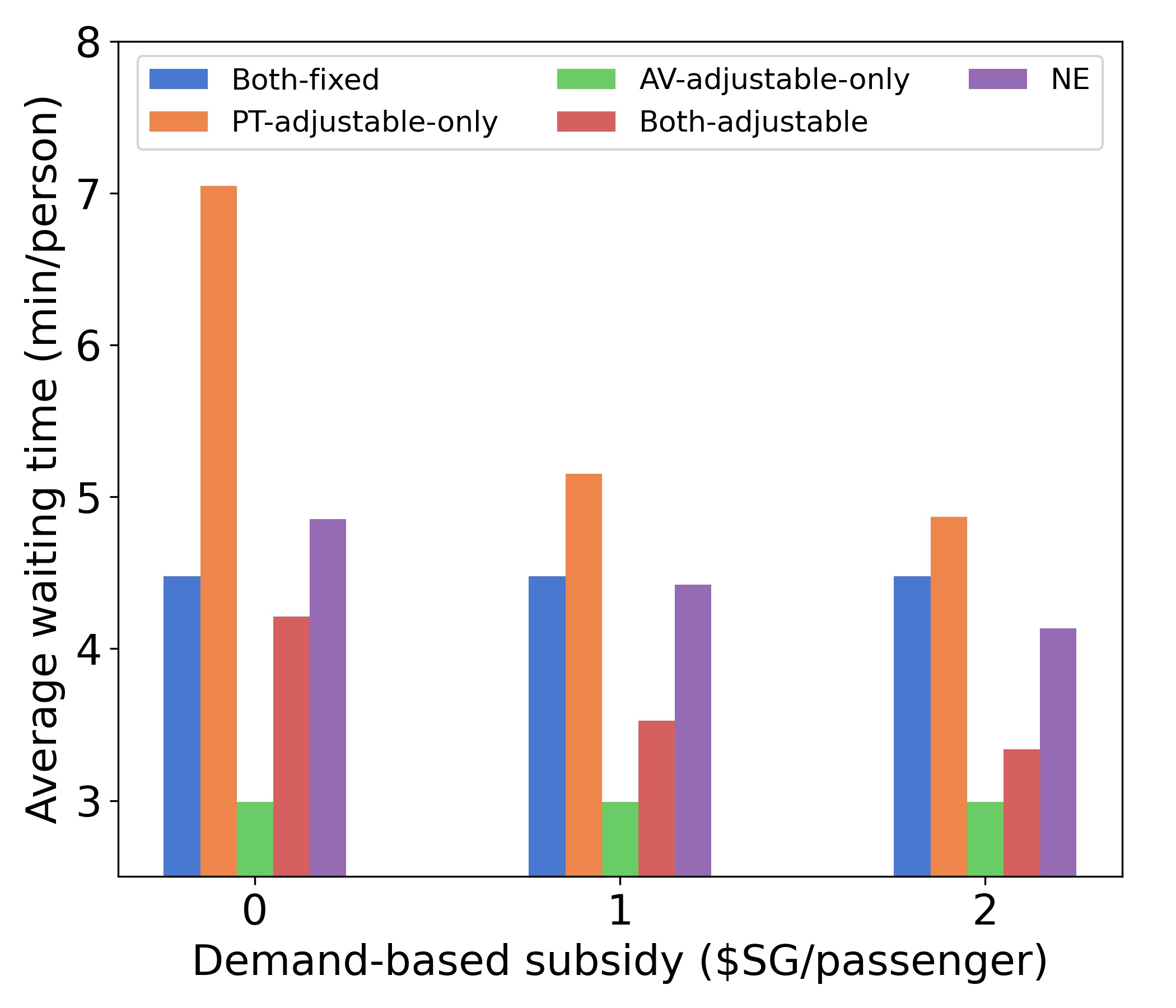}\label{fig_pax_demand_waiting_time}}
\hfil
\subfloat[Generalized travel cost]{\includegraphics[width=0.4\textwidth]{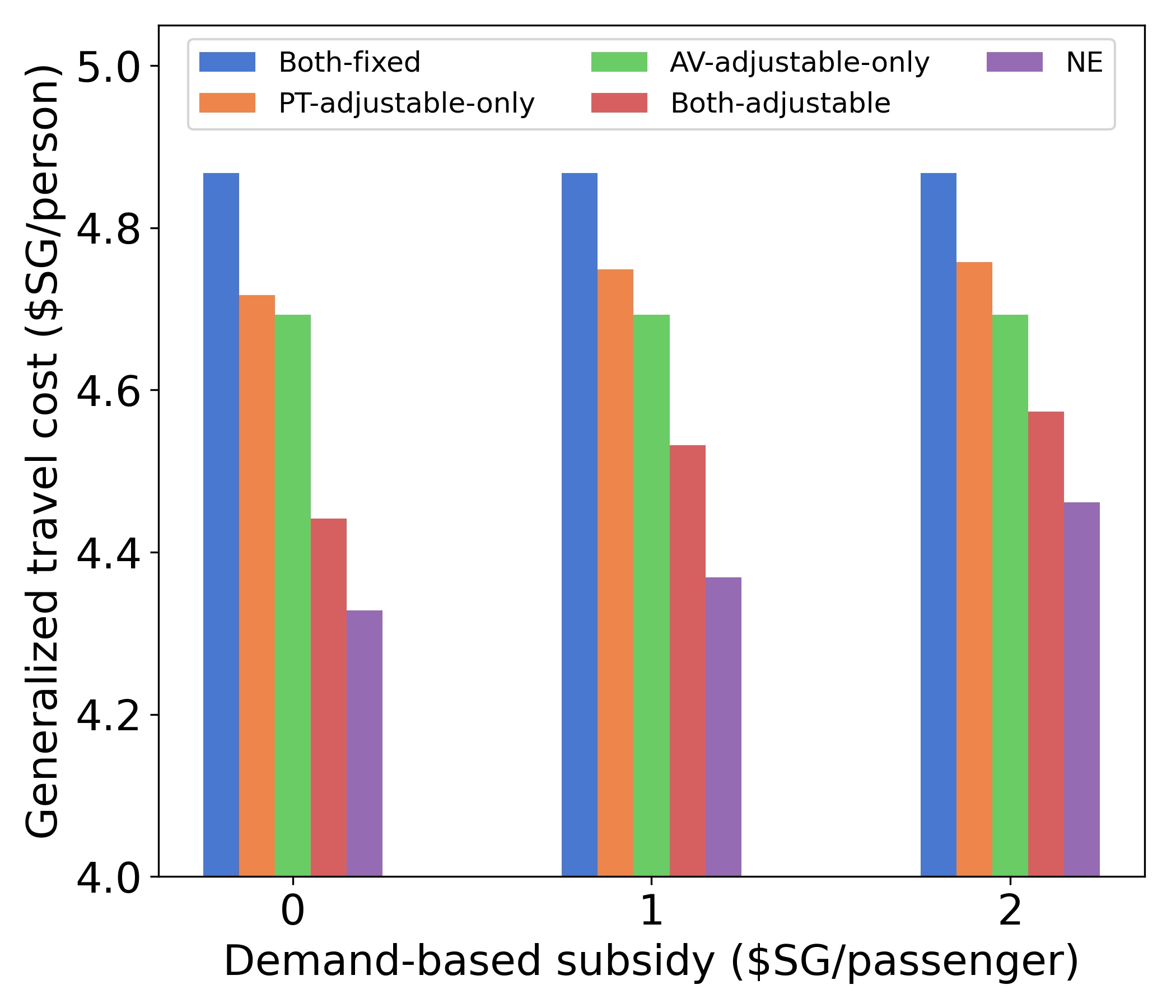}\label{fig_pax_demand_generalized_cost}}
\caption{Impact of demand-based subsidies on passengers}
\label{fig_pax_demand_based_subsidy}
\end{figure}

From the passenger's perspective, we plot travel cost, travel time, waiting time, and generalized travel cost averaged across all modes. For the travel costs (Figure \ref{fig_pax_demand_travel_cost}), we observe a slight decrease in \emph{both-adjustable} and \emph{NE} scenarios, but a slight increase in the \emph{PT-adjustable-only} scenario. In the \emph{both-adjustable} and \emph{NE} scenarios, the subsidy allows PT to attracts more passengers from AV, thus decreasing the travel cost averaged over all modes. But in the \emph{PT-adjustable-only} scenario where AV is not allowed to adjust its supply, the better PT service mostly attracts walking passengers, which increases the travel cost average over all modes. The total travel time  increases with the PT subsidy in \emph{both-adjustable} and \emph{NE} scenarios (Figure \ref{fig_pax_demand_travel_time}) again because PT subsidies attract some passengers from AV to bus. We also observe a decrease in waiting time with the increase in subsidy (Figure \ref{fig_pax_demand_waiting_time}), suggesting that the effect of increased bus supply on waiting time dwarfs that of decreased AV supply. 

\begin{figure}[H]
\centering
\subfloat[Generalized cost]{\includegraphics[width=0.4\textwidth]{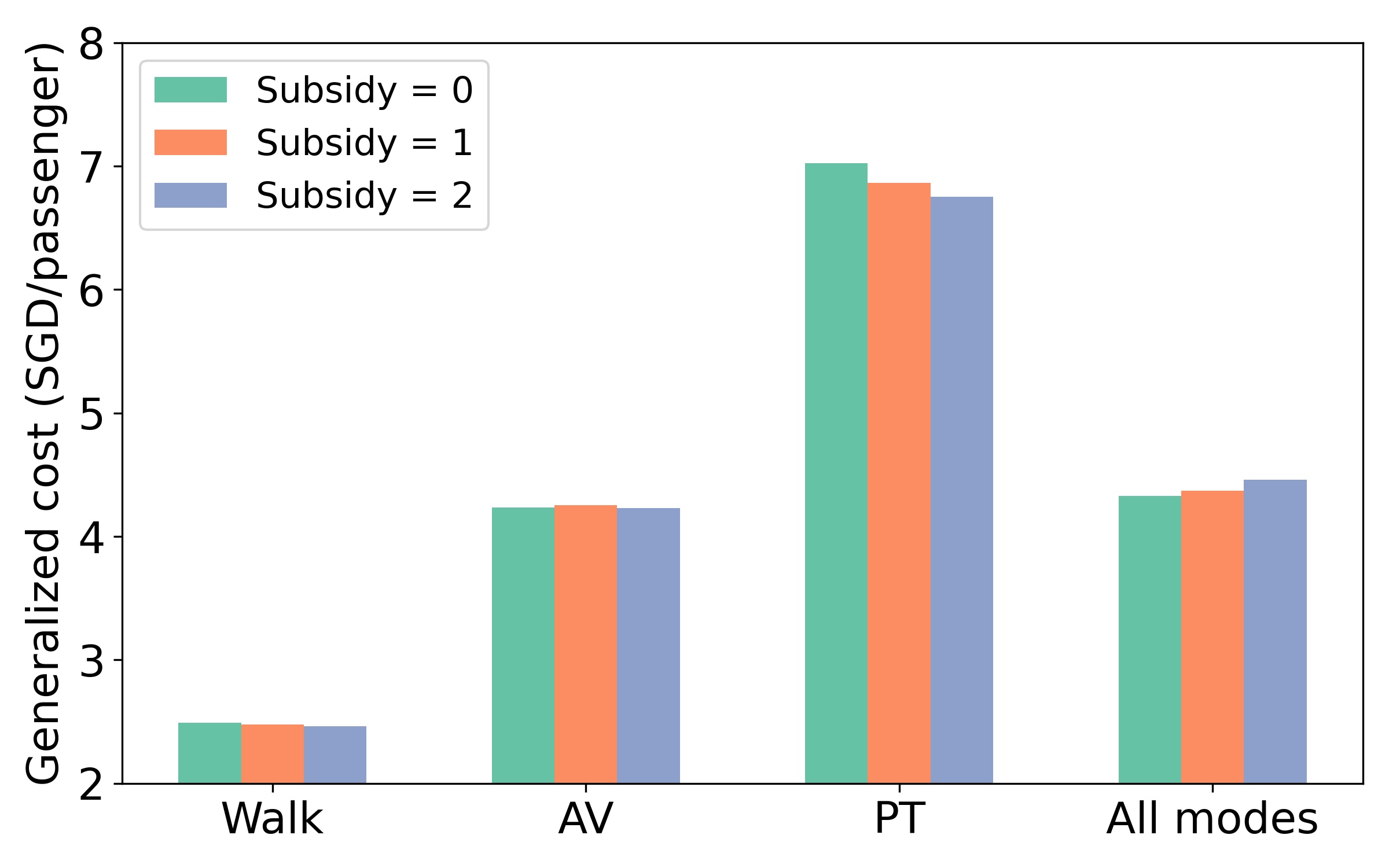}}
\hfil
\subfloat[Demand]{\includegraphics[width=0.4\textwidth]{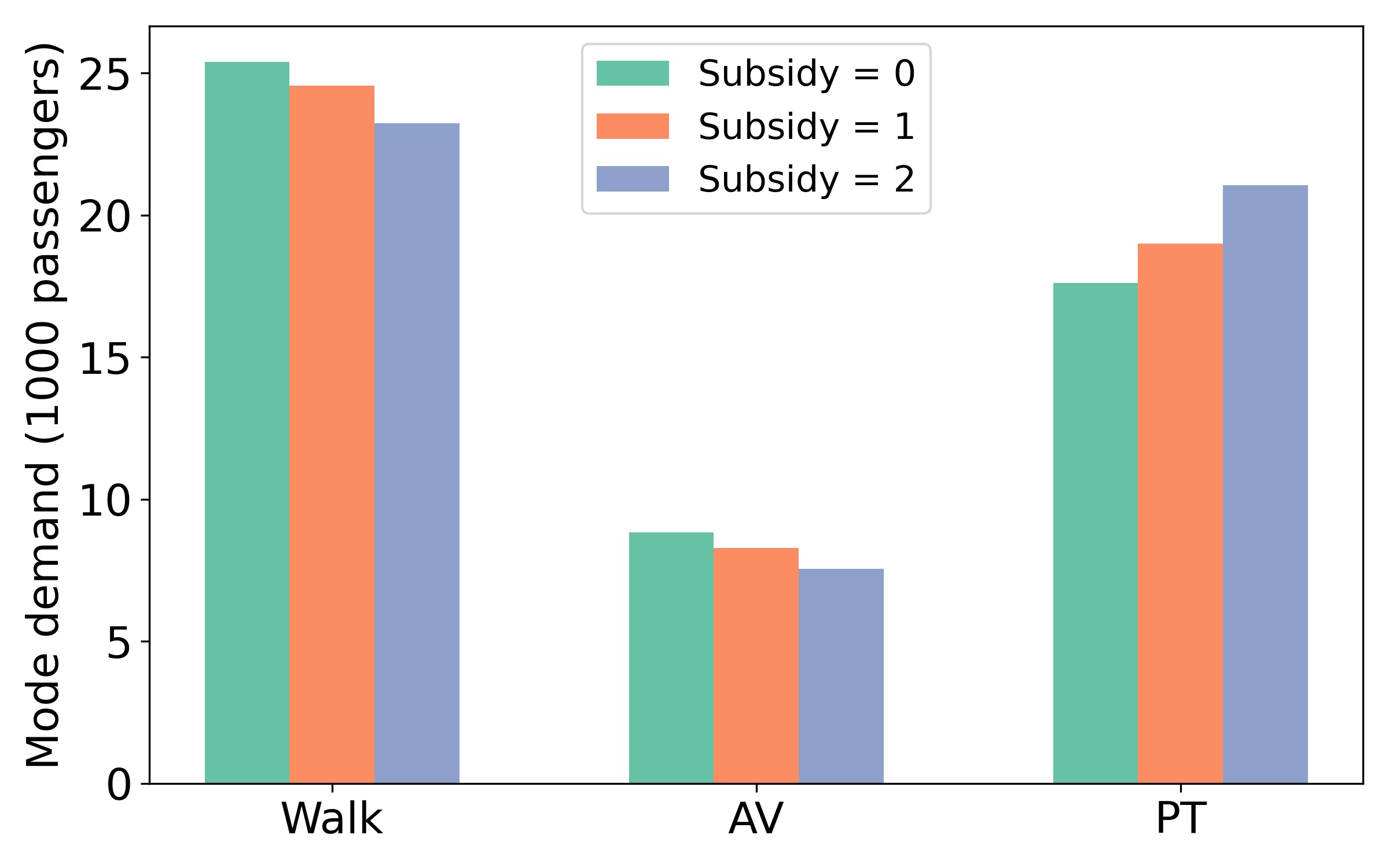}}
\caption{Generalized cost and demand by different modes for the \emph{NE} scenario}
\label{fig_by_mode_demand_based_subsidy}
\end{figure}

What's most intriguing is that passengers' generalized travel costs averaged over all modes increased (Figure \ref{fig_pax_demand_generalized_cost}) with higher PT subsidies. Figure \ref{fig_by_mode_demand_based_subsidy} separates the generalized costs and demands by the mode passengers choose in the \emph{NE} scenario. With increased subsidies, the generalized cost for PT users decreases as expected, while the generalized cost for AV users does change too much. This is because two effects on AV users' generalized cost roughly offset each other: first, some prior AV users with high generalized cost switch to PT when subsidized, decreasing the remaining AV users' average generalized cost; and second, AV decreases its supply when PT are subsidized , increasing AV users' generalized cost. Even though the PT generalized cost decreased because of the subsidy, it is still much higher than that of AV and walking. Therefore when the subsidies attract more passengers to PT, the total generalized cost averaged over all modes increase. In this particular case, the PT subsidies crowd out some of the AV supply, increasing the passengers' overall generalized cost.

\subsubsection{Transport authority perspective}
\begin{figure}[H]
\centering
\subfloat[AV average load]{\includegraphics[width=0.33\textwidth]{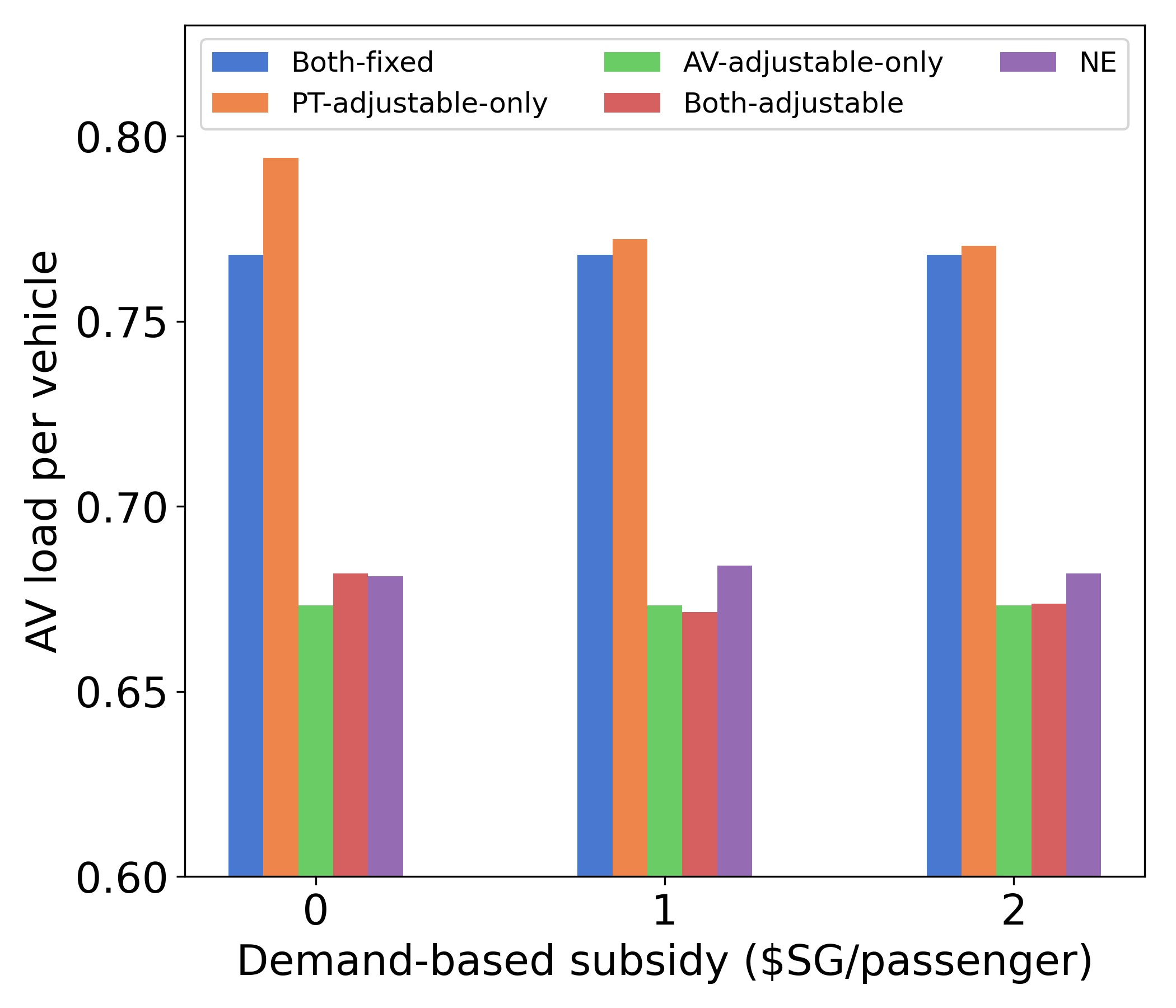}\label{fig_auth_demand_av_load}}
\hfil
\subfloat[Bus average load]{\includegraphics[width=0.33\textwidth]{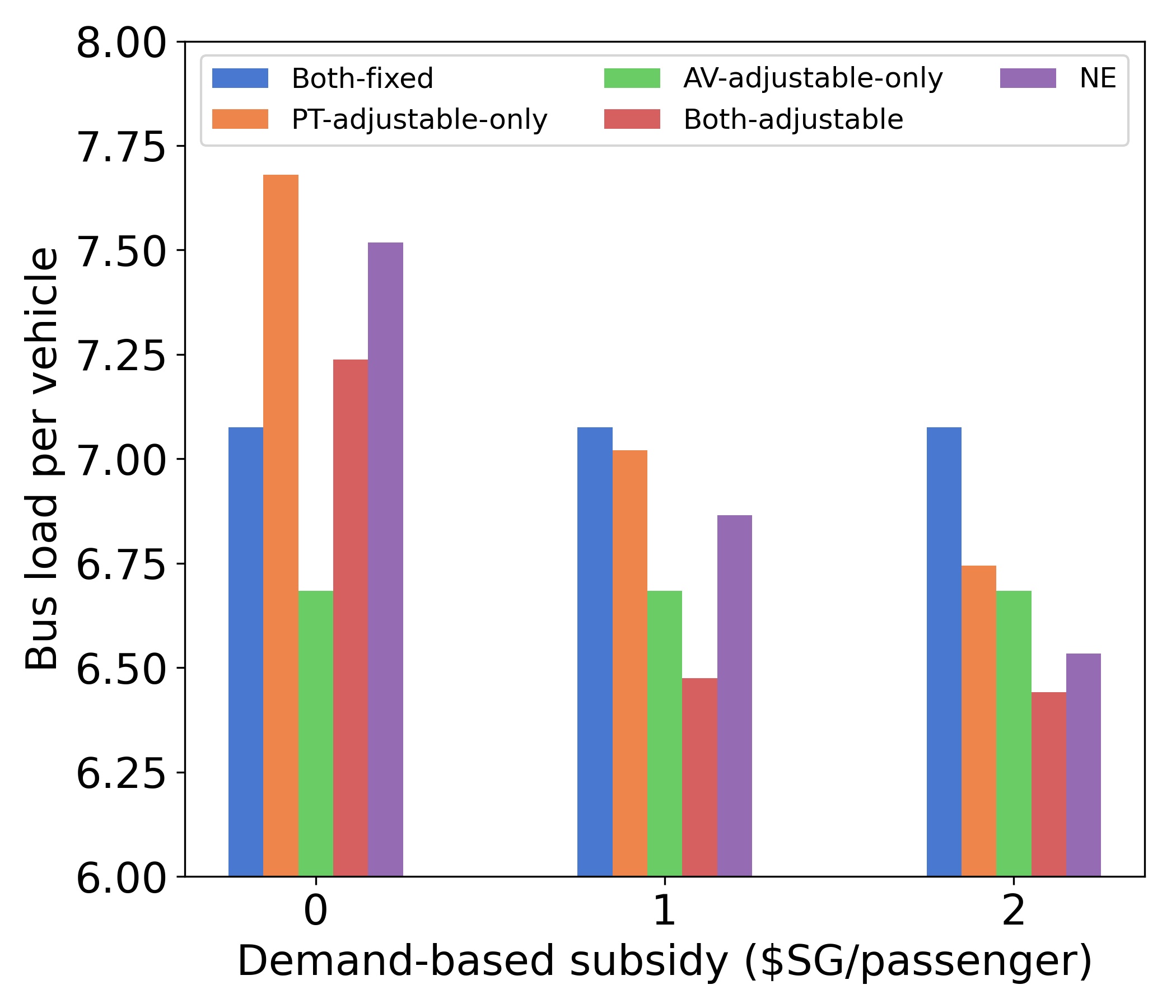}\label{fig_auth_demand_bus_load}}
\hfil
\subfloat[Total PCE (bus + AV)]{\includegraphics[width=0.33\textwidth]{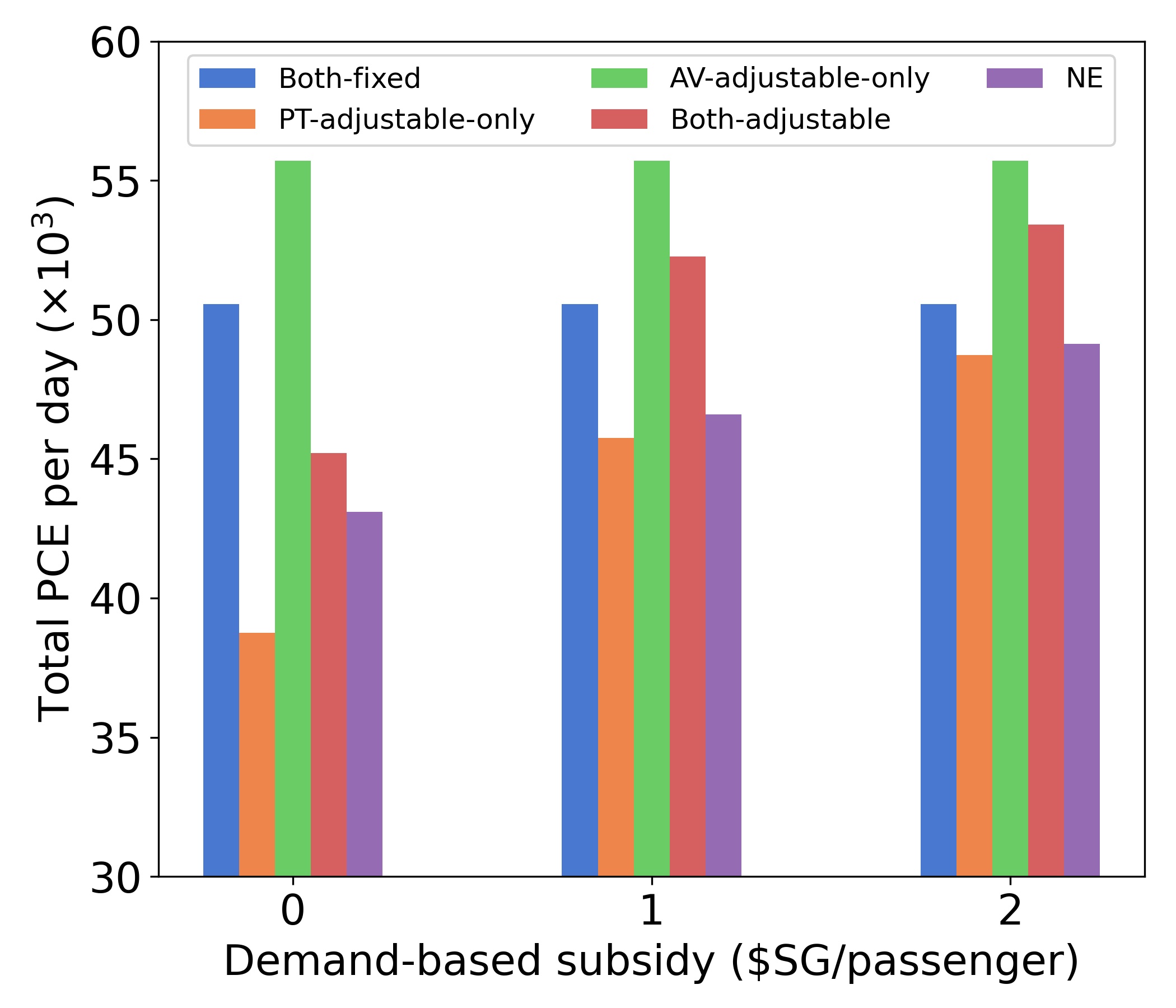}\label{fig_auth_demand_pce}}
\caption{Impact of demand-based subsidies: transport authority perspective}
\label{fig_TA_demand_based_subsidy}
\end{figure}

Figure \ref{fig_TA_demand_based_subsidy} shows the impact of the demand-based subsidy on AV average load, PT average load, and total PCE. For the AV average load (Figure \ref{fig_auth_demand_av_load}), \emph{both-adjustable} and \emph{NE} scenarios do not show significant changes, while a decrease pattern is observed for the \emph{PT-adjustable-only} scenario. This may be because when AV is allowed to adjust its supply, it preserves a relatively stable average load so that the profit is maximized. When AV cannot adjust supplies (i.e. in the \emph{PT-adjustable-only} scenario), PT's increased competitiveness pushes down both AV's market share and average load. The PT subsidy decreases the bus average load (Figure \ref{fig_auth_demand_bus_load}), reducing the operational efficiency of bus services. The total PCE increases with the PT subsidy due to the increase in bus supply as shown in Figure \ref{fig_auth_demand_pce}.

\section{Discussion} \label{con}
We simulated and evaluated the interaction between AV and PT from a competitive perspective---assuming that both AV and PT are profit-oriented and improve their own profit by supply adjustment. The first-mile market in Tampines, Singapore was selected for a case study. Five scenarios with different policy constraints were evaluated. 

\subsection{Summaries of major results and findings} 
\begin{figure}[!htp]
\centering
\subfloat[\emph{AV-adjustable-only}]{\includegraphics[width=0.6\textwidth]{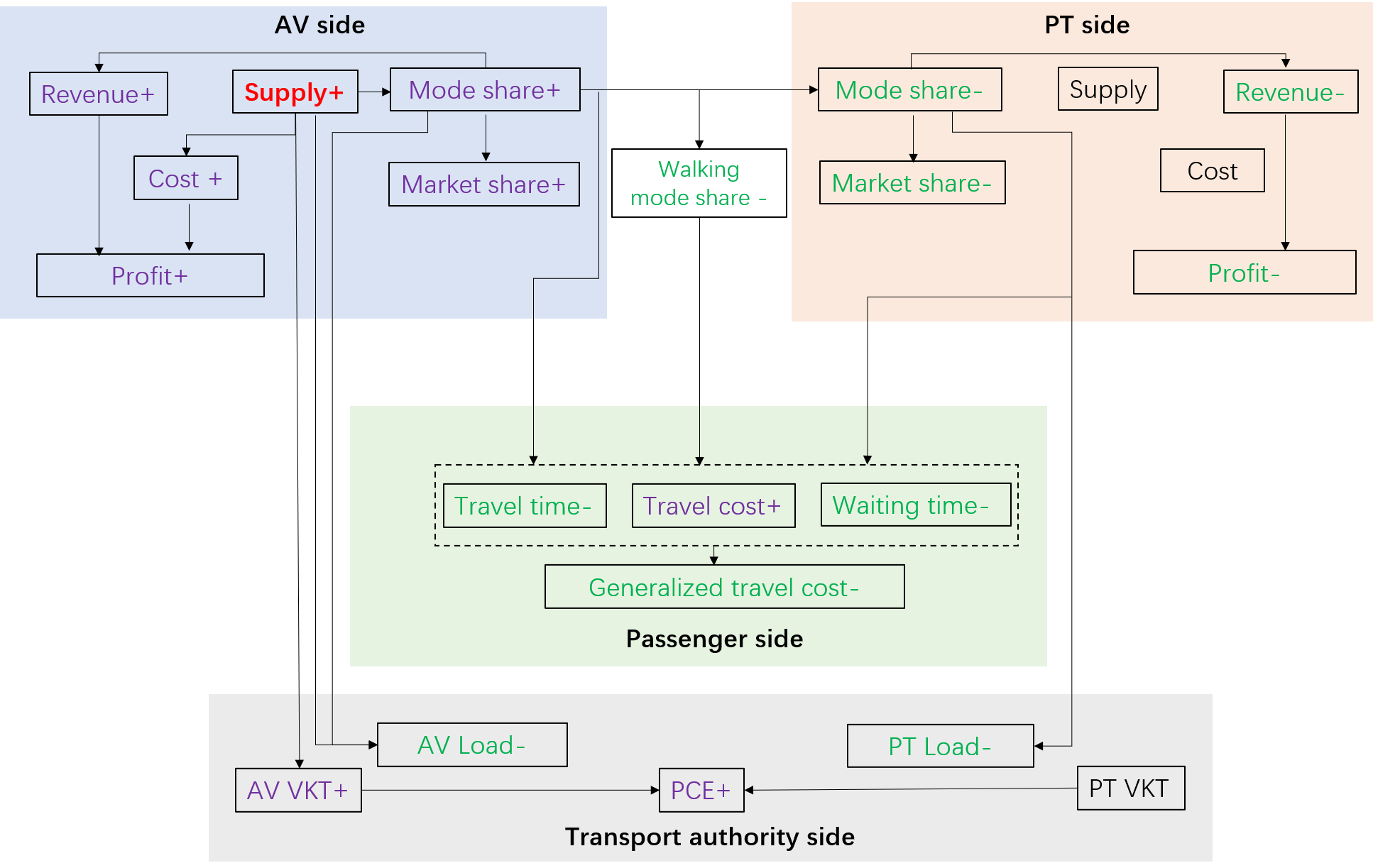}}
\hfil
\subfloat[\emph{PT-adjustable-only}]{\includegraphics[width=0.6\textwidth]{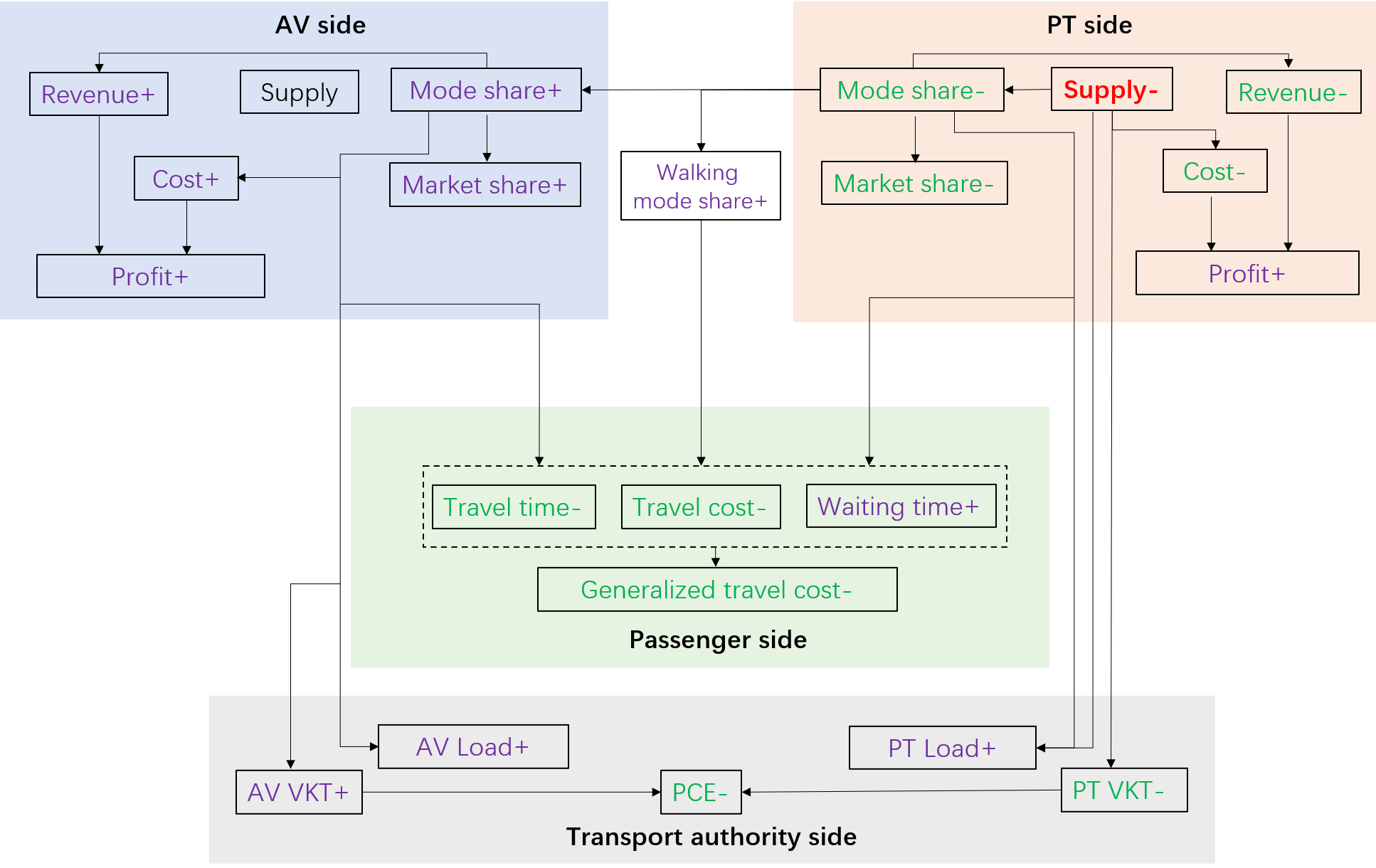}}
\hfil
\subfloat[\emph{Both-adjustable} and \emph{NE}]{\includegraphics[width=0.6\textwidth]{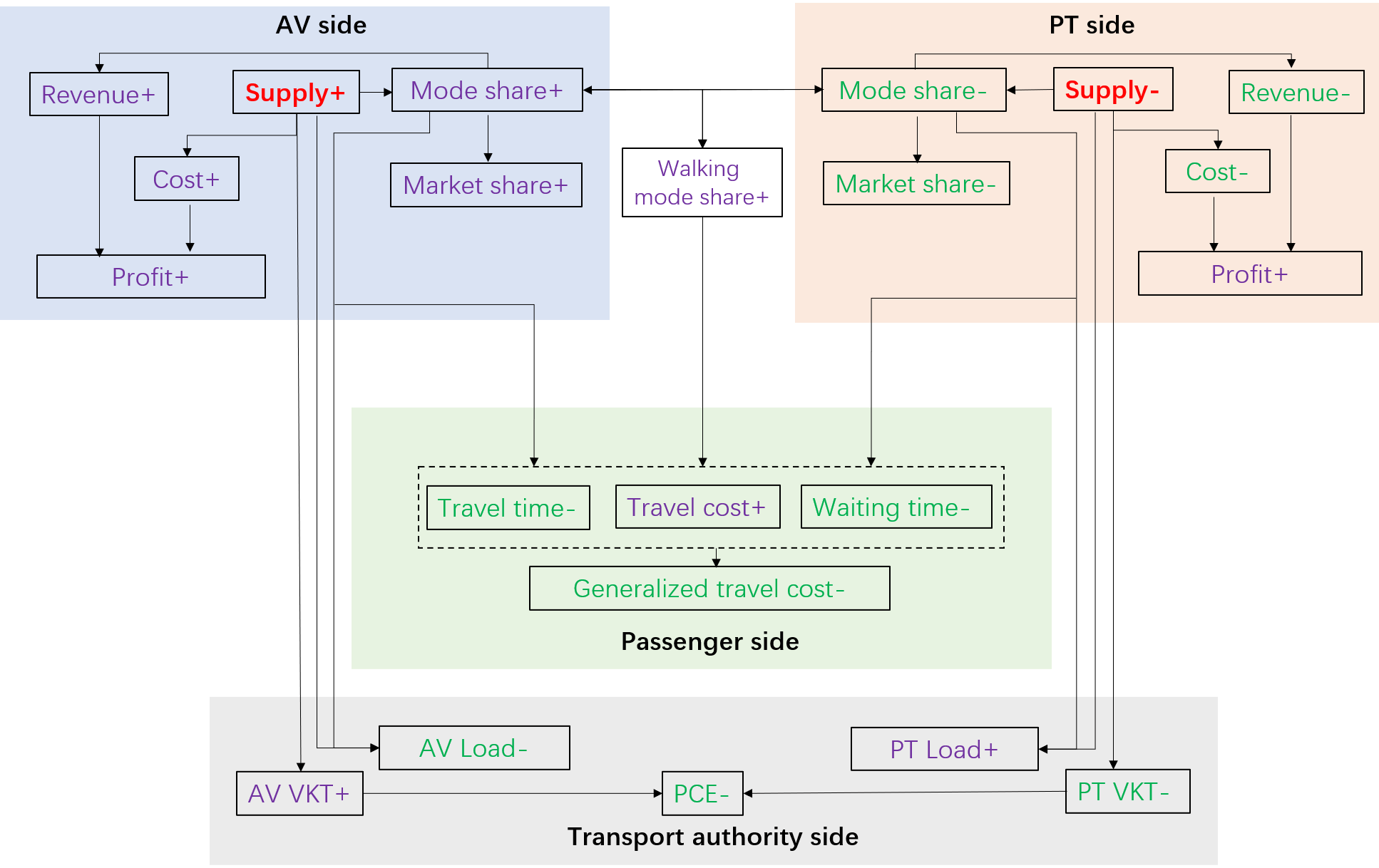}\label{fig_change_summary_both}}
\caption{Summary of the change of indicators for the non-subsidy scenario. The ``triggers'' are highlighted by red color, and the arrows represent the causal relations. The increase (purple) and decrease (green) of different indicators are shown by ``$+$'' and ``$-$'', respectively, compared to the \emph{both-fixed} scenario.}
\label{fig_sum}
\end{figure}

To better illustrate the changes in the interests of different stakeholders due to the competition, the results of non-subsidy scenarios are summarized in Figure \ref{fig_sum}. The \emph{AV-adjustable-only} scenario is beneficial to AV (profit$+$) and passengers (generalized travel cost$-$), but may not be preferred by PT (profit$-$) and transport authorities (PCE$+$). The \emph{PT-adjustable-only}, \emph{both-adjustable} and \emph{NE} scenarios are both beneficial to all stakeholders. While the \emph{NE} and \emph{both-adjustable} scenarios show larger benefits. The change patterns of \emph{NE} and \emph{both-adjustable} scenarios are similar (Figure \ref{fig_change_summary_both}), while the changes in \emph{NE} scenario are generally more prominent.

The main findings are summarized below.
\begin{itemize}

\item Overall, allowing profit-oriented bus and AMoD services to compete by adjusting supply can improve profits for both while benefiting the public and the transport authorities. Such competition forces bus operators to reduce the frequency of inefficient routes and allow AVs to fill in the gaps in the service coverage. 

\item During the competition, both AV and buses redistribute their supplies spatially and temporally. For buses, the supply of longer sinuous routes is reduced. The supply-increased routes are typically the short ones crossing the residential areas and connecting directly to the subway station. Temporarily, both AV and buses concentrate their supplies to the morning and evening peak hours and reduce the supplies in off-peak hours. 

\item The competition between AV and PT can decrease the passengers' travel time but increase their travel costs. The generalized travel cost is still reduced considering the value of time.

\item In terms of people's mode choices, bus demand decreases while AV and walking demands increase. People who live near MRT stations switch to walking more, and people who live far from MRT stations turn to AV more. Higher-income people tend to change their choices from buses to AV, but there is no evidence that low-income people are more likely to walk. 

\item The bus supply reduction and concentration increase the average bus load and reduce total PCE while the AV supply adjustment shows the opposite effect.

\item Comparing different scenarios, \emph{AV-adjustable-only} scenario is beneficial to AV and passengers but will not be preferred by PT and transport authorities. PT-only, \emph{both-adjustable}, and \emph{NE} scenarios are beneficial to all stakeholders. And the benefits to passengers and transport authorities in the \emph{NE} scenario are the largest. 

\item Providing subsidies to PT results in higher PT supply and lower AV supply compared to the non-subsidy situation. Consequently, PT's profit and market share are increased while AV's profit and market share are decreased. However,  passenger's generalized cost and total system PCE are increased. 

\end{itemize}

\subsection{Policy implications}
The comparison of the five scenarios indicates that it is possible for competition to result in a win--win situation under certain policy regulations. Our findings can help authorities design future AV--PT regulations when the two modes are both operated by for-profit companies. 

The transportation authorities can consider allowing PT operators to optimize their supply strategies to improve efficiency. However, as passengers' waiting time and travel cost may increase due to PT's supply adjustment, the authorities need to examine how much leeway should be given to PT operators to balance efficiency, total VKT, and passenger level of service. One possibility is to set specific operational constraints within which private PT operators can optimize supply, such as bounds for headway, waiting time, ridership, and passenger satisfaction scores. 

AV supply increases may reduce passenger travel time, but also reduce efficiency and increase VKT. Policies targeting AV operators can manage their possible negative externality on the system by delineating the number of licenses, operation time, vacancy rate, and service areas. Such AV policies need to consider their relations with the PT operations. For example, the government can encourage AV to serve areas and time periods with low PT frequency, which makes AV more likely to complement the PT services, especially if PT is allowed to decrease its supply in some low-profit areas.

The impact of the AV and PT competition on passengers is uneven. Authorities may need to support people who suffer from higher travel costs or longer travel times by providing discounts or other feeder modes (e.g., bike share and E-scooter share). 

The five scenarios considered in this study correspond to different degrees of regulatory intervention. In our specific case study, the \emph{NE} scenario, representing the least regulations, is plausibly the best for decreasing passenger's generalized cost and total PCE, pointing to a freer competition to better redistribute resources and improve service quality. However, this may well be the function of the specific context of the Singapore case study, future research is needed to generalize the results and characterize the exact conditions under which this finding will hold. 

\subsection{Limitations and future research}
This paper can be improved in the following aspects.
1) The results are based on the assumptions and settings in Singapore. They may not be representative of other countries or cities. This is a typical problem in many simulation-based works \citep{liu2017tracking,loeb2019fleet}. However, the objective of this paper is to provide a \emph{general framework of analysis}, which are extendable to other cities. Future research can systematically explore the impact of scenario settings and assumptions on the results. 2) Methodologically, several technical components of this study can be improved. First, the heuristic supply updating algorithm may not converge to the maximum-profit points for AV and PT. Future research can apply more advanced algorithms (e.g., reinforcement learning \citep{wen2017rebalancing}, simulation-based optimization \citep{mo2021calibrating,mo2020calibrating}) to enhance it. The supply updating process can be implemented as a multi-agent learning process, which allows more flexibility in operators' competing behaviors. Second, we assume that the total demand of passengers and its spatial and temporal distribution is fixed (even though we do represent the modal shift between AV, PT, and walking), which is a simplification. Future research can introduce a supply-demand interaction module \citep{wen2018transit} to relax this assumption. 

We point out a few directions for future research. The first is to evaluate the impact of pricing, either from the operators' perspective or from the authorities' perspective. For example, a multi-variable optimization algorithm may be developed to allow operators to adjust fare and supply at the same time. This is an extension of the current framework but relaxing the fixed-fare constraints. Second, since we assume that the total demand and environment do not evolve over time, the AV PT competition in this study is a static game. Future studies may incorporate the changes of demand patterns and infrastructure layouts to model the long-term AV PT competition as a dynamic game. Third, the paper only examines the first-mile market. Future research could explore the proposed AV--PT competition framework in a larger urban network and for broader trip purposes which requires computationally more efficient methods. Lastly, from the authorities' perspective, it is important to conduct the comparative study of different AV--PT governance structures as in \citep{shen2018integrating}. Prior studies on AV--PT interactions are typically based on specific cities assuming one specific regulatory context. A generalizable and mechanism-design-oriented model would be valuable to extend our understanding of the competition between AV and PT.

\section*{Acknowledgment}

\noindent This research is supported by the National Natural Science Foundation of China (71901164), Natural Science Foundation of Shanghai (19ZR1460700), and the Fundamental Research Funds for the Central Universities (22120180569). The research is also supported by the National Research Foundation in Singapore Prime Minister's Office under the CREATE program, and Singapore--MIT Alliance for Research and Technology Future Urban Mobility IRG. 

\section*{CRediT author statement}
\textbf{Baichuan Mo}: Conceptualization, Methodology, Software, Formal analysis, Data Curation, Writing - Original Draft, Writing - Review \& Editing, Visualization. \textbf{Zhejing Cao}: Data Curation, Visualization, Writing - Original Draft. \textbf{Hongmou Zhang}: Formal analysis, Writing - Review \& Editing. \textbf{Yu Shen}: Conceptualization, Resources, Funding acquisition. \textbf{Jinhua Zhao}: Conceptualization, Formal analysis, Writing - Review \& Editing, Supervision, Project administration, Funding acquisition.

\section*{References}

\bibliography{main}

\appendix
\appendixpage
\section{Agent behaviors}
\label{agent_behavior}
\subsection{Passenger behavior}
Passengers are assumed to enter the system every minute during the simulation day. To mimic the AV and PT competition process, the simulation needs to be run for a sufficiently long time period (one year in this study). 

In this study, we assume the passenger demand to be fixed (Figure \ref{fig_demand}). A Voronoi diagram based on the location of bus stops was used to assign aggregate travel demand to the building level. 

Passengers' mode choice is modeled based on the result of a mixed logit model \citep{shen2019built}. The trip-specific variables (travel cost, waiting time, on-vehicle time, walking time) and sociodemographic variables (e.g., income) were considered. This demand model was estimated using the results of an AV preference survey in Singapore, which matches this study. Besides, the mixed logit model allows us to capture the preference heterogeneity among people. Each individual was assigned a set of choice coefficients drawn from the predetermined distributions. The detailed model coefficients are shown in \ref{choicemodel}. 

When a passenger is generated, we first calculate the corresponding trip-specific variables based on his/her location. Since routing is not the focus of this study, we assume that passengers only choose the closest bus stop. The corresponding walking time, waiting time, and in-vehicle time are then calculated. Specifically, for the AMoD waiting time, we first find the nearest available AV (see Section \ref{av_behavior} for details). Based on the location of the passenger and the selected vehicle, the expected waiting time for the passenger is calculated. When there is no AV available at passengers' departure time, we search for the next AV that will finish the trip and assign it for the waiting time calculation (the expected waiting time will be added by the remaining time for the AV to finish its trip). AMoD's in-vehicle time is calculated based on the distance from the passenger's location to the MRT station (i.e., destination). In terms of the PT waiting time, we assume passengers can access the real-time bus locations and obtain expected bus arrival times and his/her arrival time at the bus stop. Then the PT waiting time is calculated as the arrival time of the first bus that the passenger can catch minus his/her arrival time at the stop. Similarly, the PT in-vehicle time can be calculated based on the bus route and the boarding station. To reflect the randomness in waiting time calculation and passengers' perceptional heterogeneity, a normally distributed zero-mean random error is added to the decision-making waiting time for each passenger. When the passenger departure time is not within a bus route's operating time, the bus mode is set as unavailable. 

Passengers' incomes were drawn from the distribution derived from Singapore household travel survey data \citep{HITS2012}. The mixed logit model was used to calculate the probability of choosing different modes. During the simulation process, people's mode choice changes with the change in AV and PT's supply. Considering the imperfect information transfer, a lag effect was added to people's behavior. The \emph{model-derived probability} of passenger $n$ choosing mode $i$ on day $d$ is $\hat{P}^{(d)}_{n,i}$. The actual probability used for simulation was calculated in Eq. \ref{eq_lag}.
\begin{flalign}
P^{(d)}_{n,i} = \alpha P^{(d-1)}_{n,i}+(1-\alpha) \hat{P}^{(d)}_{n,i} \text{,} &&
\label{eq_lag}
\end{flalign}
where $\alpha$ is the lag factor, which represents how much people's behavior will be lagged by the previous day. After calculating the choice probability, passengers will be assigned a specific travel mode accordingly. 
 
When walking is chosen as the travel mode, the passenger walks directly to the MRT station along the footpath. When AV is chosen as the travel mode, the passenger starts to call for the ride repeatedly until the system successfully books a car for him/her. Once a vehicle is assigned, he/she then moves to the pick-up location to wait for the AV. As the density of bus stops in Singapore is very high, the pick-up location was set as the closest bus stop from the passenger. To prevent a passenger from waiting too long when no AV is available, the maximum waiting time was set to 10 min in this study, beyond which he/she will forgo the AV request and travel by bus if available; otherwise, he/she will travel by walking. 
 
When buses are chosen, the passenger walks directly to the bus stop to wait for the next bus. Since we assume that the bus agent can update its headway to maximize its profit, people may wait for a long time. Therefore, we set a maximum waiting time (30 min) for the bus as well. People will switch to AV or walking based on the choice probability when the waiting time is beyond the threshold. 

\subsection{AMoD behavior} \label{av_behavior}
The concept of the mobility-on-demand system was studied in the 1970s and called Dial-a-Ride Transit (DART) \citep{wilson1976advanced}. In this study, the AMoD service with ride-sharing resembles a carpooling system \citep{galland2014multi,martinez2015agent}, for which only the first-mile service is provided and restricted to the service area, Tampines. This arrangement is consistent with the proposed usage of the AV prototype in Singapore \citep{chong2011autonomous}. 

We assume that the routing behaviors of all AVs are identical and all AVs follow the shortest path. As AVs and buses in the first-mile market only account for a small proportion of traffic flows on roads, we consider road congestion as an exogenous factor. The speed of AVs on a specific road is time-dependent and is calculated from Google Map API. AVs are assumed to fully comply with the central controller and never reject the service requests from customers. Each AV allows a maximum of four passengers to share the ride. AV's initial supply is approximated using Singapore Taxi data. Since there is no first-mile demand during the MRT close hours, we adjusted the initial AV supply to zero during that time period. New AVs are generated from several car clubs (Figure \ref{fig_study_area}). When an AV is called to stop the service by the control center, it will return to the nearest car club and be removed from the simulation. AVs will stay idle without moving when there is no request. 

The operating cost of the AMoD agent includes fixed cost and variable cost, both are calculated per hour, which is the temporal resolution used in this study, for AV. The fixed cost is the cost of operating one AV during this hour, such as depreciation cost and parking cost. We use the lowest car renting fare in Singapore (4 SGD/hour$\cdot$veh) to approximate the fixed cost. The variable cost is calculated based on the vehicle travel distance and gasoline fare. It is set as 0.12 SGD/km in this study.

Ride-sharing only happens under specific conditions. Given the heterogeneous ride-sharing preference among the population \citep{nazari2018shared}, we define a ride-sharing-agreement ratio (50\% in this study). That is, only half of the people are willing to share trips with others. For short-distance first-mile trips, sharing with others is not a good choice given the potential increase in detour distance and travel time \citep{SCHREIECK2016272}. Therefore, we assume that all passengers prefer to ride alone if possible, even for those who agree to share the ride. Thus, when a passenger calls for a ride, the system first scans all empty AVs and then assigns the closest available AV to pick up the passenger. Once the AV and the passenger are matched, a notice is sent to the passenger to request a meet-up at the pick-up point. If there are no empty AVs, the system searches for all occupied AVs. Passengers who have agreed to share the ride will be considered to take a shared ride. An AV is shareable only if 1) it has available seats, and 2) the incremental travel time to on-vehicle passengers due to picking up new passengers does not exceed a predetermined threshold. Detailed calculation methods are available in \citet{shen2018integrating}, which shares the same parameter settings in the ride-sharing module with us. 

The fare structure of AMoD is identical to that of the taxi service in Singapore, with a base fare within first $d_b$ km and a distance-based fare beyond $d_b$ km. The difference is that we add a ride-sharing discount for the AMoD service. The total travel fare for passenger $i$, $F_i$ is formulated in Eq. \ref{eq_av_price}.
\begin{flalign}
\label{eq_av_price}
F_i =
\begin{cases}
f_b\cdot(1-(r_i)^\lambda) & \text{if $d_i \leq d_b$} \\
[f_b+f \cdot (d_i-d_b)]\cdot(1-(r_i)^\lambda) & \text{otherwise}
\end{cases}
&&
\end{flalign}
where $d_b$ is the base distance, which equals to 1 km in this study; $f_b$ is the base fare within $d_b$; $f$ is the distance-based fare per km after $d_b$; $d_i$ is the direct travel distance of passenger $i$ in km; $D_i$ is the detoured actual travel distance in km;
$r = \frac{D_i}{d_i}-1$ is the detour ratio; $\lambda$ is the discount degree offered due to the detour.

In terms of competition, the AMoD enterprise can update the hourly supply to improve its profit. Due to the resource limitation, a maximum AV supply per hour is introduced. Given the complex interaction between agents, it is very difficult to propose an algorithm that ensures that every adjustment is optimal for the AMoD enterprise. Moreover, such an algorithm is beyond the scope of this research. In this study, we propose a heuristic supply updating algorithm for AMoD. Instead of finding the optimal adjustment, we assume a fixed step size for each update. The sign of the update (i.e., increase or decrease) is determined by supply and profit in the previous steps, which aims to make the profit higher. This concept is also in line with the reality, in which information is incomplete and every adjustment is based on previous experience. A detailed description of the algorithm is available in Section \ref{sim_platform}.

\subsection{PT behavior} \label{pt_behavior}
Bus services are schedule-based and operated based on given routes and timetables. The bus routes and initial schedules are provided by the LTA. Once a bus is dispatched, it follows the predetermined route and passes through a sequence of stops. Similar to the AMoD, the speed of buses is time-dependent and calculated from the Google Map API to reflect exogenous congestion. The spatial distribution of bus stops is illustrated in Figure \ref{fig_study_area}. Upon arrival at a bus stop, each bus dwells for a certain amount of time to pick up passengers. The dwelling time is set as 30 seconds in this study. 

The operation cost of the bus agent is purely distance-based, which is 2.71 SGD/km. This value was calculated using the annual financial reports of two major bus companies in Singapore \citep{SBST2017Annual, SMRT2016Annual}, which involves the labor cost, depreciation cost, fuel cost, and maintenance costs. Note that only the distant-based operation cost is considered because in this study the bus operator is assumed to rent vehicles from the government, instead of owning the vehicles. In Singapore, bus companies rent vehicles from LTA. LTA owns the properties and is in charge of buying new vehicles. While bus companies are only responsible for operating. Therefore, we assume there is no investment cost for bus operators (i.e., increasing bus fleet size is equivalent to rent more vehicles from the government and only adds the operation cost). 

The bus fare structure is set based on the real-world scenario. All passengers are charged a fixed fee of 0.77 SGD per trip. 
 
To compete with AMoD, the bus operator is allowed to adjust its supply strategies---the headways---to increase its profits as well. The adjusting algorithm is similar to AMoD's, while the difference is that the adjusting frequency is lower than that of AMoD. 

To ensure the basic service of PT, we set an upper bound for the adjusted headway. Every step of the headway adjustment cannot exceed the upper bound; otherwise, passengers will have to wait for a long time. On the contrary, to avoid the bus dispatching rate being too high, a headway lower bound was introduced. 

Since we calculate the revenue and cost based on the first-mile demand, it is important to ensure that the supply adjustment does not affect the market outside the first-mile market. There are two types of bus routes in this study: MRT feeder buses for which the supply adjustment only applies to the first-mile market and bus routes that pass through the study region (passing-by routes). If we directly adjust the headway of the second type of routes, it will affect not only the demand for first-mile trips but also other passengers who are just passing the study region. Thus, we need to redefine the supply adjustment for these routes. Here, we assume that the supply reduction (i.e., increase of headway) for the passing-by routes is equivalent \emph{as if some buses of these routes will not stop in the study region}. Instead, they directly drive through the region along the shortest path. For example, as shown in Figure \ref{fig_rerouting}, after the decrease in the supply, some of the buses of Route 21 will be rerouted to the shorter green path and not stop in the study area. The corresponding reduction in the operating cost is calculated by the decreased travel distance. Note that the decreased operating cost may be small under this setting. 

We assume that the supply increase (i.e., decrease in headway) for the passing-by bus routes is equivalent to adding new MRT feeder bus routes with the same stations in the study region. The corresponding increase in the operating cost is calculated as the cost of running this new route. This way, we can isolate the first-mile market, where the supply adjustment for all bus routes will not affect the outside areas. 
\setcounter{figure}{0}
\begin{figure}[!htp]
\centering
\includegraphics[width=3 in]{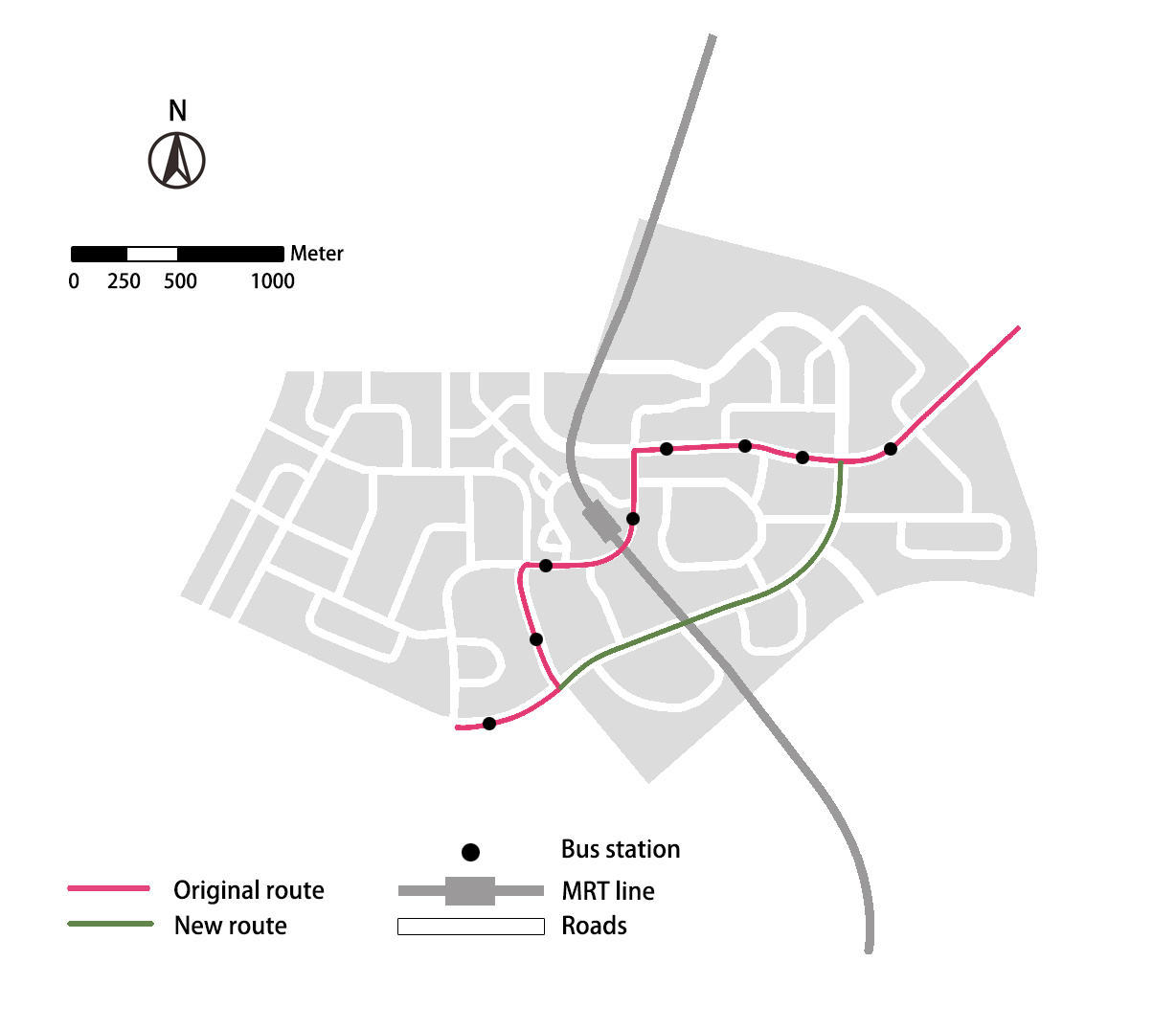}
\caption{Rerouting example of Route 21}
\label{fig_rerouting}
\end{figure}

\section{Convergence of the IBR algorithm} \label{convergence}
\begin{figure}[H]
\centering
\subfloat[Various initial AV supplies]{\includegraphics[width=0.4\textwidth]{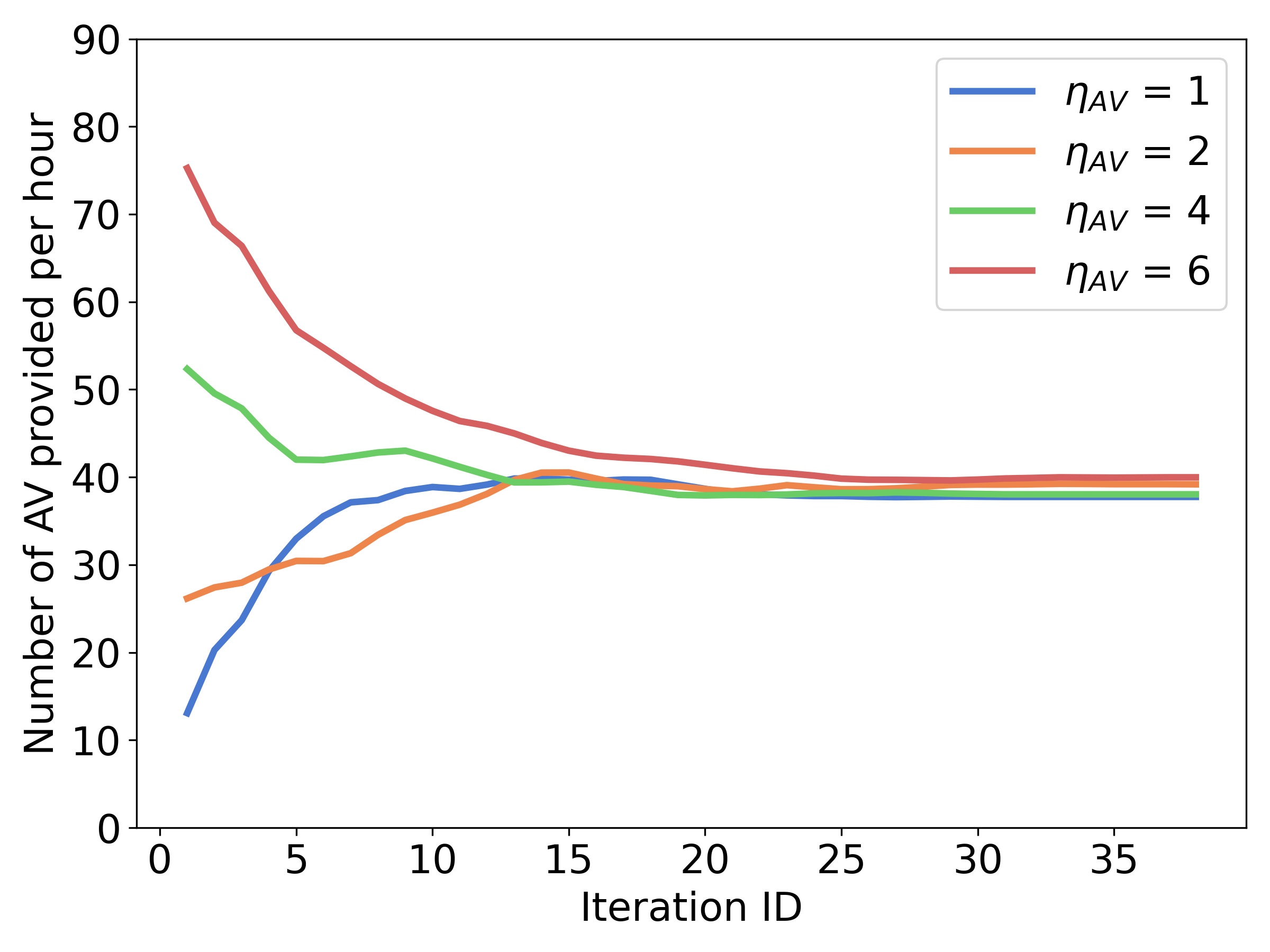}}
\hfil
\subfloat[Various initial PT supplies]{\includegraphics[width=0.4\textwidth]{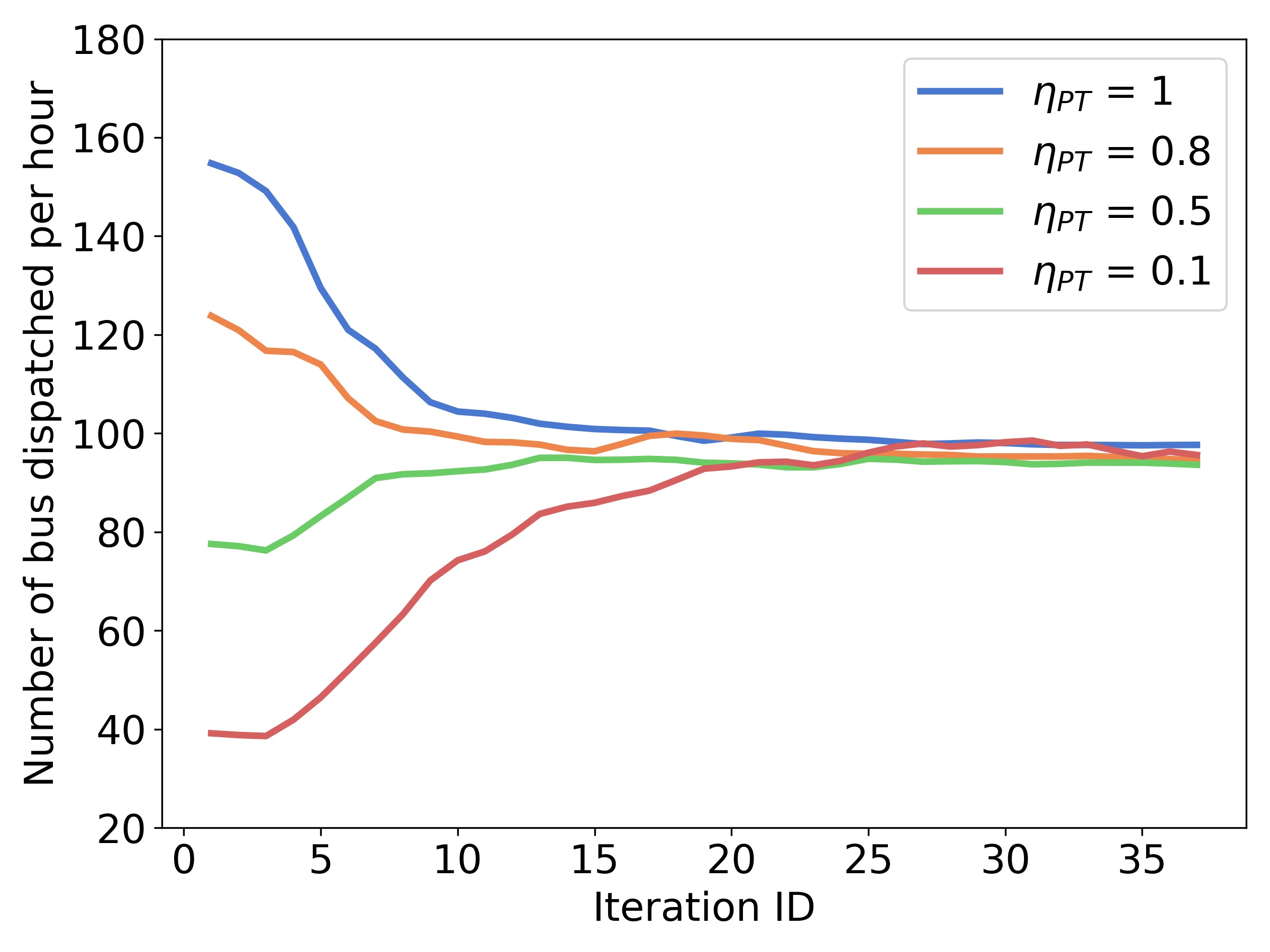}}
\caption{Convergence of the iterated best response algorithm}
\label{fig_IBR_convergence}
\end{figure}

As it is hard to analyze the convergence of the IBR algorithm theoretically in the simulation context, a numerical test with respect to different initial supply levels is conducted. Specifically, the initial fleet size for AV is set as $\eta_{A}\boldsymbol{S}_A^{(0)}$, where $\eta_{A}$ is the AV supply factor with values of $\{1, 2, 4, 6\}$. The initial bus headway is set as $\boldsymbol{S}_B^{(0)}/\eta_{B}$, where $\eta_{B}$ is the bus supply factor with values of $\{1, 0.8, 0.5, 0.1\}$. Note that the headway divided by $\eta_{B}$ is equivalent to the number of dispatched buses multiplied by $\eta_{B}$. Figure \ref{fig_IBR_convergence} shows that the system can always converge to a similar supply level despite different initial values. The converged supply can be seen as an approximated pure NE strategy. This validates the effectiveness of the IBR algorithm in this study. 

\section{Results of PT supply-based subsidies} \label{sec_supply_subsidy}
We consider three levels of supply-based subsidies (0, 0.5, and 1 SGD per km operated). These values are chosen because they have a similar scale of current operating cost (2.71 SGD per km) and are less than it.

\begin{figure}[H]
\centering
\subfloat[PT profit]{\includegraphics[width=0.33\textwidth]{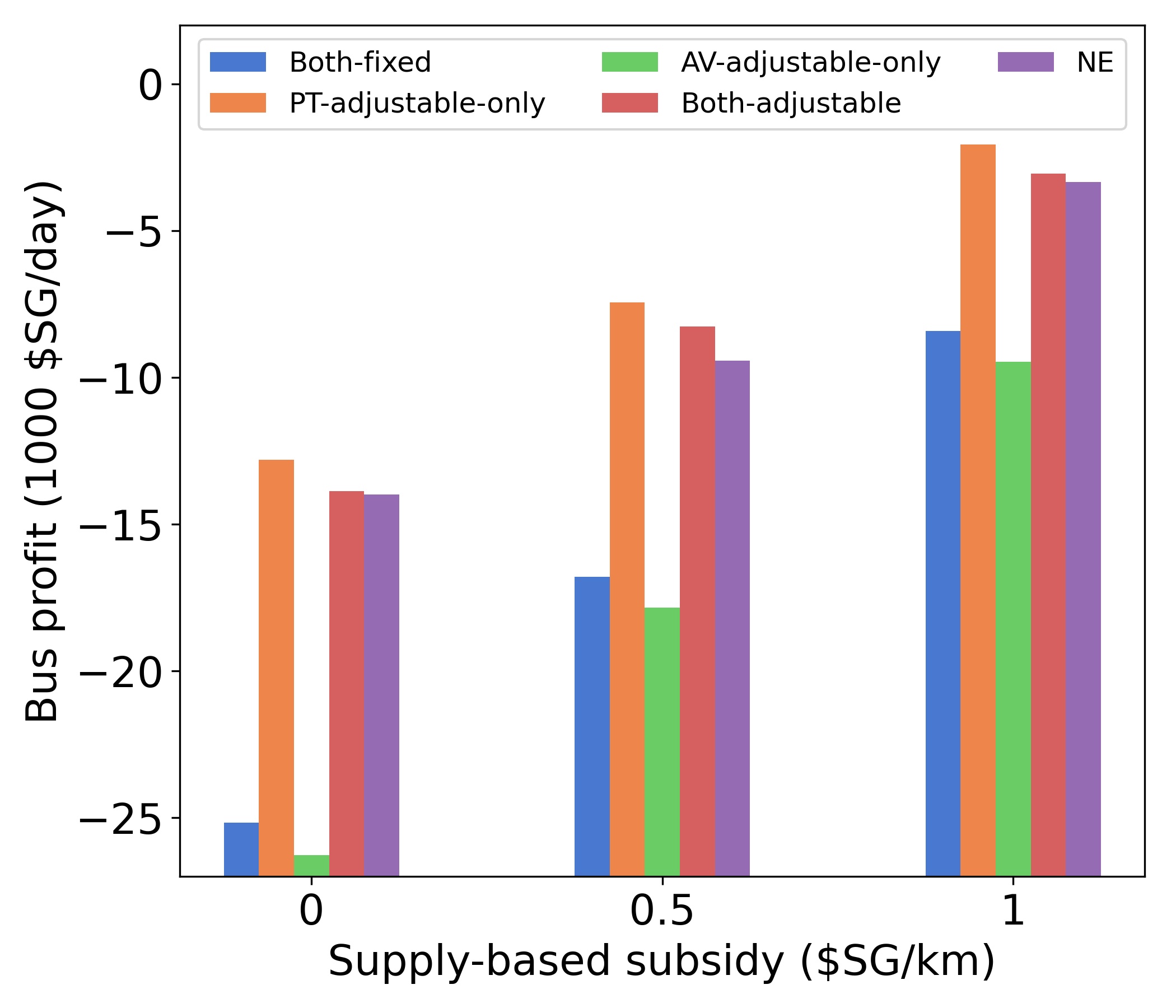}}
\hfil
\subfloat[PT supply]{\includegraphics[width=0.33\textwidth]{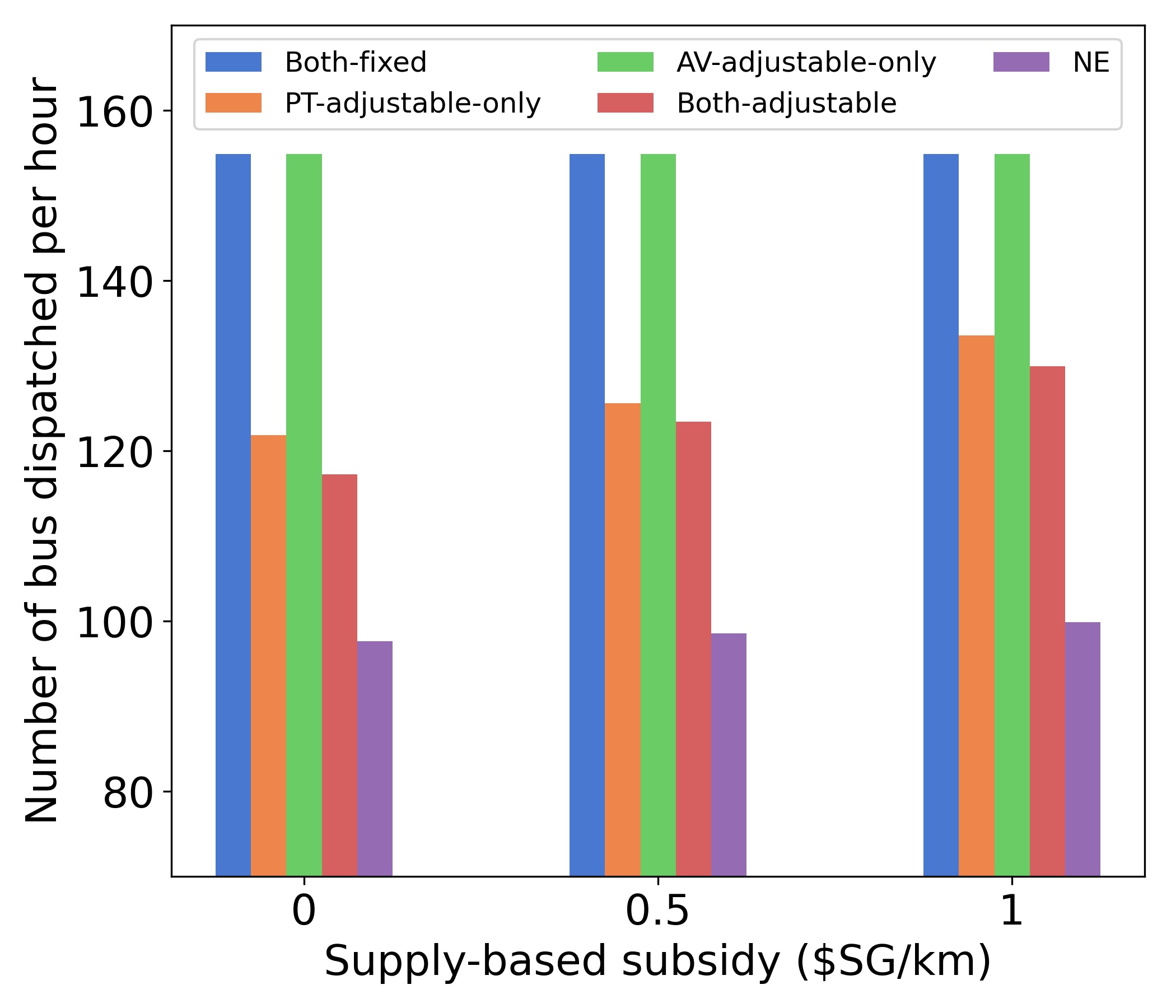}}
\hfil
\subfloat[PT market share]{\includegraphics[width=0.33\textwidth]{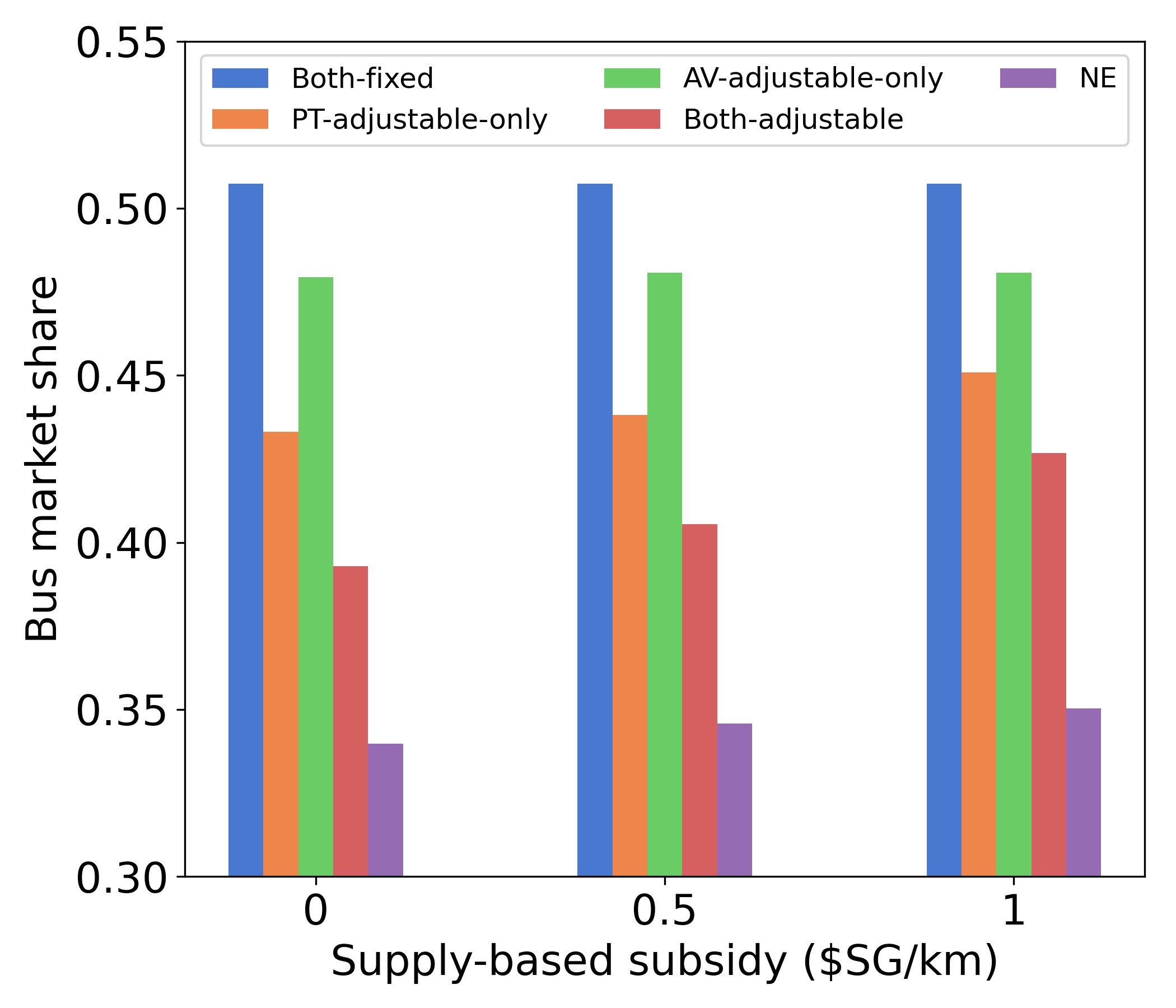}}
\caption{Impact of supply-based subsidies on PT operators.}
\label{fig_PT_supply_based_subsidy}
\end{figure}

\begin{figure}[H]
\centering
\subfloat[AMoD profit]{\includegraphics[width=0.33\textwidth]{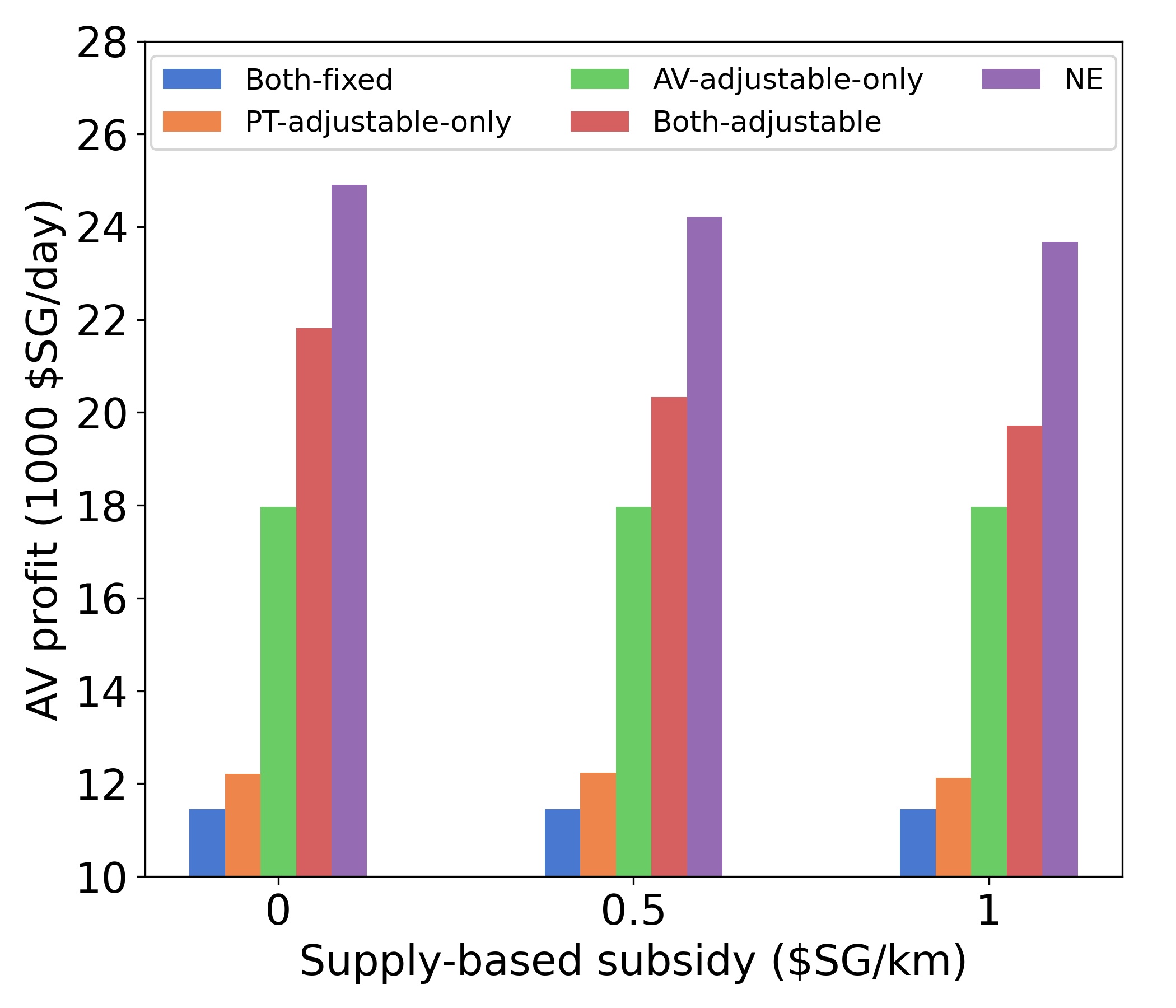}}
\hfil
\subfloat[AMoD supply]{\includegraphics[width=0.33\textwidth]{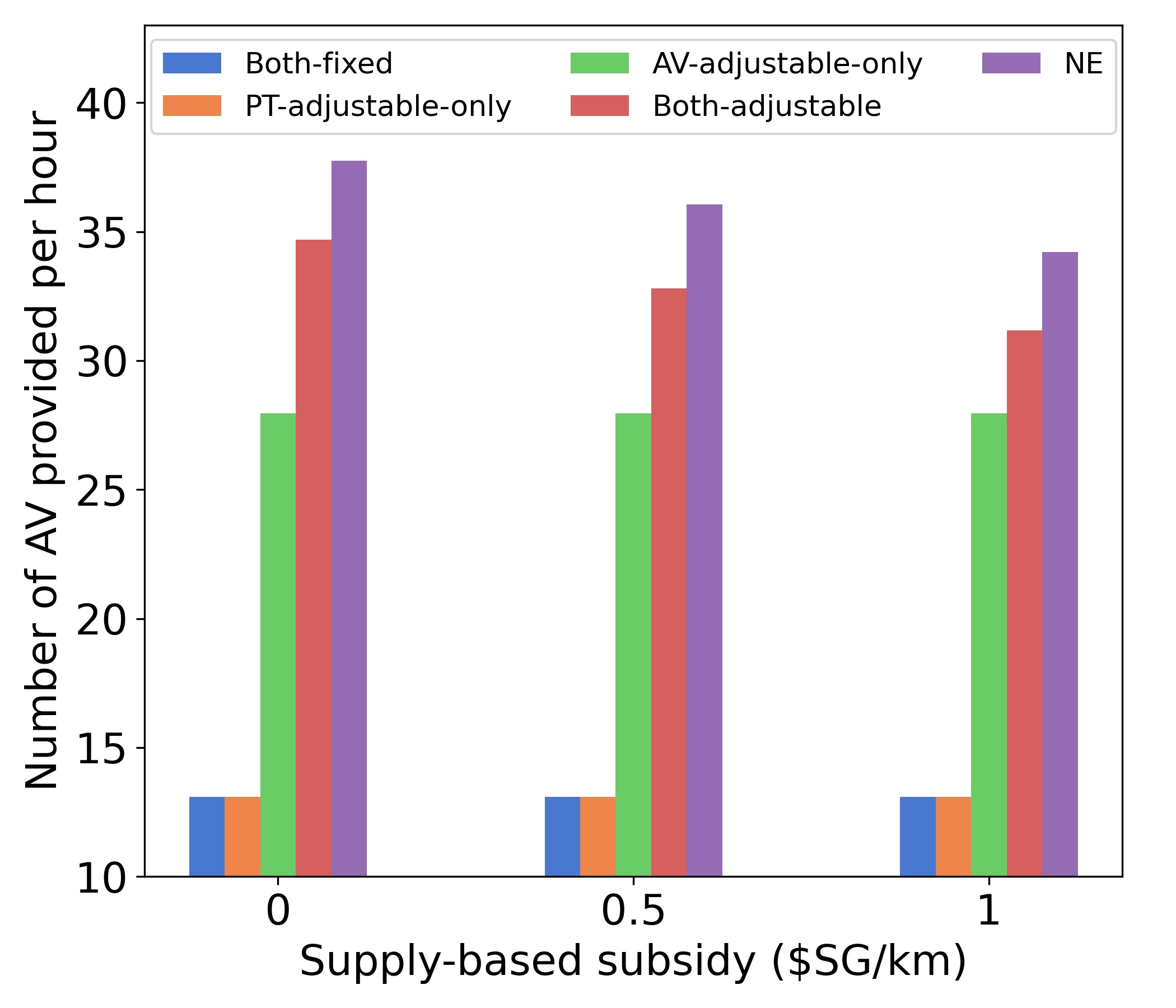}}
\hfil
\subfloat[AMoD market share]{\includegraphics[width=0.33\textwidth]{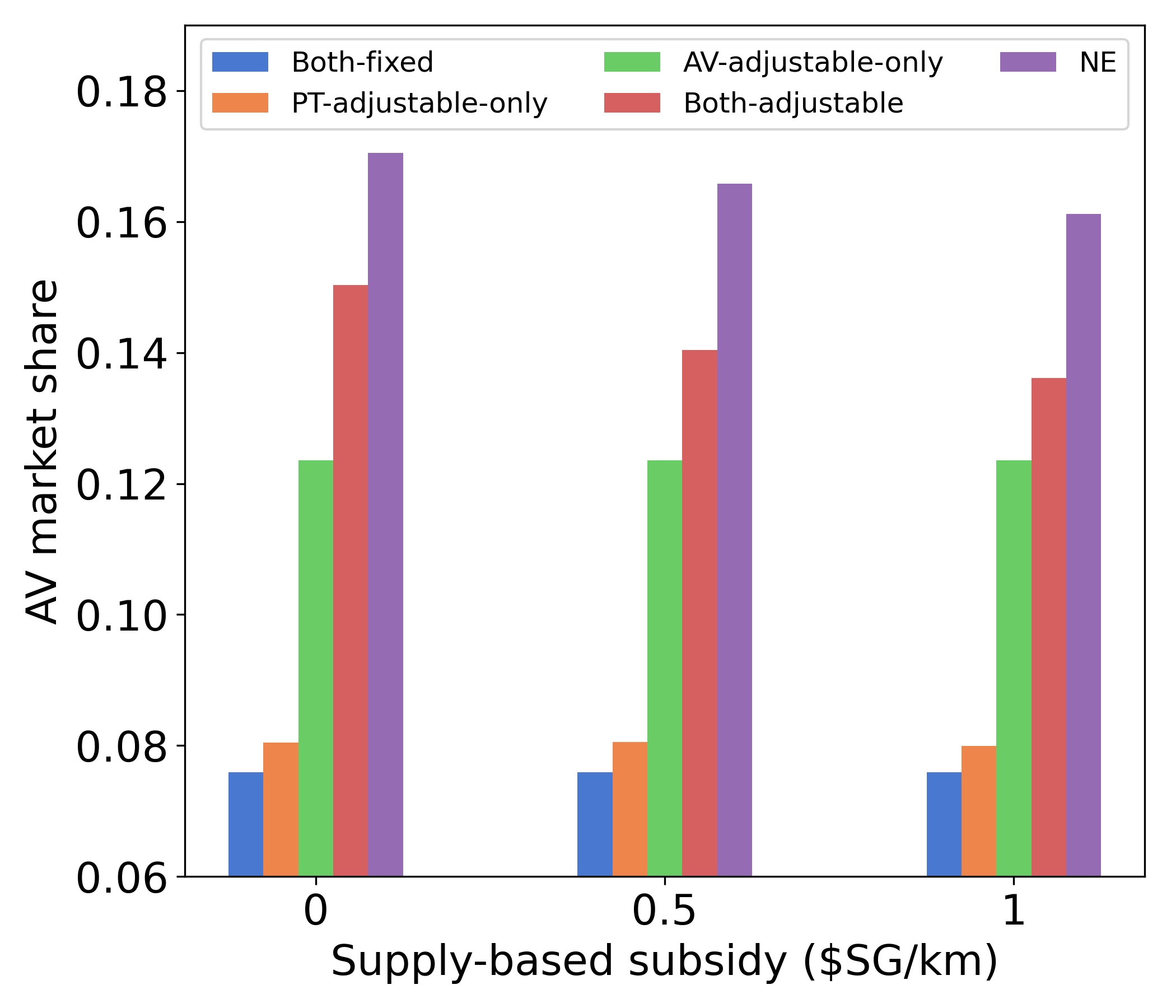}}
\caption{Impact of supply-based subsidies on AMoD operators}
\label{fig_AV_supply_based_subsidy}
\end{figure}

\begin{figure}[H]
\centering
\subfloat[Travel cost]{\includegraphics[width=0.4\textwidth]{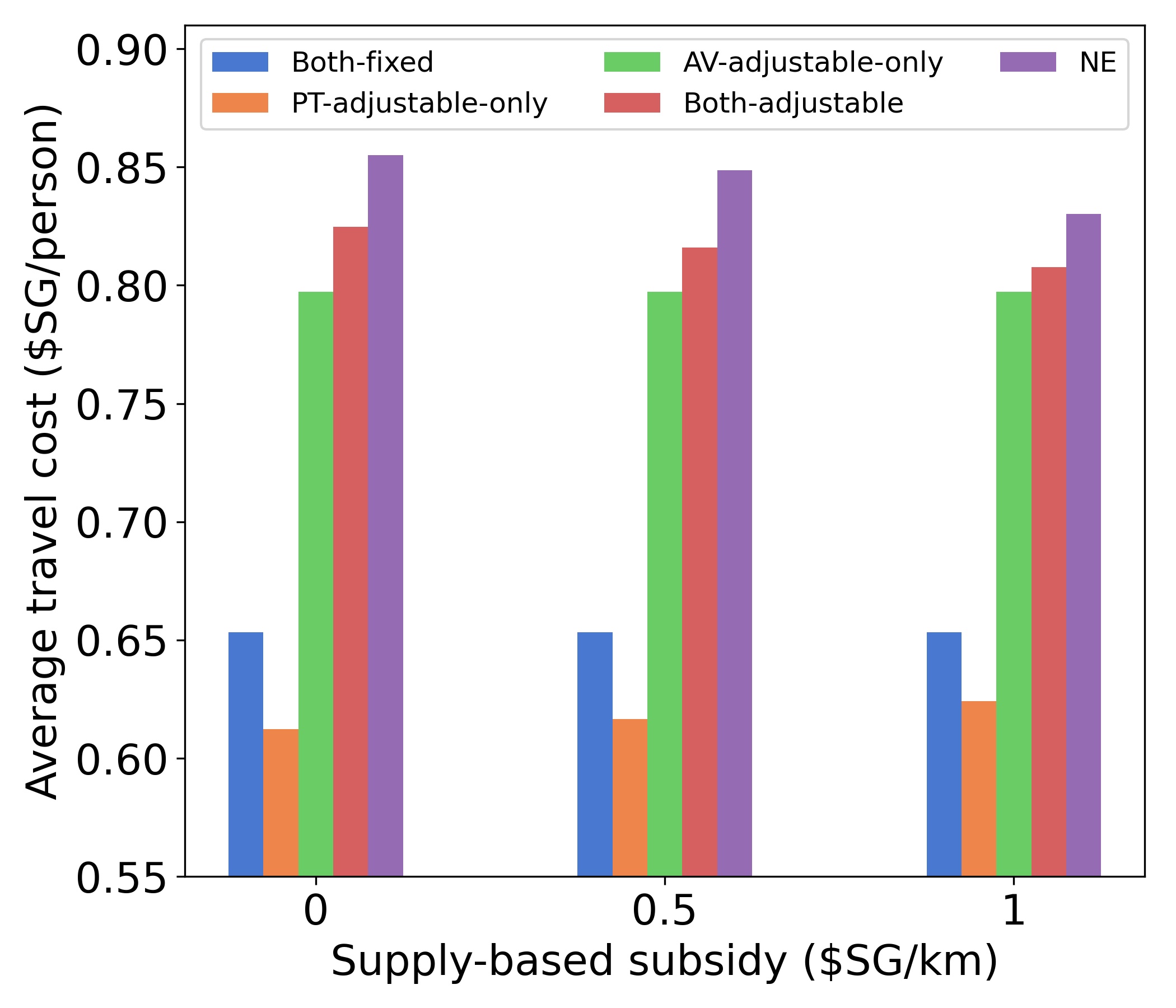}}
\hfil
\subfloat[Total travel time]{\includegraphics[width=0.4\textwidth]{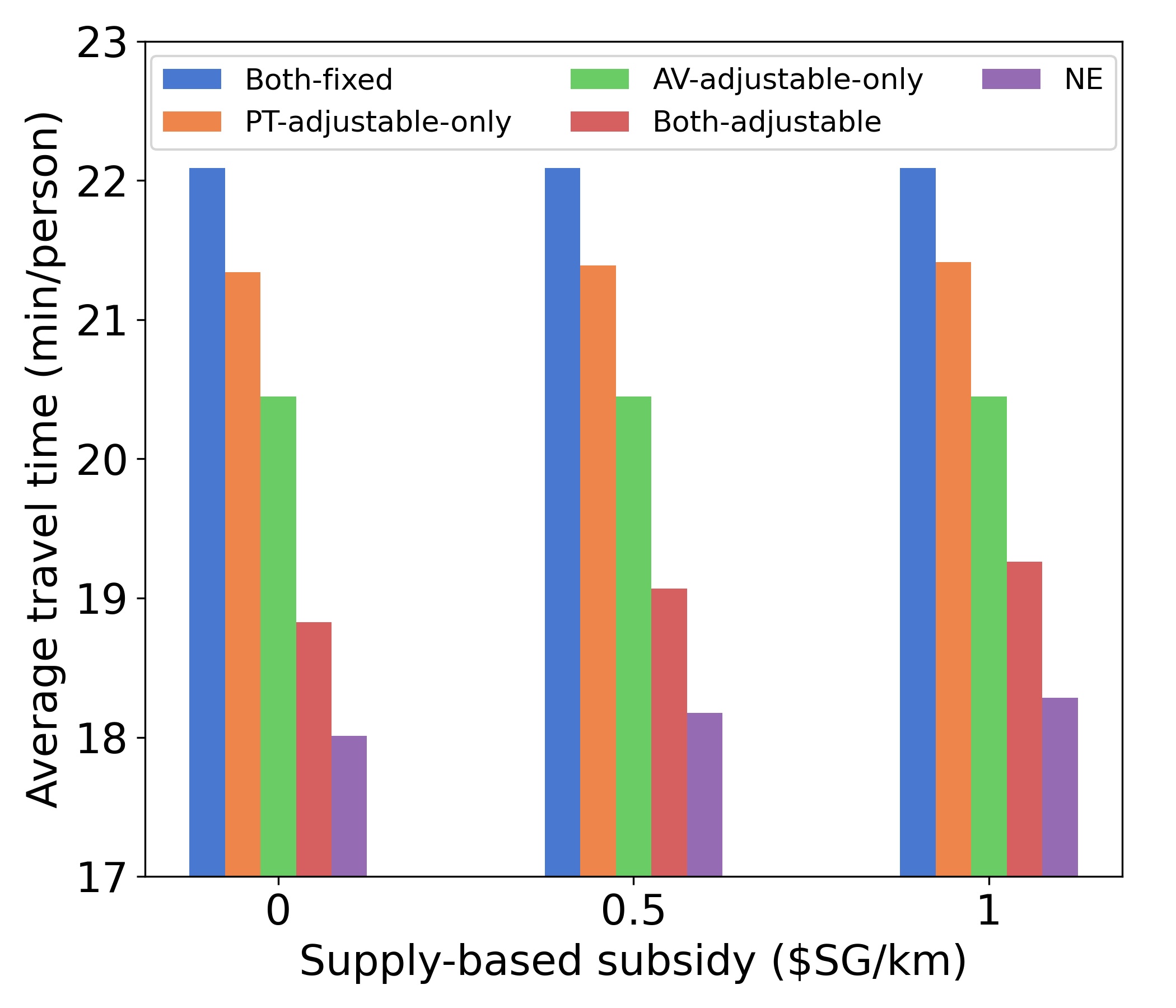}}
\hfil
\subfloat[Waiting time]{\includegraphics[width=0.4\textwidth]{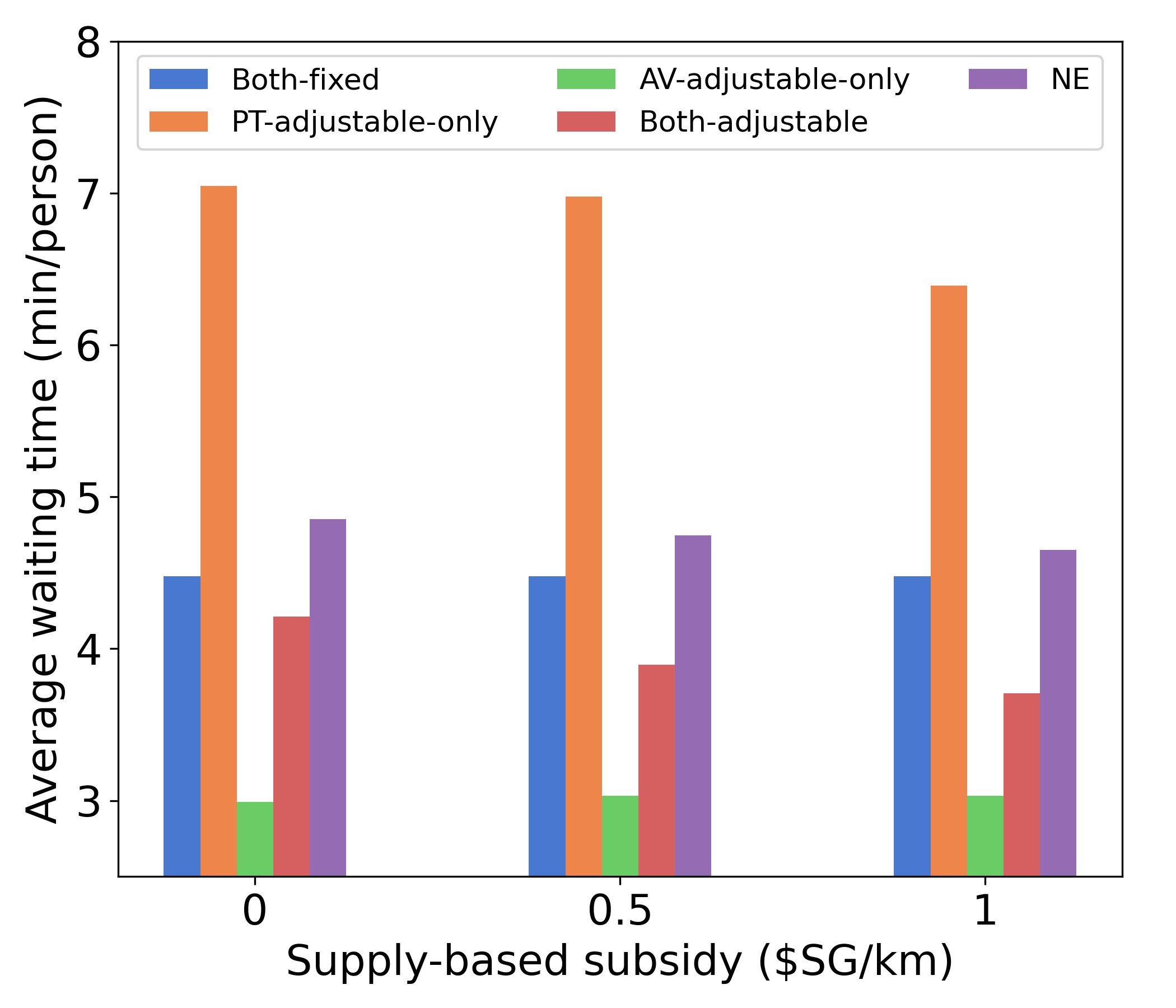}}
\hfil
\subfloat[Generalized travel cost]{\includegraphics[width=0.4\textwidth]{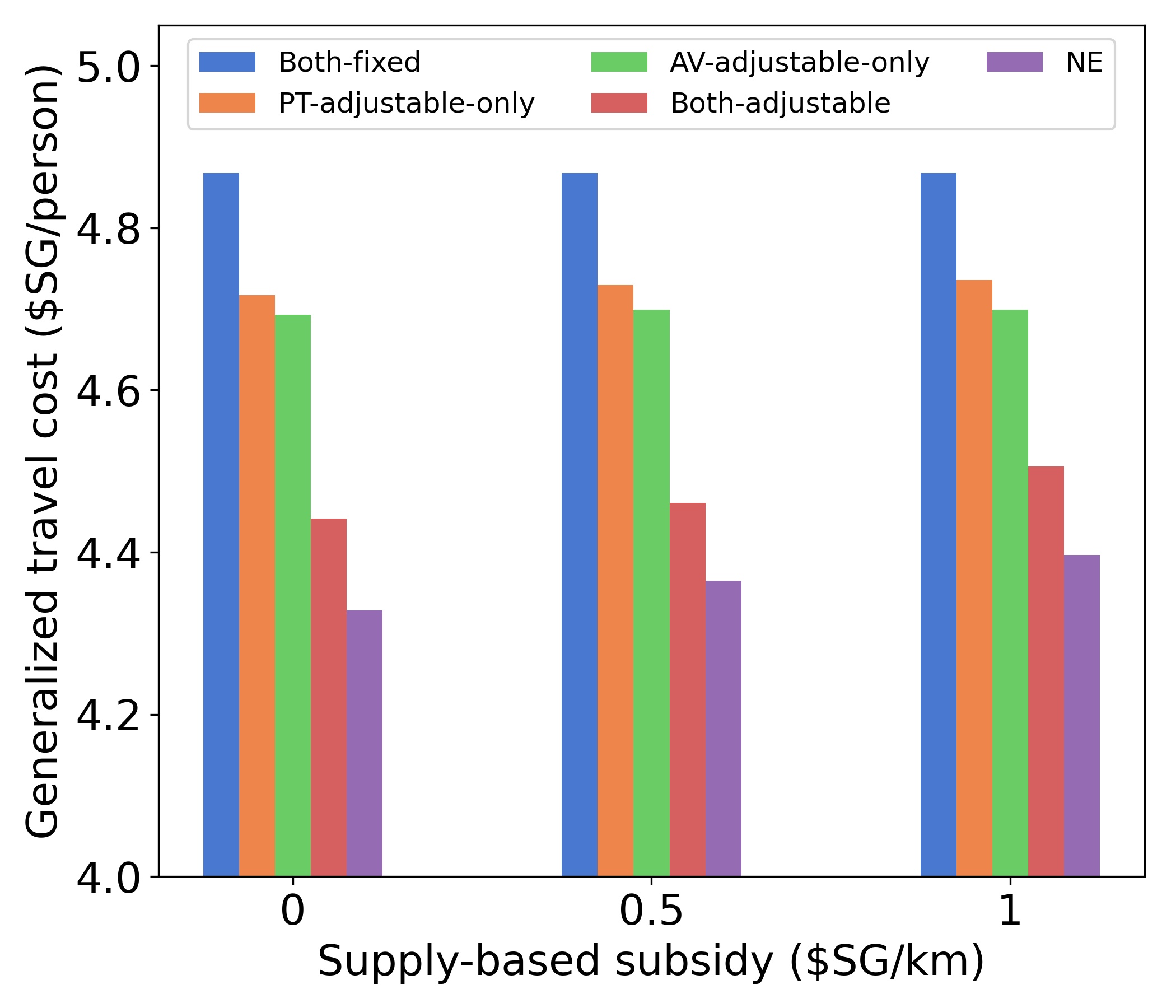}}
\caption{Impact of supply-based subsidies on passengers}
\label{fig_pax_supply_based_subsidy}
\end{figure}

\begin{figure}[H]
\centering
\subfloat[AV average load]{\includegraphics[width=0.33\textwidth]{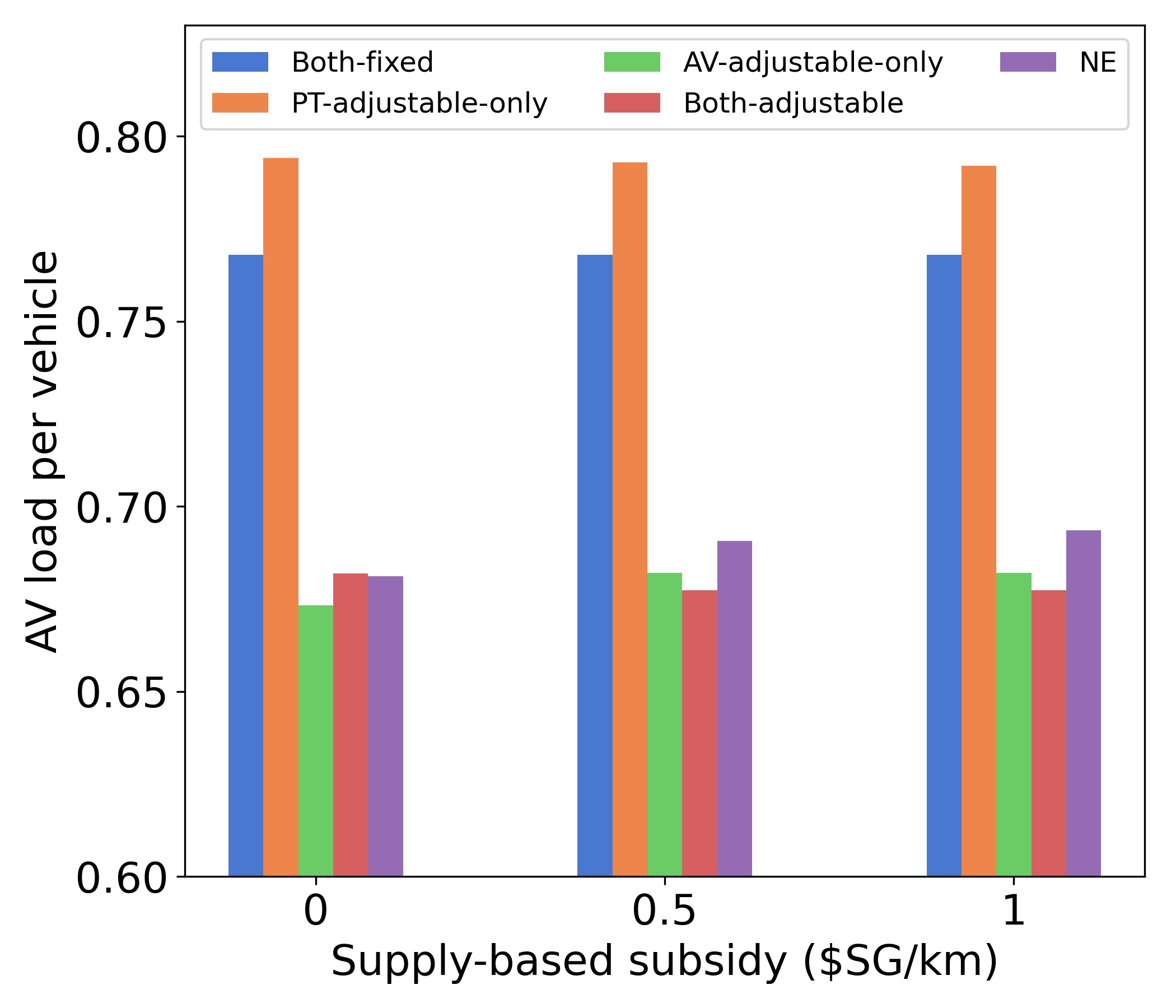}}
\hfil
\subfloat[Bus average load]{\includegraphics[width=0.33\textwidth]{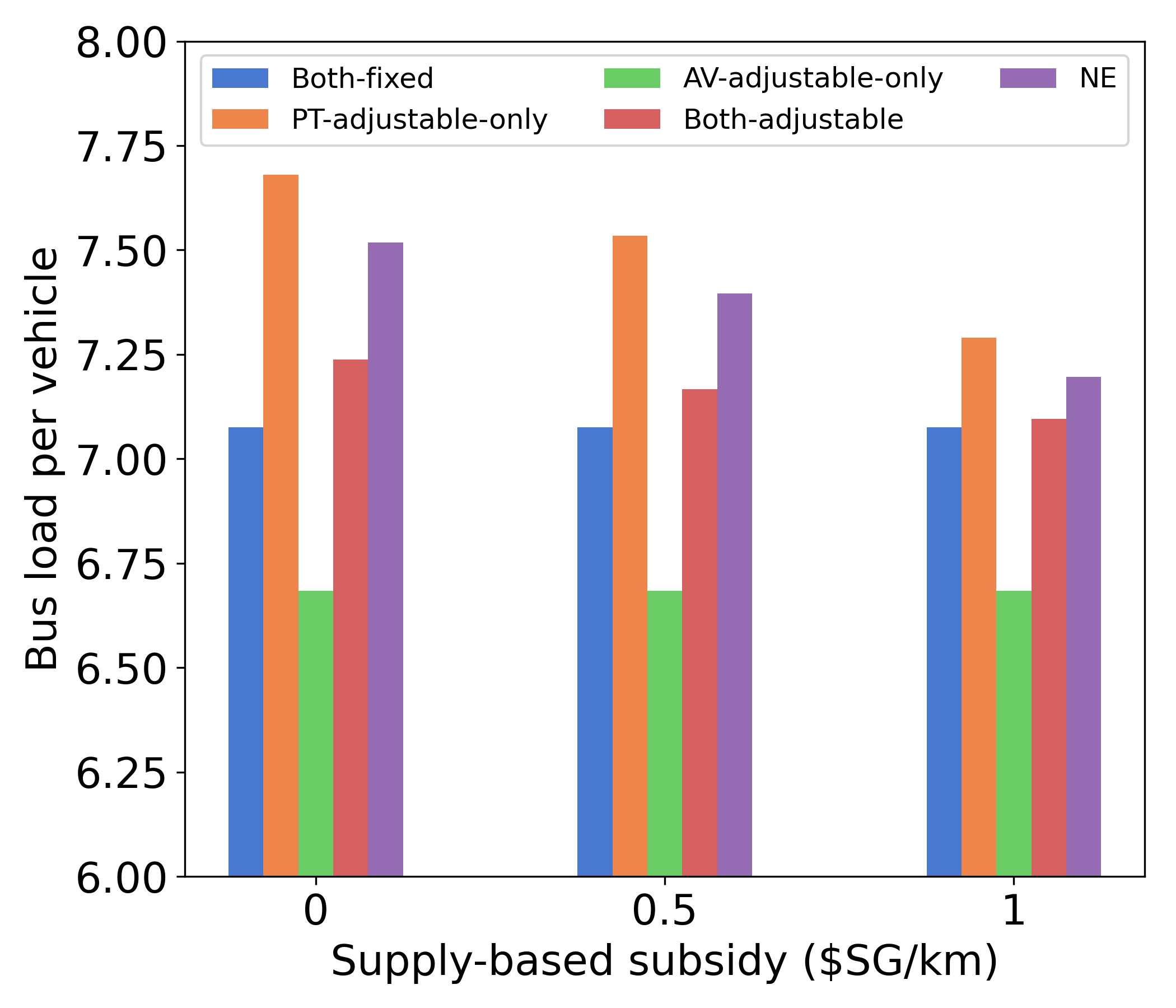}}
\hfil
\subfloat[Total PCE (bus + AV)]{\includegraphics[width=0.33\textwidth]{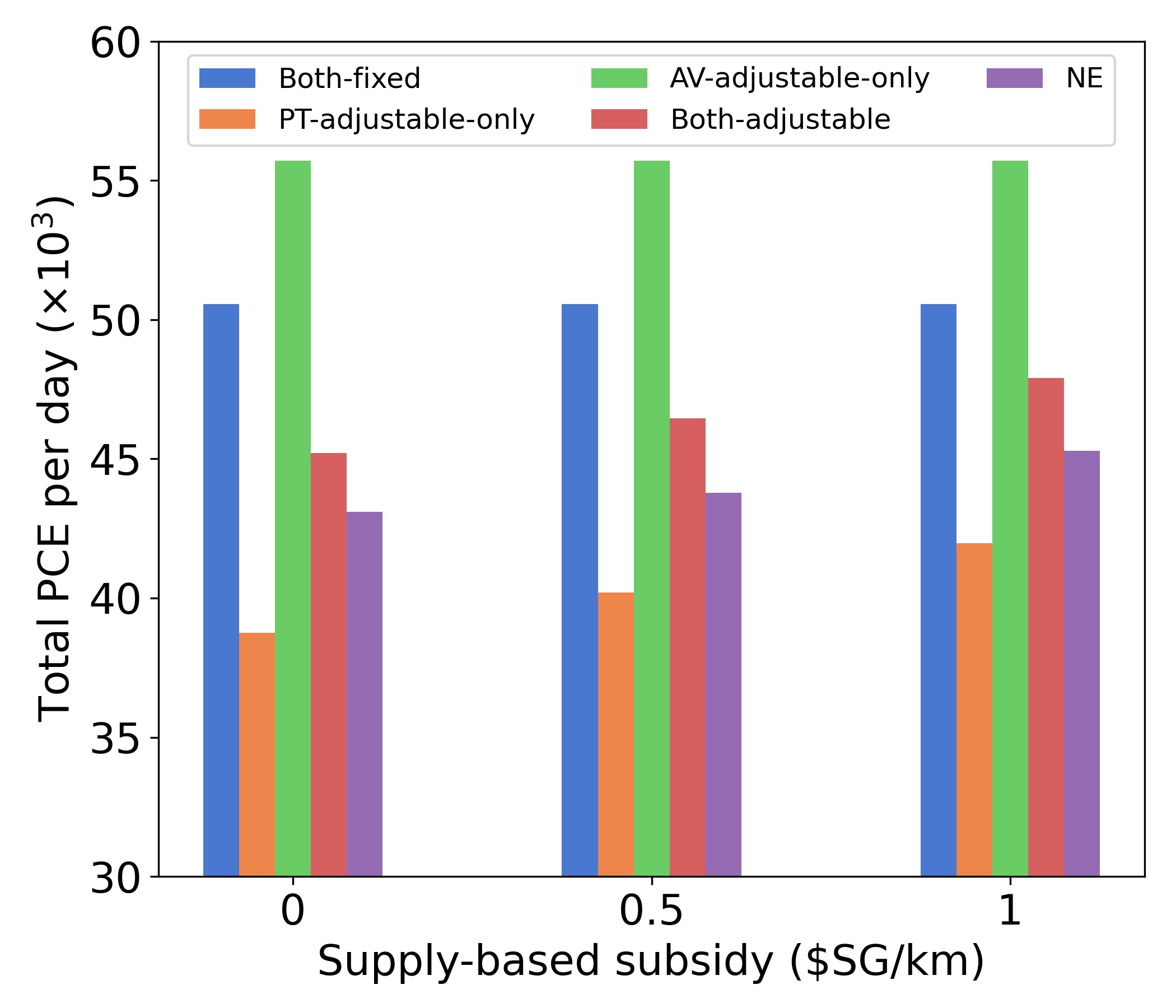}}
\caption{Impact of supply-based subsidies: transport authority perspective}
\label{fig_TA_supply_based_subsidy}
\end{figure}

\section{Mixed logit model for passenger mode choice} \label{choicemodel}
The mixed logit model is conducted based on an AV preference survey in Singapore \citep{shen2019built}. The deterministic part of the utility function $V_{njt}$ for individual $n$ choosing alternative $j$ in choice situation $t$ is given as
\begin{flalign}
V_{njt}=\alpha_{nj}+\beta_{nj}T_{njt}+\delta_{nj}X_n,
\label{eq_choice_uti}
\end{flalign}
where $T_{njt}$  is the vector of trip specific attributes of mode $j$ for individual $n$ in situation $t$; $X_n$ is the vector of sociodemographic variables of individual $n$; $\alpha_{nj}$ is the alternative specific constant to estimate the inherent preference of individual $n$ on mode $j$; $\beta_{nj}$ and $\delta_{nj}$ are the corresponding coefficients to be estimated. The situation $t$ is introduced because the survey contains panel questions.

The results for the mixed logit model are shown in Table \ref{tab_mixed_logit}, where for each parameter, we estimate both the mean and the standard deviation. 
\setcounter{table}{0}
\begin{table}[htb]
\caption{Estimation Results of Mixed Logit Model}
\label{tab_mixed_logit}
\centering
\footnotesize
\begin{threeparttable}
\begin{tabular}{l|lrrl} \hline
Variables & Parameters & value & t-test &\\\hline
\emph{Alternative Specific Constant} & & & &\\
Walk & Mean & fixed at 0 & &\\
 &     Std. & fixed at 0 & &\\
Bus & Mean & -0.569  & -2.11 & **\\
 &     Std. & 0.818 & 1.89 & *\\
On-demand AV &     Mean &     -0.568  & -2.56 & **\\
 &     Std. & 0.758 &  3.72 & ***\\ \hline
 \emph{Generalized travel cost} & & & &\\
Walk: Walking time (min) &         Mean &         -0.363 &     -28.20 &     ***\\
 &         Std. & 0.171  &     22.26 &     ***\\
Bus: Travel cost (SGD) &    Mean &    -1.14  &-8.86 &***\\
 &     Std. &     0.436 &0.16&\\
Bus: In-vehicle time (min) &     Mean &     -0.212  & -12.10 & ***\\
 &     Std. &     0.174  & 9.16 & ***\\
Bus: Waiting time (min)     & Mean     & -0.271  & -10.27 & ***\\
 &     Std. &     0.223  & 5.31 & ***\\
Bus: Walking time to bus stop (min)     & Mean     & -0.214  & -10.61 & ***\\
 &     Std. &     0.140  & 4.14 & ***\\
On-demand AV: Travel cost (SGD) &     Mean &     -0.984  & -18.56 & ***\\
 &     Std. &     0.465  & 13.54 & ***\\
On-demand AV: In-vehicle time (min) &     Mean &     -0.195  & -11.0 & ***\\
 &     Std. &     0.0288  & 1.16 & \\
On-demand AV: Waiting time (min) &     Mean &     -0.222 & -8.50 & ***\\
 &     Std. &     0.0310  & 0.58 & \\
\hline
 \emph{Sociodemographic variables} & & & &\\
 \multirow{2}{*}{\tabincell{l}{On-demand AV: household monthly income lower than SGD 4,000}} &     Mean &     -0.497  & -2.81 & ***\\
&    Std.&    0.300& 0.62 &    \\
\hline
\emph{Statistical summary} & & & &\\
Number of individuals    &&    1,242&&\\
Number of observations    &&    8,689&&\\
Number of random draws    &&    5,000&&\\
Initial log-likelihood at zero    &&    -10832.448&&\\
Final log-likelihood    &&    -6581.302&&\\
Adjusted McFadden $\rho^2$    &&    0.390&&\\
\hline
\end{tabular}
\begin{tablenotes}\footnotesize
\item *: $p<0.1$; **: $p<0.05$; ***: $p<0.01$. Std.: standard deviation.
\end{tablenotes}
\end{threeparttable}
\end{table}

\end{document}